\documentclass[11pt,a4paper]{article}
\usepackage{jcappub}
\pdfoutput=1

\usepackage{bm}
\usepackage{amsfonts}
\usepackage{subfigure}

\renewcommand\({\left(}
\renewcommand\){\right)}
\renewcommand\[{\left[}
\renewcommand\]{\right]}

\newcommand{\bv}{{\bf v}}
\newcommand{\br}{{\bf r}}
\newcommand{\bk}{{\bf k}}
\newcommand{\bx}{{\bf x}}
\newcommand{\ba}{{\bf a}}
\newcommand{\bb}{{\bm\beta}}

\newcommand{\GF}{G_{\rm F}}
\newcommand{\bmax}{\beta_{\rm max}}
\newcommand{\half}{{\textstyle\frac{1}{2}}}

\long\def\dump#1{}

\begin{document}

%%%%%%%%%%%%%%%%%%%%%%%%%%%%%%%%%%%%%%%%%%%%%%%%%%%%%%%%%%%%%%%%%%%%%%%%%%%%%%
% Frontpage %%%%%%%%%%%%%%%%%%%%%%%%%%%%%%%%%%%%%%%%%%%%%%%%%%%%%%%%%%%%%%%%%%
%%%%%%%%%%%%%%%%%%%%%%%%%%%%%%%%%%%%%%%%%%%%%%%%%%%%%%%%%%%%%%%%%%%%%%%%%%%%%%

\title{Self-induced flavor conversion of supernova neutrinos on small scales}

\author[a]{S.~Chakraborty,}
\author[b]{R.~S.~Hansen,}
\author[a]{I.~Izaguirre,}
\author[a]{and G.~G.~Raffelt}

\affiliation[a]{Max-Planck-Institut f{\"u}r Physik
(Werner-Heisenberg-Institut),\\
F{\"o}hringer Ring 6, 80805 M{\"unchen}, Germany}

\affiliation[b]{Department of Physics and Astronomy,
University of Aarhus,
8000 Aarhus C, Denmark}

\emailAdd{sovan@mpp.mpg.de}
\emailAdd{rshansen@phys.au.dk}
\emailAdd{izaguirr@mpp.mpg.de}
\emailAdd{raffelt@mpp.mpg.de}

\abstract{Self-induced flavor conversion of supernova (SN) neutrinos
  is a generic feature of neutrino-neutrino dispersion. The
  corresponding run-away modes in flavor space can spontaneously break
  the original symmetries of the neutrino flux and in particular can
  spontaneously produce small-scale features as shown in recent
  schematic studies. However, the unavoidable ``multi-angle matter
  effect'' shifts these small-scale instabilities into regions of
  matter and neutrino density which are not encountered on the way out
  from a SN. The traditional modes which are uniform on the largest
  scales are most prone for instabilities and thus provide the most
  sensitive test for the appearance of self-induced flavor conversion.
  As a by-product we clarify the relation between
  the time evolution of an expanding neutrino gas and the radial
  evolution of a stationary SN neutrino flux. Our results depend on several
  simplifying assumptions, notably stationarity of the solution, the
  absence of a ``backward'' neutrino flux caused by residual
  scattering, and global spherical symmetry of emission.}

\maketitle

\vskip1.0cm

%%%%%%%%%%%%%%%%%%%%%%%%%%%%%%%%%%%%%%%%%%%%%%%%%%%%%%%%%%%%%%%%%%%%%%%%%%%%%%
\section{Introduction}
%%%%%%%%%%%%%%%%%%%%%%%%%%%%%%%%%%%%%%%%%%%%%%%%%%%%%%%%%%%%%%%%%%%%%%%%%%%%%%

Core-collapse supernovae (SNe) or neutron-star mergers are powerful neutrino
sources and probably the only astrophysical phenomena where these elusive
particles are dynamically important and crucial for nucleosynthesis
\cite{Janka:2012wk, Burrows:2012ew}. The low energies of some tens of MeV
imply that $\beta$ reactions of the type $\nu_e+n\leftrightarrow p+e^-$ and
$\bar\nu_e+p\leftrightarrow n+e^+$ are the dominant charged-current
processes. Heavy-lepton neutrinos $\nu_\mu$, $\bar\nu_\mu$, $\nu_\tau$, and
$\bar\nu_\tau$, in this context often collectively referred to as $\nu_x$,
interact only by neutral-current processes. Therefore, neutrino energy
transfer and the emitted fluxes depend on flavor and one may think that
flavor oscillations are an important ingredient. However, the large matter
effect in this dense environment implies that eigenstates of propagation and
those of interaction very nearly coincide \cite{Wolfenstein:1977ue,
Wolfenstein:1979ni}. In spite of large mixing angles, flavor oscillations are
irrelevant except for MSW conversion when neutrinos pass the resonant density
as they stream away \cite{Mikheev:1986gs, Mikheev:1986if}. Therefore, the
neutrino signal from the next galactic SN may carry a detectable imprint of
the yet unknown neutrino mass hierarchy \cite{Dighe:1999bi, Dighe:2003jg,
Borriello:2012zc, Abbasi:2011ss, Serpico:2011ir}.

This picture can fundamentally change when the refractive effect of neutrinos
on each other is included \cite{Pantaleone:1992eq, Pantaleone:1994ns,
Sigl:1992fn}. The mean field representing background neutrinos can have
flavor off-diagonal elements (``off-diagonal refractive index'') due to
flavor coherence caused by oscillations and can lead to strong flavor
conversion effects \cite{Samuel:1993uw, Kostelecky:1993dm, Samuel:1995ri,
Sawyer:2005jk, Duan:2005cp, Duan:2006an, Hannestad:2006nj, Fogli:2007bk,
Balantekin:2006tg, Raffelt:2007yz, EstebanPretel:2007yq,
EstebanPretel:2008ni, Dasgupta:2009mg, Chakraborty:2009ej, Duan:2010bg,
Friedland:2010sc, Raffelt:2010za, Banerjee:2011fj, Dasgupta:2011jf,
Galais:2011gh, Pehlivan:2011hp, Raffelt:2011yb, Chakraborty:2011nf,
Chakraborty:2011gd, Sarikas:2011am, Cherry:2012zw, Cherry:2013mv,
Sarikas:2012vb, Sarikas:2012ad, Duan:2014mfa, Raffelt:2013rqa,
Raffelt:2013isa, Mirizzi:2013rla, Mirizzi:2013wda, Chakraborty:2014nma,
Hansen:2014paa, Mangano:2014zda, Duan:2014gfa, Mirizzi:2015fva,
Mirizzi:2015hwa}. (We ignore additional effects that would arise from
non-standard neutrino interactions \cite{Blennow:2008er}, spin-flip effects
caused by neutrino magnetic dipole moments~\cite{Giunti:2014ixa,
deGouvea:2012hg, deGouvea:2013zp}, by refraction in inhomogeneous or
anisotropic media~\cite{Studenikin:2004bu, Vlasenko:2013fja,
Cirigliano:2014aoa, Vlasenko:2014bva}, or the role of neutrino-antineutrino
pair correlations~\cite{Volpe:2013uxl, Vaananen:2013qja, Serreau:2014cfa,
Kartavtsev:2015eva}.) Self-induced flavor conversion preserves the global
flavor content of the ensemble, but re-shuffles it among momentum modes or
between neutrinos and antineutrinos. The simplest example would be a gas of
$\nu_e$ and $\bar\nu_e$ converting to $\nu_\mu$ and $\bar\nu_\mu$, leaving
the overall flavor content unchanged. The interacting modes of the neutrino
field can be seen as a collection of coupled oscillators in flavor space. The
eigenmodes of this interacting system include collective harmonic
oscillations, but can also include run-away solutions (instabilities) which
lead to self-induced flavor conversion \cite{Banerjee:2011fj}. Under which
physical conditions will instabilities occur, how can we visualize them, and
how will they affect the flavor composition of neutrinos propagating in the
early universe or stream away from a SN core?

To study these questions, many simplifications were used and especially
symmetry assumptions were made to reduce the dimensionality of the problem.
However, symmetry assumptions suppress those unstable solutions which break
the assumed symmetry. Therefore, when instabilities are the defining feature
of the dynamics, symmetry assumptions about the solutions can lead to
misleading conclusions because, even if the system was set up in a symmetric
state, the interacting ensemble can break this symmetry spontaneously. This
behavior is analogous to the hydrodynamical aspects of SN physics which
cannot show convective overturn if the simulation is spherically symmetric,
yet such 3D effects are now understood to be crucial for SN physics
\cite{Blondin:2002sm, Nordhaus:2010uk, Hanke:2011jf, Dolence:2012kh,
Murphy:2012id, Couch:2013kma, Couch:2013coa, Couch:2014kza, Takiwaki:2013cqa,
Tamborra:2013laa, Tamborra:2014hga, Tamborra:2014aua, Lentz:2015nxa,
Melson:2015tia, Melson:2015spa, Chakraborty:2014lsa}.

Our present concern is the question of ``spatial spontaneous symmetry
breaking'' in self-induced flavor conversion. In previous studies, the flavor
content of neutrinos streaming away from a SN core was taken to remain
uniform in the transverse directions. However, recent studies of simplified
systems suggest that this symmetry can be spontaneously
broken~\cite{Duan:2014gfa, Mangano:2014zda, Mirizzi:2015fva,
Mirizzi:2015hwa}.  To avoid the complication of global spherical coordinates,
it is sufficient to model the neutrino stream at some distance with
plane-parallel geometry, i.e., we can use wave vectors $\bk$ in the
transverse plane to describe small-scale spatial variations. In this
terminology, traditional studies only considered $\bk=0$ (global spherical
symmetry). In analogy to this $\bk=0$ case it was found that for any $\bk$
there is some range of effective neutrino densities where unstable solutions
exist~\cite{Duan:2014gfa, Mangano:2014zda, Mirizzi:2015fva, Mirizzi:2015hwa}.
We usually express the neutrino density in terms of an effective
neutrino-neutrino interaction energy $\mu=\sqrt{2}\GF n_{\nu_e}(r)\,(R/r)^2$,
where $n_{\nu_e}(r)$ is the $\nu_e$ density at distance $r$ and $R$ is some
reference radius playing the role of the neutrino sphere. The parameter $\mu$
varies with $r^{-4}$ because the neutrino density decreases as $r^{-2}$ with
distance. In this terminology, for any $\bk$ there is a range $\mu_{\rm
min}<\mu<\mu_{\rm max}$ where the system is unstable. For larger $\bk$
(smaller spatial scales), the instability range shifts to larger $\mu$, i.e.,
to regions closer to the SN core.  This finding suggests that the neutrino
stream is never stable because at any neutrino density there is some range of
unstable ${\bf k}$ modes.

\begin{figure}[ht]
\centering
\includegraphics[width=0.80\textwidth]{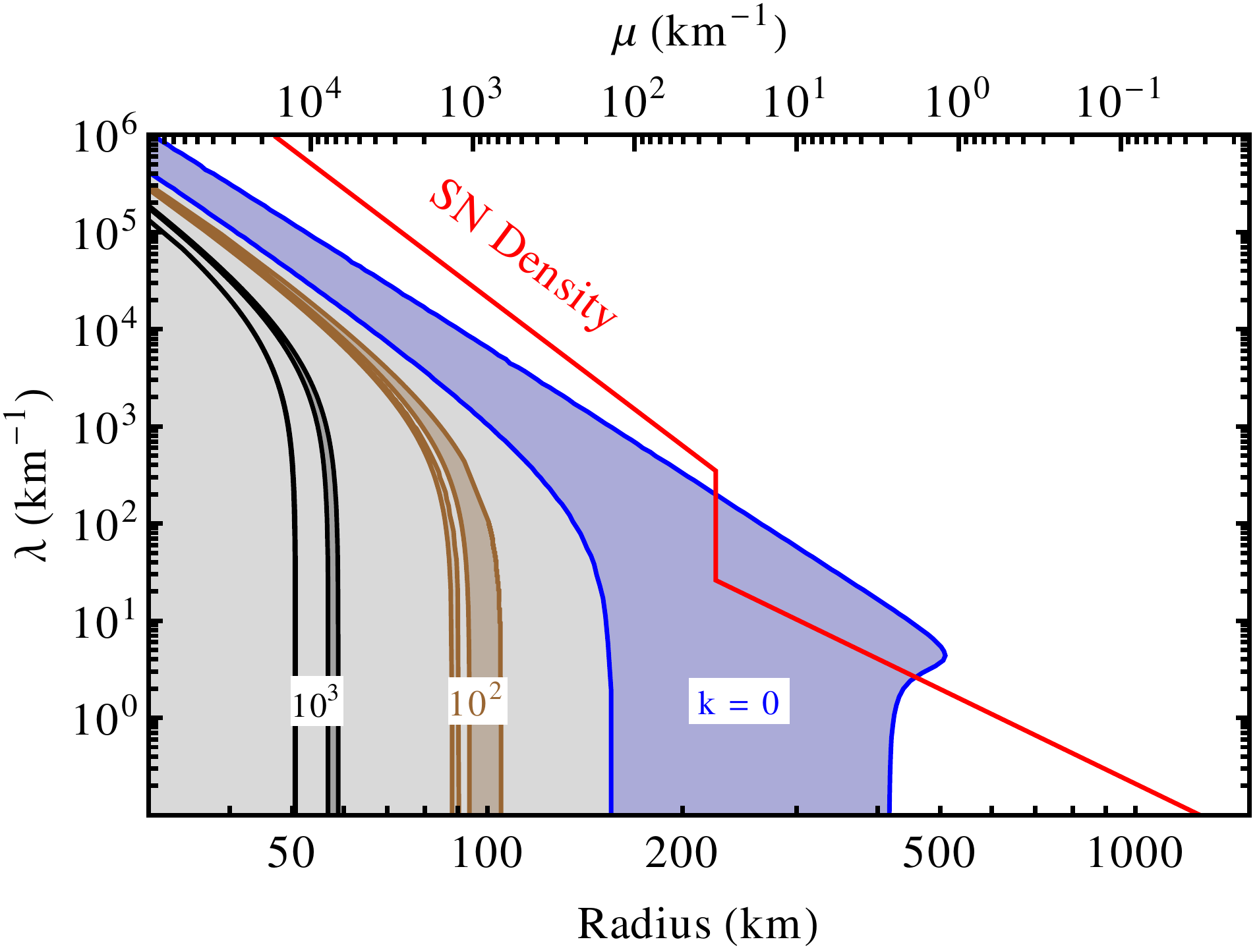}
\caption{Footprint of the MAA instability region
in the parameter space of effective neutrino density
$\mu=\sqrt{2}\GF n_{\nu_e}(R/r)^2$, where $R$ is the neutrino-sphere
radius, and matter density $\lambda=\sqrt{2}\GF n_{_e}(R/r)^2$
for the schematic SN model described in the text. Because $\mu\propto r^{-4}$,
the horizontal axis is equivalent to the distance from the SN as indicated
on the lower horizontal axis.
We also show a representative schematic
SN density profile where the sharp density drop
marks the shock wave. We also show the instability footprint explicitly for
co-moving wave numbers $k=10^2$ and $k=10^3$ in units of the vacuum oscillation
frequency. Notice that for the same value of $k$ there are two separate
instability strips. The collection of all small-scale instabilities fill
the gray-shaded region below the traditional $k=0$ (blue shaded) instability
region, whereas they leave the space above untouched.} \label{fig:firstfootprint}
\end{figure}

However, such conclusions may be premature as one also needs to include the
refractive effect of matter which also tends to shift the instability to
larger $\mu$, a phenomenon termed ``multi-angle matter suppression'' of the
instability \cite{EstebanPretel:2008ni, Chakraborty:2011nf}.  We usually express the
multi-angle matter effect in terms of
the parameter $\lambda=\sqrt{2}\GF n_e(r)\,(R/r)^2$, where $n_e(r)$ is the
electron density at distance $r$. One should study the instability region in
the two-parameter space of effective matter and neutrino densities, $\lambda$
and $\mu$, which we call the ``footprint of the instability.'' In
figure~\ref{fig:firstfootprint} we show as an example the footprint of the
MAA instability for a schematic SN model (MAA stands for ``multi azimuth
angle,'' i.e., one type of instability). We always define the instability region
by the requirement that the growth rate $\kappa>10^{-2}\,\omega_0$, where $\omega_0$ is
a typical vacuum oscillation frequency.

Our schematic SN model consists of neutrinos and antineutrinos with a single
energy corresponding to a vacuum oscillation frequency $\omega_0=\Delta
m^2/2E=1~{\rm km}^{-1}$. We assume they are emitted isotropically at the
neutrino-sphere radius $R=30$~km. We consider two-flavor oscillations and,
after subtracting the $\nu_x$ flux, we are left with a $\nu_e$ and
$\bar\nu_e$ flux such that there emerge twice as many $\nu_e$ than
$\bar\nu_e$. Finally, we assume that the effective neutrino-neutrino
interaction energy, $\mu=\sqrt{2} \GF n_{\nu_e}(R/r)^2$, at the neutrino
sphere $r=R$ is $\mu_0=10^5~{\rm km}^{-1}$. Under these circumstances,
run-away solutions for $k=0$ exist within the footprint area shown in
figure~\ref{fig:firstfootprint} as a blue-shaded region. We also show, as a
solid red line, a possible density profile, corresponding to electron density
as function of radius. The sudden density drop at $r\approx200$~km is the
shock wave. In this example, the density profile does not intersect with the
instability footprint for radii below the shock wave. On the other hand, for
larger radii, collective flavor conversion would begin. We further show the
footprint for inhomogeneities with assumed wave-number $k=10^2$ and $k=10^3$,
measured in units of the vacuum oscillation frequency $\omega_0$. Notice that
we define $k$ in ``co-moving'' coordinates along the radius, i.e., a fixed
$k$ represents a fixed angular scale, not a fixed length scale. Notice also
that for the MAA instability shown here, which is relevant for normal mass
ordering, non-zero $k$ values lead to {\it two} instability regions. This
phenomenon does not arise for the bimodal instability.

The full range of all $k$-values inevitably fills the entire region below the
$k=0$ mode (blue shading) in this plot, i.e., the entire gray-shaded region
is unstable, whereas the region above the blue-shaded part remains unscathed.
In other words, essentially the largest-scale mode with $k=0$ is the ``most dangerous''
mode. If this mode is stable on the locus of the SN density profile in
figure~\ref{fig:firstfootprint}, the higher-$k$ modes are stable as well. Of
course, if any instability is encountered by the physical SN density profile,
these instabilities will span a range of scales and create complicated flavor
conversion patterns.

The rest of our paper is devoted to substantiating this main point and to
explain our exact assumptions. We stress that our simplifications may be too
restrictive to provide a reliable proxy for a realistic SN. In particular, we
assume stationary neutrino emission and that the solution is stationary as
well, i.e., we assume that the evolution can be expressed as a function of
distance from the surface alone.  We also ignore the ``halo flux'' caused by
residual scattering which can be a strong effect. Our study would not be
applicable at all in regions of strong scattering, i.e., below the neutrino
sphere. We assume that the original neutrino flux is homogeneous and
isotropic in the transverse directions, i.e., global spherical symmetry of
emission at the neutrino sphere. It has not yet been studied if this
particular assumption has any strong impact on the stability question, i.e.,
if violations of such an ideal initial state substantially change the
instability footprint, or if such disturbances would simply provide seeds for
instabilities to grow. It is impossible to understand and study all effects
at once, so here we only attempt to get a grasp of the differential impact of
including spatial inhomogeneities in the form of self-induced small-scale
flavor instabilities. All the other questions must be left for future
studies.

%%%%%%%%%%%%%%%%%%%%%%%%%%%%%%%%%%%%%%%%%%%%%%%%%%%%%%%%%%%%%%%%%%%%%%%%%%%%%%
\section{Equations of motion}
\label{sec:EOM}
%%%%%%%%%%%%%%%%%%%%%%%%%%%%%%%%%%%%%%%%%%%%%%%%%%%%%%%%%%%%%%%%%%%%%%%%%%%%%%

Beginning from the full equation of motion for the neutrino
density matrices in flavor space, we develop step-by-step the simplified
equations used in our linearized stability analysis. In particular, we
formulate the stationary, spherical SN problem, where the flavor evolution is
a function of radius, as an equivalent time-dependent 2D problem in the
tangential plane. A fixed neutrino speed in the tangential plane corresponds
to the traditional ``single angle'' treatment, whereas neutrino speeds taking
on values between
0 and a maximum, determined by the distance from the
neutrino sphere, corresponds to the traditional ``multi angle'' case.

\subsection{Setting up the system}

We describe the neutrino field in the usual way by $3{\times}3$ flavor
matrices $\varrho(t,\br,E,\bv)$, where the diagonal elements are occupation
numbers for the different flavors, whereas the off-diagonal elements contain
correlations among different flavor states of equal momentum. We follow the
convention where antineutrinos are described by negative energy $E$ and the
corresponding matrix includes a minus sign, i.e., it is a matrix of negative
occupation numbers.

We always work in the free-streaming limit, ignoring neutrino collisions. In
this case, neutrino propagation is described by the commutator equation
\cite{Sigl:1992fn, Raffelt:2013rqa}
\begin{equation}\label{eq:EOM1}
i(\partial_t+\bv\cdot{\bm\nabla}_\br)\varrho=[{\sf H},\varrho]\,,
\end{equation}
where $\varrho$ and ${\sf H}$ are functions of $t$, $\br$, $E$, and $\bv$.
The Hamiltonian matrix is
\begin{equation}
{\sf H}=\frac{{\sf M}^2}{2E}+\sqrt2\GF
\[{\sf N}_\ell+\int d\Gamma'\,
\frac{(\bv-\bv')^2}{2}\,\varrho_{t,\br,E',\bv'}\],
\end{equation}
where ${\sf M}^2$, the matrix of neutrino mass-squares, is what causes vacuum
oscillations. The matrix of charged-lepton densities, ${\sf N}_\ell$,
provides the usual Wolfenstein matter effect. The integration $d\Gamma'$ is
over the neutrino and antineutrino phase space. Because antineutrinos are
denoted with negative energies, we have explicitly $\int
d\Gamma'=\int_{-\infty}^{+\infty}dE'E'^2\int d{\bv}'/(2\pi)^3$ and the
velocity integration $d\bv'$ is over the unit sphere. Because the neutrino
speed $|\bv|=1$ we were able, for later convenience, to write the
current-current velocity factor in the unusual form
$(1-\bv\cdot\bv')=\frac{1}{2}(\bv-\bv')^2$.

Studying this 7-dimensional problem requires significant simplifications. For
neutrino oscillations in the early universe, one will usually assume initial
conditions at some time $t=0$ and then solve these equations as a function of
time. To include spatial variations, one may Fourier transform these
equations in space, replacing the spatial dependence on $\br$ by a
wave-number dependence $\bk$, whereas $\bv\cdot{\bm\nabla}_\br\to
i\bv\cdot\bk$ and the r.h.s.\ becomes a convolution of Fourier modes
\cite{Mangano:2014zda}. One can then perform a linearized stability analysis
for every mode $\bk$ and identify when modes of
different wave number are unstable and lead to self-induced flavor
conversion~\cite{Duan:2014gfa}. One can also use this representation for
numerical studies~\cite{Mangano:2014zda, Mirizzi:2015fva, Mirizzi:2015hwa}.

The other relatively simple case is inspired by neutrinos streaming from a
supernova (SN) core. One assumes that, on the relevant time scales, the
source is stationary and that the solution is stationary as well, so that
$\partial_t\to 0$. In addition, one assumes that neutrinos stream only away
from the SN so that it makes sense to ask about the variation of the neutrino
flavor content as a function of distance, assuming we are provided with
boundary conditions at some radius $R$ which we may call the neutrino sphere.
Actually, this description can be a poor proxy for a real SN because the
small ``backward'' flux caused by residual neutrino scattering in the outer
SN layers, the ``halo flux,'' can be surprisingly important for
neutrino-neutrino refraction because of its broad angular range
\cite{Cherry:2012zw, Cherry:2013mv, Sarikas:2012vb}. Here we will ignore this
issue and use the simple picture of neutrinos streaming only outward.

We stress that this simplification is the main limitation of our study and
its interpretation in the physical SN context. If (self-induced)
instabilities exist on small spatial scales, they could even exist below the
neutrino sphere where the picture of neutrinos streaming only in one
direction would be very poor. The approach taken here to reduce the
7-dimensional problem to a manageable scope may then hide the crucial
physics. Therefore, our case study leaves open important questions about a
real SN.

\subsection{Large-distance approximation}

The main point of our study is to drop the assumption of spatial
uniformity, i.e., we include variations transverse to the radial direction.
However, we are not interested in an exact description of large-scale modes.
At some distance from
the SN, outward-streaming neutrinos cannot communicate with others which
travel in some completely different direction as long as we only include
neutrino-neutrino refraction and not, for example, lateral communication by
hydrodynamical effects.  If we are only interested in relatively small
transverse scales, we may approximate a given spherical shell locally as a
plane, allowing us to use Cartesian coordinates in the transverse direction
rather than a global expansion in spherical harmonics.

We now denote with $z$ the ``radial'' direction, and use bold-faced
characters to denote vectors in the transverse plane, notably $\ba$ for the
coordinate vector in the transverse plane and $\bb$ the transverse velocity
vector. In the stationary limit, the equation of motion (EoM) becomes
\begin{equation}\label{eq:EOM2}
i(v_z\partial_z+\bb\cdot{\bm\nabla}_\ba)\varrho=[{\sf H},\varrho]\,,
\end{equation}
where $v_z=\sqrt{1-\bb^2}$ and $\varrho$ and ${\sf H}$ depend on $z$, $\ba$,
$E$, and $\bb$. If the neutrino-sphere radius is $R$, then at distance $r$
from the SN the maximum neutrino transverse velocity is $\bmax\approx
R/r\ll1$. This latter ``large distance approximation'' is the very
justification for using Cartesian coordinates in the transverse direction.

Therefore, it is self-consistent to expand the equations to order $\beta^2$.
(We need to go to quadratic order lest the neutrino-neutrino interaction term
vanishes.) The $\beta$ expansion is not necessary for our stability analysis,
but avoiding it does not provide additional precision and performing it
provides significant conceptual clarity. Numerical precision for a specific
SN model is not our goal, and in this case we would have to avoid modeling
the transverse direction as a flat space anyway, especially when considering
regions that are not very far away from the SN core.

In equation~\eqref{eq:EOM2} we multiply with $1/v_z\approx(1+\half\beta^2)$ and
notice that the gradient term remains unchanged if we expand only to order
$\beta^2$ so that
\begin{equation}\label{eq:EOM3}
i(\partial_z+\bb\cdot{\bm\nabla}_\ba)\varrho=[{\sf H},\varrho]\,,
\end{equation}
where the Hamiltonian matrix is the old one times $(1+\half\beta^2)$ or
explicitly
\begin{equation}
{\sf H}=\(1+\frac{\beta^2}{2}\)\(\frac{{\sf M}^2}{2E}+\sqrt2\GF
{\sf N}_\ell\)+\sqrt2\GF\int d\Gamma'\,\frac{(\bb-\bb')^2}{2}\,\varrho_{z,\ba,E',\bb'}\,.
\end{equation}
The flux factor under the integral in the second expression is also an expansion to
${\cal O}(\beta^2)$ in the form $1-v_zv_z'-\bb\cdot\bb'=
1-\sqrt{1-\beta^2}\sqrt{1-\beta'^2}-\bb\cdot\bb'\approx\half\beta^2+\half\beta'^2
-\bb\cdot\bb'=\half(\bb-\bb')^2$. Multiplying this expression with
$(1+\half\beta^2)$ makes no difference because it is already ${\cal O}(\beta^2)$.

As a next step, we re-label our variables and denote the radial direction $z$
as time $t$. Moreover, we rescale the transverse velocities as $\bb=\bv
\bmax$ where $\bv$ is now a 2D vector obeying $0\leq|\bv|\leq1$. Coordinate
vectors in the transverse plane are also rescaled as $\bx=\ba\bmax$, i.e.,
the new transverse coordinate vector $\bx$ is ``co-moving'' in that it
denotes a fixed angular scale relative to the SN. After these substitutions,
the EoMs are
\begin{equation}\label{eq:EOM4}
i(\partial_t+\bv\cdot{\bm\nabla}_\bx)\varrho=[{\sf H},\varrho]\,,
\end{equation}
with
\begin{equation}
{\sf H}=\(1+\frac{\bmax^2}{2}\,\bv^2\)\(\frac{{\sf M}^2}{2E}+\sqrt2\GF
{\sf N}_\ell\)+\sqrt2\GF\bmax^2
\int d\Gamma'\,\frac{(\bv-\bv')^2}{2}\,\varrho_{t,\bx,E',\bv'}\,.
\end{equation}
Of course, the neutrino phase-space integration $\int d\Gamma'$ is understood
in the new variables.

Our stationary 3D problem has now become a time-dependent 2D problem. In the
SN interpretation, the aspect ratio of the neutrino sphere shrinks with
distance and correspondingly, $\bmax$ shrinks. In other words, the physics is
analogous to neutrino oscillations in the expanding universe. In the SN case,
linear transverse scales grow as $r/R$, where $r$ is the distance to the SN,
i.e., our ``Hubble parameter'' is $R^{-1}$ where $R$ is the neutrino-sphere
radius and the scale factor grows linearly with ``time.'' The physical neutrino
density decreases with inverse-distance squared and in addition, the factor
$\bmax^2$ accounts for the decreasing value of the current-current factor in
the neutrino interaction. Therefore, the effective neutrino number density
decreases with $(\hbox{scale factor})^{-4}$ in the familiar way.

%Of course, our equations are only valid if all time scales related to
%neutrino oscillations are fast relative to the expansion speed, i.e., in the
%adiabatic limit. Moreover, transverse motions have a ``causal horizon'' in
%that different patches cannot communicate with each other if they are too
%far separated. We are aiming at a stability analysis which is local at some
%fixed distance from the SN, so these questions are not directly relevant for
%us. In this sense, we do not use an expanding space, but a static one with
%fixed choices of effective densities of neutrinos and background matter.

\subsection{Single angle vs.\ multi angle}

In the SN context, one often distinguishes between the single-angle and
multi-angle cases, referring to the zenith angle of neutrino emission at the
SN core. If all neutrinos were emitted with a fixed zenith angle, their
transverse speeds would be $|\bb|=\bmax$ and in our new variables, $|\bv|=1$.
In this case $(1+\half\bmax^2\bv^2)\to(1+\half\bmax^2)$ is simply a small and
negligible numerical correction to the vacuum oscillation frequencies and the
matter effect. Also, we can revert to the traditional form of the flux factor
$\half(\bv-\bv')^2=(1-\bv\cdot\bv')$. Therefore, the SN single-angle case is
equivalent, without restrictions, to a 2D neutrino gas evolving in time.
Therefore, neutrino oscillations in an expanding space (``early universe'')
is exactly equivalent to the single-angle approximation of neutrinos
streaming from a SN core with properly scaled effective neutrino and matter
densities.

It has been recognized a long time ago that in the single-angle case, the
ordinary matter effect has no strong impact on self-induced flavor conversion
\cite{Duan:2005cp}. As usual, one can go to a rotating coordinate system in
flavor space. In this new frame, the matrix of vacuum oscillation
frequencies, ${\sf M}^2/2E$, has fast-oscillating off-diagonal elements and,
in a time-averaged sense, it is diagonal in the weak-interaction basis. These
fast-oscillating terms are what kick-starts the instabilities at the
beginning of self-induced flavor conversion but are otherwise irrelevant. For
a larger matter effect, more $e$-foldings of exponential growth of the
instability are needed to ``go nonlinear.'' In this sense, matter has a
similar effect concerning the onset of the instability that would be caused
by reducing the mixing angle. These effects concern the perturbations which
cause the onset of instabilities, not the existence and properties of the
unstable modes themselves.

We may ignore the small correction to the vacuum oscillation frequency
provided by the factor $(1+\half\bmax^2\bv^2)$. We need to keep terms of
order $\beta^2$ in the context of the matter and neutrino-neutrino term which
in the interesting case are large and, after multiplication with $\bmax^2$,
still larger than the vacuum oscillation term. Therefore, we find
\begin{equation}\label{eq:Htwoflavor}
{\sf H}=\left\langle\frac{{\sf M}^2}{2E}\right\rangle+
\sqrt2\GF{\sf N}_\ell\bmax^2\,\frac{\bv^2}{2}+\sqrt2\GF\bmax^2
\int d\Gamma'\,\frac{(\bv-\bv')^2}{2}\,\varrho_{t,\bx,E',\bv'}\,,
\end{equation}
where the first term symbolizes the time-averaged vacuum term in the fast
co-rotating frame. In the single-angle case, where $\bv^2=1$ for all modes,
the remaining matter term can be rotated away as well.

The multi-angle SN case, in this representation, corresponds to a 2D neutrino
gas with variable propagation speed $0\leq|\bv|\leq1$, i.e., the velocity
phase space is not just the surface of the 2D unit sphere (a circle with
$|\bv|=1$), but fills the entire 2D unit sphere (a disk with $|\bv|\leq 1$).
There is no counterpart to this effect in a ``normal'' neutrino gas. The
early-universe analogy does not produce multi-angle effects, but we can
include them without much ado by allowing the neutrino velocities to fill the
2D unit sphere.

If neutrinos are emitted ``black-body like'' from a spherical surface, from a
distance this neutrino sphere looks like a disk of uniform surface
brightness, in analogy to the solar disk in the sky. Therefore, this
assumption corresponds to the neutrino transverse velocities filling the 2D
unit sphere uniformly. In earlier papers of our group, we have used the
variable $u=\bv^2$ with $0\leq u\leq1$ as a co-moving transverse velocity
coordinate, representing the neutrino zenith angle of emission. In terms of
this variable, the black-body like case corresponds to the familiar top-hat
$u$ distribution on the interval $0\leq u\leq1$.

Our overall set-up was inspired by that of Duan and Shalgar
\cite{Duan:2014gfa}, except that they consider only one transverse dimension
with single-angle emission at the SN. In other words, their system is
equivalent to two colliding beams evolving in time and allowing spatial
variations. A numerical study of this case, in both single and multi angle,
was very recently performed by
Mirizzi, Mangano and Saviano \cite{Mirizzi:2015fva}, going beyond the
linearized case. Not unexpectedly, the outcome of the nonlinear evolution
is found to be flavor decoherence. In our approach,
multi-angle effects in this colliding-beam system
can be easily included by using a 1D velocity distribution that fills
the entire interval $-1\leq v\leq+1$ and not just the two values
$v=\pm 1$. A black-body like zenith-angle distribution corresponds to a
uniform velocity distribution on this interval.

\subsection{Two-flavor system}

As a further simplification we limit our discussion to a two-flavor system
consisting of $\nu_e$ and some combination of $\nu_\mu$ and $\nu_\tau$ that
we call $\nu_x$, following the usual convention in SN physics. We are having
in mind oscillations driven by the atmospheric neutrino mass difference and
by the small mixing angle $\Theta_{13}$. The vacuum oscillation frequency is
\begin{equation}
\omega=\frac{\Delta m^2}{2E}=0.63~{\rm km}^{-1}\,\(\frac{10~{\rm MeV}}{E}\)\,,
\end{equation}
where we have used $\Delta m^2=2.5\times10^{-3}~{\rm eV}^2$. Henceforth we
will describe the neutrino energy spectrum by an $\omega$ spectrum instead,
with negative $\omega$ describing antineutrinos.

The matrix of vacuum oscillation frequencies, in the fast-rotating flavor
basis, takes on the diagonal form
\begin{equation}\label{eq:omega-definition}
\left\langle\frac{{\sf M}^2}{2E}\right\rangle\to
\omega\,\begin{pmatrix}+\half&0\\0&-\half\end{pmatrix}\,,
\end{equation}
where we have removed the part proportional to the unit matrix which drops
out of commutator expressions.
We do not include the fast-oscillating
off-diagonal elements which is irrelevant for the stability
analysis. The matter effect appears in a similar form,
\begin{equation}\label{eq:mattereffect1}
\sqrt2\GF{\sf N}_\ell\bmax^2\,\frac{\bv^2}{2}\to
\frac{\lambda \bv^2}{2}\,\begin{pmatrix}+\half&0\\0&-\half\end{pmatrix}\,,
\end{equation}
where again we have removed the piece proportional to the unit matrix.
The parameter $\lambda$ describing the multi-angle matter effect
at distance $r$ from the SN with neutrino-sphere radius $R$ and using
$\bmax=R/r$ is
\begin{equation}
\lambda=\sqrt2\GF n_e(r) \frac{R^2}{r^2}
=3.86\times10^8~{\rm km}^{-1}~\frac{Y_e(r)\,\rho(r)}{10^{12}~{\rm g}~{\rm cm}^{-3}}\,
\frac{R^2}{r^2}\,,
\end{equation}
where $n_e(r)$ is the net density of electrons minus positrons, $\rho(r)$ the
mass density, and $Y_e(r)$ the electron fraction per baryon, each at radius
$r$. The matter density drops steeply outside the neutrino sphere and
jumps downward by an order of magnitude at the shock-wave radius. Therefore,
we need to consider $\lambda$ values perhaps as large as some $10^7~{\rm km}^{-1}$ all
the way to vanishingly small values.

Turning to the neutrino-neutrino term, notice that the $\varrho$ matrices
play the role of occupation numbers and that the $\int d\Gamma$ integration
includes the entire phase space of occupied neutrino and antineutrino modes.
Therefore, ${\sf N}_{\nu}=\int d\Gamma \varrho$ is a flavor matrix of net
neutrino minus antineutrino number densities, in analogy to the
corresponding charged-lepton matrix ${\sf N}_\ell$. It is less obvious,
however, how to best define an effective neutrino-neutrino interaction
strength $\mu$ which plays an analogous role to $\lambda$. If we were to
study a system that initially consists of equal number densities of $\nu_e$
and $\bar\nu_e$, the matrix ${\sf N}_\nu$ vanishes, but later develops
off-diagonal elements. Therefore, we rather use the number density of $\nu_e$
without subtracting the antineutrinos and define
\begin{equation}
\mu=\sqrt2\GF n_{\nu_e}(r) \frac{R^2}{r^2}
=4.72\times10^5~{\rm km}^{-1}~\frac{L_{\nu_e}}{4{\times}10^{52}~{\rm erg/s}}
\frac{10~{\rm MeV}}{\langle E_{\nu_e}\rangle}\,
\(\frac{30~{\rm km}}{R}\)^2\,\(\frac{R}{r}\)^4\,,
\end{equation}
where $L_{\nu_e}$ is the $\nu_e$ luminosity and $\langle E_{\nu_e}\rangle$ their
average energy. More precisely, $n_{\nu_e}$ is the $\nu_e$ density at radius $r$ that we
would obtain in the absence of flavor conversions after emission at radius
$R$. Previously we have sometimes normalized $\mu$ to $n_{\bar\nu_e}$
instead, or to the difference between the $\bar\nu_e$ and $\bar\nu_x$
densities. However, in our schematic studies we assume that initially we have
only a gas consisting of $\nu_e$ and $\bar\nu_e$, again obviating the need
for these fine distinctions. The exact definition of $\mu$ has no physical
impact because it always appears as a product with the density matrices.

In previous papers~\cite{Raffelt:2013rqa, Banerjee:2011fj}, a further factor
$1/2$ was included in the definition of the multi-angle $\lambda$ and $\mu$.
We have kept this factor explicitly in equation~\eqref{eq:Htwoflavor} both in the
matter term and in the flux factor $\half(\bv-\bv')^2$ to maintain its
traditional form. In this way, the equations can be directly applied to a
traditional ``early universe'' system. To make contact with previous SN
discussions, one can always absorb this factor in the definition of $\lambda$
and $\mu$.

As a next step, we project out the trace-free part of the density matrices
and normalize them to account for the above normalization of the effective
neutrino-neutrino interaction strength $\mu$,
\begin{equation}
\varrho_{t,\bx,\omega,\bv}=\frac{{\rm Tr}(\varrho_{t,\bx,\omega,\bv})}{2}
+\frac{n_{\nu_e}}{2}\,{\sf G}_{t,\bx,\omega,\bv}\,.
\end{equation}
With these definitions, the two-flavor EoMs finally become
\begin{equation}\label{eq:EOM5}
i(\partial_t+\bv\cdot{\bm\nabla}_\bx){\sf G}_{t,\bx,\omega,\bv}
=[{\sf H}_{t,\bx,\omega,\bv},{\sf G}_{t,\bx,\omega,\bv}]\,,
\end{equation}
with the Hamiltonian matrix
\begin{equation}\label{eq:two-flavor-Hamiltonian}
{\sf H}_{t,\bx,\omega,\bv}=\(\omega+\lambda_\bx\,\half\bv^2\)
\begin{pmatrix}+\half&0\\0&-\half\end{pmatrix}
+\mu\int d\Gamma'\,\frac{(\bv-\bv')^2}{2}\,\frac{{\sf G}_{t,\bx,\omega',\bv'}}{2}\,,
\end{equation}
where we have included a possible spatial dependence of the electron density
in the form of $\lambda_\bx$ depending on location in the 2D space. The neutrino
velocity domain of integration is determined by the dimensionality of the
chosen problem and if multi-angle effects are to be considered.

\subsection{Mass ordering}

In a two-flavor system, one important parameter for matter effects in general and for
self-induced flavor conversion in particular is the mass ordering. In our context the
question is if the dominant mass
component of $\nu_e$ is the heavier one (inverted ordering) or the lighter one (normal
ordering). Traditionally ``mass
ordering'' is also termed ``mass hierarchy'' and we denote the two cases as IH (inverted
hierarchy) and NH (normal
hierarchy).
We are concerned with 1-3-mixing, the corresponding mixing angle is not large, and so it is
clear what we mean with the
``dominant mass component.''

Our equations are formulated such that they apply to IH, the traditional case where
self-induced
flavor conversion is important in
the form of the bimodal instability.
Of course, it has become clear that NH is actually the more interesting case. For NH,
$\Delta
m^2$ is negative, but we prefer
to consider
$\Delta m^2$ a positive parameter. Therefore, NH is achieved by including explicitly a minus
sign on the r.h.s.\ of
equation~\eqref{eq:omega-definition}. This change of sign translates into a minus sign for
$\omega$ in the first bracket in
equation~\eqref{eq:two-flavor-Hamiltonian}.

For flavor conversion, it is irrelevant if neutrinos oscillated ``clockwise'' or ``counter
clockwise'' in flavor space, i.e.,
in equation~\eqref{eq:EOM5} we may change $i\to -i$ or
${\sf H}\to -{\sf H}$ without changing physical results. However, the {\em relative\/} sign
between $\omega$ and $\lambda$
and $\mu$ is crucial. Therefore, switching the hierarchy is achieved by
\begin{equation}
\hbox{IH${}\to{}$NH:}
\qquad
\lambda\to-\lambda
\quad\hbox{and}\quad
\mu\to-\mu\,.
\end{equation}
In our stability analysis we will consider the parameter range $-\infty<\mu<+\infty$ and
$-\infty<\lambda<+\infty$ as these are simply formal mathematical parameters. Physically
both parameters being positive corresponds to IH, whereas the quadrant of both parameters
being negative corresponds to NH.

\subsection{Linearization}

As a next step, we linearize the EoMs in the sense that the complex
off-diagonal element of every ${\sf G}$ is supposed to be very small compared
to its diagonal part. We write these matrices explicitly as
\begin{equation}
{\sf G}=\begin{pmatrix}g&G\\G^*&-g\end{pmatrix}
\end{equation}
where $g$ is a real and $G$ a complex number and all quantities carry indices
$(t,\bx,\omega,\bv)$. To linear order in $G$ we then find the EoMs
\begin{subequations}
\begin{eqnarray}
i(\partial_t+\bv\cdot{\bm\nabla}_\bx)\,g_{t,\bx,\omega,\bv}\,\,&=&0\,,\\[4pt]
i(\partial_t+\bv\cdot{\bm\nabla}_\bx)\,G_{t,\bx,\omega,\bv}&=&
\[\omega+\lambda_\bx\,\half\bv^2+
\mu\int d\Gamma'\,\half(\bv-\bv')^2\,g_{t,\bx,\omega',\bv'}\]G_{t,\bx,\omega,\bv}
\nonumber\\
&-&g_{t,\bx,\omega,\bv}\,
\[\mu\int d\Gamma'\,\half(\bv-\bv')^2\,G_{t,\bx,\omega',\bv'}\]\,.
\label{eq:Gintegral}
\end{eqnarray}
\end{subequations}
Up to normalization, the ``spectrum'' $g_{t,\bx,\omega,\bv}$ is essentially
the phase-space density of all neutrinos. It is not affected by flavor
conversion, but evolves by free-streaming if it is not homogeneous.

\subsection{Homogeneous neutrino and electron densities}

We are primarily interested in self-induced instabilities. Disturbances in
the neutrino density and/or the electron density will certainly exist in a
real SN and can play the role of seeds for growing modes. However, if these
disturbances are small, it is unlikely that they will be responsible for
instabilities themselves. Henceforth we will assume that the neutrino and
electron densities do not depend on the transverse coordinate $\bx$, although
the flavor content may well depend on $\bx$. Free streaming does not change
the density if it is uniform. As a consequence, $g_{t,\bx,\omega,\bv}$ does
not depend on $\bx$ or $t$ and likewise, $\lambda_\bx$ does not depend on
$\bx$.

With this assumption, $g_{\omega,\bv}$ describes the initially prepared
neutrino distribution, i.e., their density in the phase space spanned by
$\omega$ and $\bv$. Inspecting the first integral in
equation~\eqref{eq:Gintegral}, we may write the three independent terms as
\begin{equation}
\epsilon=\int d\Gamma\,g_{\omega,\bv}\,,
\quad
{\bm\epsilon}_1=\int d\Gamma\,g_{\omega,\bv}\,\bv\,,
\quad\hbox{and}\quad
\epsilon_2=\int d\Gamma\,g_{\omega,\bv}\,\bv^2\,.
\end{equation}
Here, $\epsilon$ represents the ``asymmetry'' between neutrinos and
antineutrinos. The second term, ${\bm\epsilon}_1$, represents a neutrino current
which exists if their distribution is not isotropic and not
symmetric between neutrinos and antineutrinos. Overall, the first term in
square brackets becomes
\begin{equation}
\omega+\half\lambda\bv^2+\half\epsilon\mu\bv^2
-\mu{\bm\epsilon}_1\cdot\bv+\half\epsilon_2\mu\,.
\end{equation}
The last term is simply a constant and can be removed by changing the overall
frequency of the rotating frame. Defining
\begin{equation}
\bar\lambda=\lambda+\epsilon\mu
\end{equation}
the term in square brackets effectively becomes
$\omega+\half\bar\lambda\bv^2- \mu\,{\bm\epsilon}_1\cdot\bv$. We may also
return to the notation used in our previous papers and define
\begin{equation}
S_{t,\bx,\omega,\bv}=\frac{G_{t,\bx,\omega,\bv}}{g_{\omega,\bv}}\,.
\end{equation}
The linearized EoM then takes on the more familiar form
\begin{equation}\label{eq:EOM6}
i(\partial_t+\bv\cdot{\bm\nabla}_\bx)\,S_{t,\bx,\omega,\bv}=
\(\omega+\half\bar\lambda\bv^2-\mu\,{\bm\epsilon}_1\cdot\bv\)S_{t,\bx,\omega,\bv}
-\mu\int d\Gamma'\,\half(\bv-\bv')^2\,g_{\omega',\bv'}S_{t,\bx,\omega',\bv'}\,.
\end{equation}
This equation corresponds to equation~(6) of reference~\cite{Raffelt:2013rqa}. Besides
the streaming term (the gradient term on the l.h.s.) that we have now
included to deal with self-induced inhomogeneities, we have also found the
additional term $\mu{\bm\epsilon}_1\cdot\bv$ which is unavoidable in a
non-isotropic system, irrespective of the question of homogeneity. This
neutrino flux term is missing in reference~\cite{Raffelt:2013rqa}. The presence
of this term modifies the eigenvalue equation for a non-isotropic system.

\subsection{Spatial Fourier transform}

We can now perform the spatial Fourier transform of our linearized EoM of
equation~\eqref{eq:EOM6}. It simply amounts to replacing the spatial dependence on
$\bx$ of $S$ by it dependence on the wave vector $\bk$ and
$\bv\cdot{\bm\nabla}_\bx\to i\bv\cdot\bk$, leading to
\begin{equation}\label{eq:EOM7}
i\dot S_{t,\bk,\omega,\bv}=
\(\omega+\half\bar\lambda\bv^2+\bar\bk\cdot\bv\)S_{t,\bk,\omega,\bv}
-\mu\int d\omega'\int d\bv'\,\half(\bv-\bv')^2\,g_{\omega',\bv'}S_{t,\bk,\omega',\bv'}\,,
\end{equation}
where $\bar\bk=\bk-\mu{\bm\epsilon}_1$. Therefore, the wave vector $\bk$ has
the same effect as a neutrino current. Including an electron current would
appear in a similar way. However, in the following studies of explicit cases
we will focus on isotropic distributions and worry primarily about
self-induced anisotropies, not about the modifications caused by initially
prepared anisotropies.

In equation~\eqref{eq:EOM7} we have spelled out the meaning of the phase-space
integral $\int d\Gamma'=\int d\omega'\int d\bv$. Notice that the meaning of
$\int d\Gamma'$ has changed in the course of changing variables that describe
the neutrino modes. All phase-space factors and Jacobians have been absorbed
in the definition of the effective neutrino-neutrino interaction strength
$\mu$ as well as the normalization of the ``spectrum'' $g_{\omega,\bv}$. In
particular, if we begin with an ensemble consisting of only $\nu_e$ and
$\bar\nu_e$ and no $\nu_x$ or $\bar\nu_x$, then our normalizations mean that
$\int_0^\infty d\omega\int d\bv\, g_{\omega,\bv}=1$. In this latter integral,
we have only included positive frequencies (neutrinos, no antineutrinos) so
that this normalization coincides with our definition that $\mu$ is
normalized to $n_{\nu_e}$.

\subsection{Oscillation eigenmodes}

In order to find unstable modes we seek solutions of our linearized EoM of
the form $S_{t,\bk,\omega,\bv}=Q_{\Omega,\bk,\omega,\bv}e^{-i\Omega t}$,
leading to an EoM in frequency space of the form
\begin{equation}\label{eq:master0}
\(\half\bar\lambda\,\bv^2+\bar\bk\cdot\bv+\omega-\Omega\)Q_{\Omega,\bk,\omega,\bv}
=\mu\int d\omega'\int
d\bv'\,\half(\bv-\bv')^2\,g_{\omega',\bv'}\,Q_{\Omega,\bk,\omega',\bv'}\,.
\end{equation}
Eigenvalues $\Omega=\gamma+i\kappa$ with a positive imaginary part represent
unstable modes with the growth rate $\kappa$.

\subsection{Monochromatic and isotropic neutrino distribution}

In our explicit examples we will always consider monochromatic neutrinos with
some fixed energy, implying a spectrum of two oscillation frequencies
$\omega=\pm\omega_0$. Assuming that the energy and velocity distribution
factorize, we may write the spectrum in the form
\begin{equation}
g_{\omega,\bv}=h_\omega\,f_\bv\,.
\end{equation}
The monochromatic energy spectrum is
\begin{equation}\label{eq:monochromaticspectrum}
h_{\omega}=-\alpha\,\delta(\omega+\omega_0)+\delta(\omega-\omega_0)\,,
\end{equation}
meaning that we have $\alpha$ antineutrinos (frequency $\omega=-\omega_0$)
for every neutrino ($\omega=\omega_0$). The spectral asymmetry is
$\epsilon=1-\alpha$.

We will consider isotropic velocity distributions which, in addition, are
uniform, corresponding to blackbody-like angular emission in the SN context.
In this case, $f_\bv=1/\Gamma_\bv$, where $\Gamma_\bv$ is the volume of the
velocity phase space. The eigenvalue equation~\eqref{eq:master0} finally
simplifies to the form in which we will use it,
\begin{equation}\label{eq:master}
\(\half\bar\lambda\,\bv^2+\bk\cdot\bv+\omega-\Omega\)Q_{\Omega,\bk,\omega,\bv}
=\mu\int d\omega'\,h_{\omega'}\,\frac{1}{2\Gamma_\bv}\int d\bv'
\,(\bv-\bv')^2\,Q_{\Omega,\bk,\omega',\bv'}\,.
\end{equation}
We now consider systematically different cases of velocity distributions.

%%%%%%%%%%%%%%%%%%%%%%%%%%%%%%%%%%%%%%%%%%%%%%%%%%%%%%%%%%%%%%%%%%%%%%%%%%%%%%
\section{One-dimensional system}
\label{sec:1D}
%%%%%%%%%%%%%%%%%%%%%%%%%%%%%%%%%%%%%%%%%%%%%%%%%%%%%%%%%%%%%%%%%%%%%%%%%%%%%%

As a first case study we consider a 1D system, i.e., the toy model of
``colliding beams'' that has been used in the recent literature as a simple
case where one can easily see the impact of spontaneous spatial symmetry
breaking \cite{Raffelt:2013isa, Duan:2014gfa, Mangano:2014zda,
Mirizzi:2015fva}. We go beyond previous studies in that we include the
multi-angle matter effect and study the ``footprint'' of the various
instabilities in the two-dimensional parameter space $-\infty<\mu<+\infty$
and $-\infty<\lambda<+\infty$. This schematic study already leads to the
conclusion that essentially
the largest-scale instabilities are ``most dangerous'' in the
context of SN neutrino flavor conversion.

\subsection[Single angle ($v=\pm1$)]{Single angle (\boldmath{$v=\pm1$})}

\subsubsection{Eigenvalue equation}

We begin with 1D systems, i.e., colliding beams of neutrinos and
antineutrinos with different velocity distributions. The first case is what
we call ``single angle,'' a nomenclature which refers to the zenith-angle
distribution of SN neutrinos. As we have explained, in our way of writing the
equations, ``single angle'' means that the neutrino velocity distribution has
$|\bv|=1$. In our first 1D case this means we consider two colliding beams
with $v=\pm1$. Matter effects can be rotated away.

The eigenfunction $Q_{\Omega,k,\omega,v}$ now consists of four discrete
components. We denote these four amplitudes with the complex numbers $R$ for
right-moving ($v=+1$) neutrinos ($\omega=+\omega_0$), $\bar R$ for
right-moving antineutrinos, and analogous $L$ and $\bar L$ for left movers.
Our master equation~\eqref{eq:master} then reads
\begin{equation}
\[\begin{pmatrix}
\omega_0+k&0&-\mu&\mu\alpha\\
0&-\omega_0+k&-\mu&\mu\alpha\\
-\mu&\mu\alpha&\omega_0-k&0\\
-\mu&\mu\alpha&0&-\omega_0-k
\end{pmatrix}-\Omega\]
\begin{pmatrix}
R\\
\bar R\\
L\\
\bar L
\end{pmatrix}
=0\,,
\end{equation}
corresponding to the equivalent result of Duan and Shalgar
\cite{Duan:2014gfa}. The eigenvalues $\Omega$ are found from equating the
determinant of the matrix in square brackets with zero. This condition
can be written in the form
\begin{equation}\label{eq:quartic}
\(\frac{1}{-k+\omega_0-\Omega}-\frac{\alpha}{-k-\omega_0-\Omega}\)
\(\frac{1}{k+\omega_0-\Omega}-\frac{\alpha}{k-\omega_0-\Omega}\)\mu^2=1\,.
\end{equation}
This expression depends only on $\mu^2$ and therefore yields identical
eigenvalues for positive and negative $\mu$, i.e., for both neutrino mass
hierarchies, as noted by Duan and Shalgar. It is also even under $k\to-k$ as
it must because the system was set up isotropically, so the eigenvalues
cannot depend on the orientation of $k$.

\subsubsection[Homogeneous mode ($k=0$)]{Homogeneous mode (\boldmath{$k=0$)}}
\label{sec:1D-beam-homogeneous}

\begin{figure}[ht]
\centering
\includegraphics[width=0.6\textwidth]{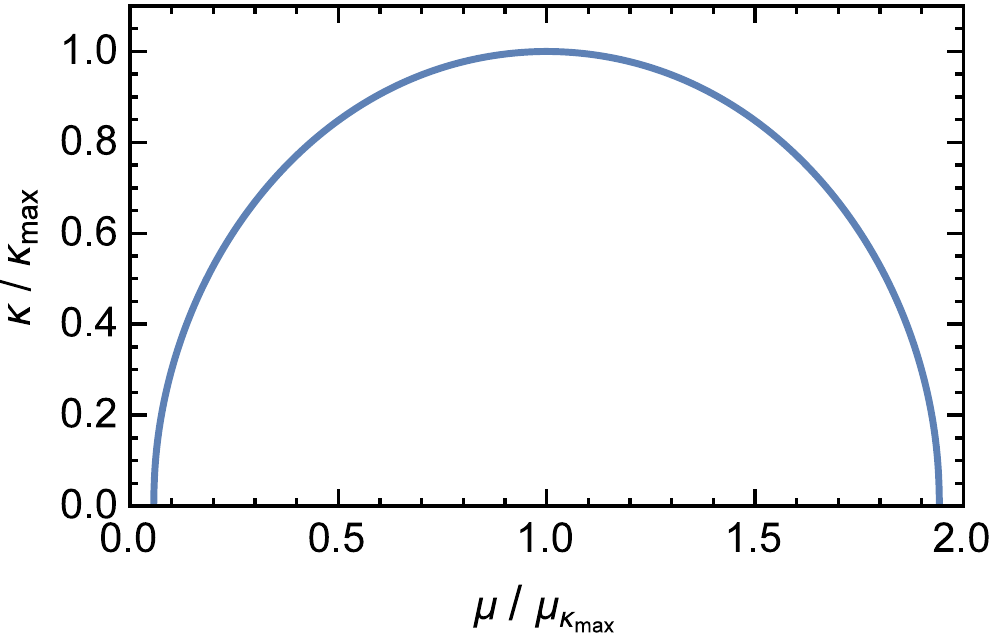}
\caption{Growth rate $\kappa$ for the unstable mode in the homogeneous ($k=0$) 1D case
for $\alpha=1/2$.
The maximum growth rate $\kappa_{\rm max}$ is given in equation~\eqref{eq:kappa-max},
the corresponding interaction strength $\mu_{\kappa_{\rm max}}$ in
equation~\eqref{eq:mu-kappa-max}.}
\label{fig:kappa-1D-hom}
\end{figure}

For the homogeneous mode, $k=0$, this eigenvalue equation simplifies
considerably. We already know that we have the same solution for positive and
negative $\mu$, where the latter is the left-right symmetry breaking solution
discovered in reference~\cite{Raffelt:2013isa}. We here limit ourselves to $\mu>0$
and need to solve the quadratic equation
\begin{equation}\label{eq:quadratic1}
\omega_0^2-\Omega^2-\mu \bigl[(1+\alpha)\omega_0+(1-\alpha)\Omega\bigr]=0\,.
\end{equation}
It has the solutions
\begin{equation}
\Omega=-\frac{(1-\alpha)\mu}{2}\pm\sqrt{\[\omega_0+\frac{(1-\alpha)\mu}{2}\]^2-2\mu\omega_0}\,.
\end{equation}
Unstable solutions exist for
\begin{equation}
\frac{2}{(1+\sqrt\alpha)^2}<\frac{\mu}{\omega_0}<\frac{2}{(1-\sqrt\alpha)^2}\,,
\end{equation}
which for $\alpha=1/2$ is the range $12-8\sqrt2<\mu/\omega_0<12+8\sqrt2$ or
numerically $0.6863\lesssim\mu/\omega_0\lesssim23.31$. The maximum growth
rate is
\begin{equation}\label{eq:kappa-max}
\kappa_{\rm max}=\frac{2\sqrt\alpha}{1-\alpha}\,\omega_0\,.
\end{equation}
For $\alpha=1/2$ this is $\kappa_{\rm max}/\omega_0=2\sqrt2\approx2.828$. The
maximum growth rate occurs at the interaction strength
\begin{equation}\label{eq:mu-kappa-max-k0}
\mu_{\kappa_{\rm max}}=\frac{2\,(1+\alpha)}{(1-\alpha)^2}\,.
\end{equation}
We show the growth rate normalized to its maximum in figure~\ref{fig:kappa-1D-hom}
as a function of $\mu/\mu_{\kappa_{\rm max}}$.
For this normalization, the unstable range is
\begin{equation}\label{eq:unstablerange}
1-\frac{2\sqrt{\alpha}}{1+\alpha}<\frac{\mu}{\mu_{\kappa_{\rm
max}}}<1+\frac{2\sqrt{\alpha}}{1+\alpha}\,.
\end{equation}
If we use $\alpha=1-\epsilon$ and expand to lowest order in $\epsilon$, this
range is $\epsilon^2/8<\mu/\mu_{\kappa_{\rm max}}<2-\epsilon^2/8$. Therefore,
even if $\epsilon$ is not very small ($\epsilon=1/2$ in
figure~\ref{fig:kappa-1D-hom}), the unstable range is close to its maximum
range from 0 to 2.

\subsubsection[Inhomogeneous modes ($k>0$)]{Inhomogeneous modes (\boldmath{$k>0$)}}

The quartic eigenvalue equation~\eqref{eq:quartic} is not easy to
disentangle. However, for large $k$ it simplifies and can be solved. We may
guess that, for large $k$, the real part of $\Omega$ is approximately $-k$
and, without loss of generality, we may go to a rotating frame such that
$\Omega=\tilde\Omega-k$. Moreover, based on numerical studies, Duan and
Shalgar \cite{Duan:2014gfa} have conjectured that for large $k$, the unstable
$\mu$-range scales with $\sqrt{k}$. This observation motivates us to write
$\mu=m\sqrt{\omega_0 k}$ and, without loss of generality, the eigenvalue
equation reads
\begin{equation}\label{eq:quartic2}
\(\frac{\omega_0}{\omega_0-\tilde\Omega}-\frac{\alpha\omega_0}{-\omega_0-\tilde\Omega}\)
\(\frac{k}{2k+\omega_0-\tilde\Omega}-\frac{\alpha k}{2k-\omega_0-\tilde\Omega}\)m^2=1\,.
\end{equation}
Now we can take the limit $k\to\infty$ and approximate the second bracket as
$(1-\alpha)/2$. The resulting quadratic equation is now easily solved and
yields
\begin{equation}
\frac{\tilde\Omega}{\omega_0}
=-\frac{\epsilon^2 m^2}{4}\pm\frac{\sqrt{16-8m^2(2-\epsilon)\epsilon+\epsilon^4m^4}}{4}\,.
\end{equation}
One can try the same exercise with the opposite rotating frame
$\Omega=\tilde\Omega+k$ and finds a similar-looking result, where however the
argument of the square-root is always positive, i.e., there is no unstable
mode with this property.

From equation~\eqref{eq:quartic} one finds that the maximum growth rate
$\kappa_{\rm max}$ is the same as in the homogeneous case of
equation~\eqref{eq:kappa-max}, i.e., for large $k$ the maximum growth rate does
not depend on $k$ and is the same as for $k=0$. However,  it now occurs at
the interaction strength
\begin{equation}\label{eq:mu-kappa-max}
\mu^2_{\kappa_{\rm max}}=\frac{4\,(1+\alpha)}{(1-\alpha)^3}\,\omega_0 k\,.
\end{equation}
The unstable range is the same as given in equation~\eqref{eq:unstablerange} if we
substitute $\mu/\mu_{\kappa_{\rm max}}$ with $(\mu/\mu_{\kappa_{\rm max}})^2$.
The growth rate as a function of $\mu$ is the same as shown in
figure~\ref{fig:kappa-1D-hom} if we interpret the horizontal axis
as $(\mu/\mu_{\kappa_{\rm max}})^2$ with our new $\mu_{\kappa_{\rm max}}$.

For intermediate values of $k$, the maximum growth rate deviates slightly
from the two extreme cases. It is somewhat surprising that the structure of
the eigenvalue equation is such that the maximum possible growth rate depends
only on the vacuum oscillation frequency $\omega_0$ and $\alpha$, but not on
the potentially large scale $k$.

\subsection[Multi-angle effects ($0\leq v\leq1$)]
{Multi-angle effects (\boldmath{$0\leq v\leq1$})}

\subsubsection{Eigenvalue equation}

We may now study the multi-angle impact of matter, in the spirit of the SN
system, in our 1D model by extending the velocity integration over the entire
interval $-1\leq v\leq 1$ instead of using only the modes $v=\pm1$. One
approach is to introduce $n_v$ discrete velocities on the interval $0<v\leq
1$, i.e., $2n_v$ modes on the interval $-1\leq v\leq 1$, leading to a total
of $4n_v$ discrete equations. One may then proceed to find numerically the
eigenmodes. This approach is marred by the appearance of spurious
instabilities and one may need a large number of modes to obtain physical
results \cite{Sarikas:2012ad}.

Therefore, we usually avoid discrete velocities and represent the
eigenfunctions $Q_{\Omega,\bk,\omega,\bv}$ as continuous functions of their
variables. Equation~\eqref{eq:master} reads for our specific case
\begin{equation}\label{eq:master2}
\(\half\bar\lambda\,v^2+k\,v+\omega-\Omega\)Q_{\Omega,k,\omega,v}
=\mu\int_{-\infty}^{+\infty} d\omega'\,h_{\omega'}\,\frac{1}{4}
\int_{-1}^{+1} dv' \,(v-v')^2\,Q_{\Omega,k,\omega',v'}\,.
\end{equation}
Since our system was prepared ``isotropic'' in the sense of left-right
symmetry, the instabilities cannot depend on the sign of $k$ so that it is
enough to consider $0\leq k<\infty$.
The r.h.s.\ of equation~\eqref{eq:master2}, as a function of $v$, has the form
$A_0+A_1 v+A_2 v^2$. Because $1$, $v$, and $v^2$ are linearly independent
functions on the interval $-1\leq v\leq+1$, the l.h.s.\ must be of that form
as well and we may use the ansatz
\begin{equation}\label{eq:eigenfunctionansatz}
Q_{\Omega,k,\omega,v}=\frac{A_0+A_1 v+A_2 v^2}{\half\bar\lambda\,v^2+k\,v+\omega-\Omega}
\end{equation}
for the eigenfunctions. Inserting this form on both sides yields
\begin{equation}
A_0+A_1 v+A_2 v^2
=\frac{\mu}{4}\int_{-\infty}^{+\infty} d\omega'\,h_{\omega'}
\int_{-1}^{+1} dv' \,(v'^2-2vv'+v^2)\,
\frac{A_0+A_1 v'+A_2 v'^2}{\half\bar\lambda\,v'^2+k\,v'+\omega'-\Omega}
\,.
\end{equation}
This equation really consists of three linearly independent equations for the
parts proportional to different powers of $v$, so we get three equations
linear in $A_0$, $A_1$ and $A_2$. This set of linear equations is compactly
written as
\begin{equation}\label{eq:1D-matrix-equation}
\[1-\begin{pmatrix}I_2&I_3&I_4\\-2I_1&-2I_2&-2I_3\\I_0&I_1&I_2\end{pmatrix}\]
\begin{pmatrix}A_0\\A_1\\A_2\end{pmatrix}=0\,,
\end{equation}
where
\begin{equation}\label{eq:In-definition}
I_n=\frac{\mu}{4}\int_{-\infty}^{+\infty} d\omega\,h_{\omega}
\int_{-1}^{+1} dv\,\frac{v^n}{\half\bar\lambda\,v^2+k\,v+\omega-\Omega}\,,
\end{equation}
where we have dropped the prime from the integration variables.

Notice that $I_n$ is odd under $k\to-k$ if $n$ is odd, and it is even
if $n$ is even.  Moreover, in the determinant of the matrix in square
brackets in equation~\eqref{eq:1D-matrix-equation}, every term involving
$I_n$ with an odd power of $n$ involves another factor $I_m$ with $m$
odd. Therefore, the determinant is even under $k\to-k$, in agreement
with our earlier statement that without loss of generality we may
assume $k\geq 0$.

\subsubsection[Homogeneous mode ($k=0$) without matter effects
($\bar\lambda=0$)] {Homogeneous mode (\boldmath{$k=0$}) without matter
effects (\boldmath{$\bar\lambda=0$})}

\label{sec:1D-homogeneous-nomatter}

As a first simple example we consider homogeneous solutions ($k=0$) in the
absence of matter effects ($\bar\lambda=0$). The latter assumption requires
an exact cancellation $\bar\lambda=\lambda+\epsilon\mu=0$ between the matter
effects caused by the background medium and by neutrinos themselves. We
consider this case only for mathematical convenience without physical
motivation. The velocity integrals vanish for odd powers of $v$. For even
powers, and using the monochromatic frequency spectrum of
equation~\eqref{eq:monochromaticspectrum}, we find
\begin{equation}
I_n=\frac{\mu}{2(1+n)}\(\frac{\alpha}{\omega_0+\Omega}+\frac{1}{\omega_0-\Omega}\)
=\frac{\mu}{2(1+n)}\frac{(1+\alpha)\,\omega_0-(1-\alpha)\,\Omega}{\omega_0^2-\Omega^2}
\end{equation}
and the eigenvalue equation corresponds to
\begin{equation}\label{eq:simplematrix}
{\rm
det}\[\omega_0^2-\Omega^2-\frac{\mu}{2}\begin{pmatrix}\frac{1}{3}&0&\frac{1}{5}\\0&-\frac{2}{3}&0\\
1&0&\frac{1}{3}\end{pmatrix}
\[(1+\alpha)\,\omega_0+(1-\alpha)\,\Omega\]\]=0\,.
\end{equation}
Before searching for solutions, we may diagonalize the $3{\times}3$ matrix.
This leads to three independent equations of the form of
equation~\eqref{eq:quadratic1}, where we need to substitute
\begin{equation}\label{eq:mucases}
\mu\to\frac{\mu}{30}\times
\begin{cases}
5+3\sqrt{5}&>0\,,\\
-10&<0\,,\\
5-3\sqrt{5}&<0\,.
\end{cases}
\end{equation}
Therefore, we get three instabilities: One for $\mu>0$, the usual bimodal
instability (IH), and two negative-$\mu$ solutions (NH). The maximum growth
rate is the same in every case as the one that was found in
Sec.~\ref{sec:1D-beam-homogeneous} and was given in
equation~\eqref{eq:kappa-max}. The exact unstable $\mu$-ranges have changed
according to the $\mu$ scaling provided by equation~\eqref{eq:mucases}. For
our usual example with $\alpha=1/2$, the instability ranges for both
hierarchies in the colliding-beam example were $0.68<|\mu/\omega_0|<23.31$.
After $v$-integration they become explicitly
\begin{subequations}\label{eq:1Dinstabilityranges}
\begin{eqnarray}
1.76<&\mu/\omega_0&<59.74
\,,\\
-69.94<&\mu/\omega_0&<-2.06
\,,\\
-409.44<&\mu/\omega_0&<-12.05
\,.
\end{eqnarray}
\end{subequations}
We conclude that integrating over the velocity interval $-1\leq v\leq+1$
modifies the unstable $\mu$-ranges, breaks the symmetry between normal and
inverted hierarchy, and introduces another normal-hierarchy instability.

Qualitatively, these results are analogous to the three types of instability
discovered in the study of axial-symmetry breaking in the SN context
\cite{Raffelt:2013rqa}. The one inverted-hierarchy solution appearing in all
cases is the bimodal instability and corresponds to the original flavor
pendulum~\cite{Hannestad:2006nj}. The first normal-hierarchy instability is
what was termed the multi-azimuth angle (MAA) instability, although in our 1D
system we have only two ``azimuth angles,'' i.e., the two beam directions.
The third instability, appearing in normal hierarchy for a much larger
$\mu$-range, is what was termed the ``multi zenith angle'' (MZA) instability.
It requires, in the SN terminology, a nontrivial range of zenith angles,
corresponding here to a non-trivial $v$-range, i.e., anything beyond the
trivial $v=\pm1$ velocity distribution.

Notice that our 1D MAA instability corresponds to an eigenfunction which is
anti-symmetric in $v$ (it breaks the left-right symmetry) and corresponds to
the middle entry $-2/3$ in the matrix of equation~\eqref{eq:simplematrix} which
decouples from the remaining $2{\times}2$ block. The latter yields left-right
symmetric solutions (even under $v\to-v$). In this sense, it is the MZA
instability which corresponds, for normal hierarchy, to the bimodal solution.
It exists only for at least three velocity modes, and of course requires the
presence of two vacuum oscillation frequencies, here always chosen as
$\omega=\pm\omega_0$, i.e., we need at least a total of six modes, making a
simple visual interpretation more difficult.

\begin{figure}
\centering
\includegraphics[width=0.5\textwidth]{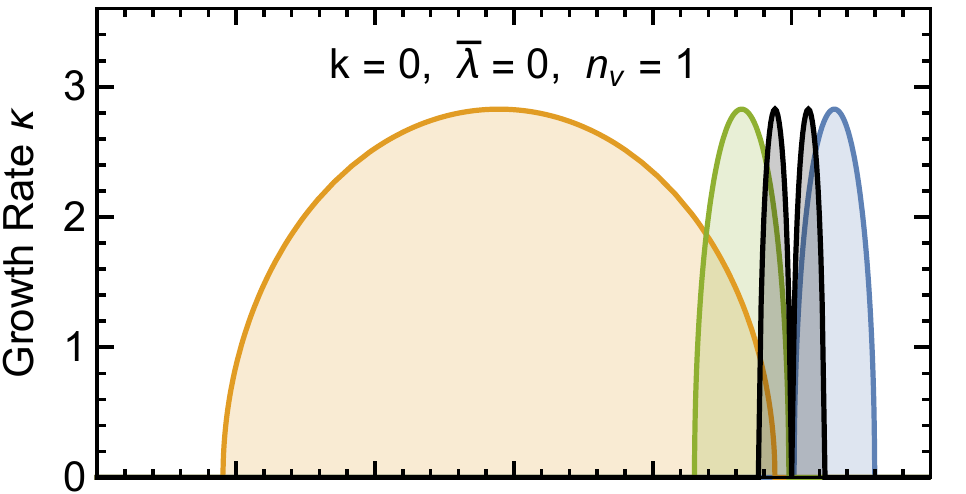}
\includegraphics[width=0.5\textwidth]{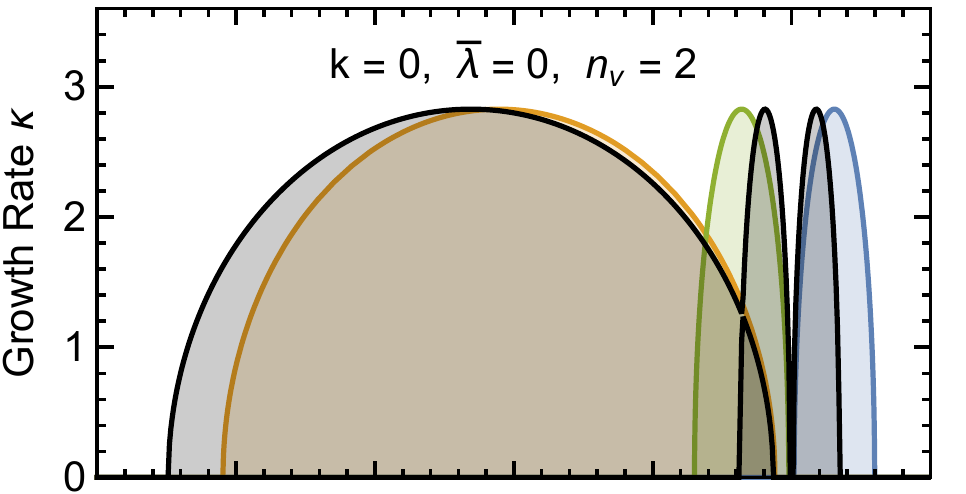}
\includegraphics[width=0.5\textwidth]{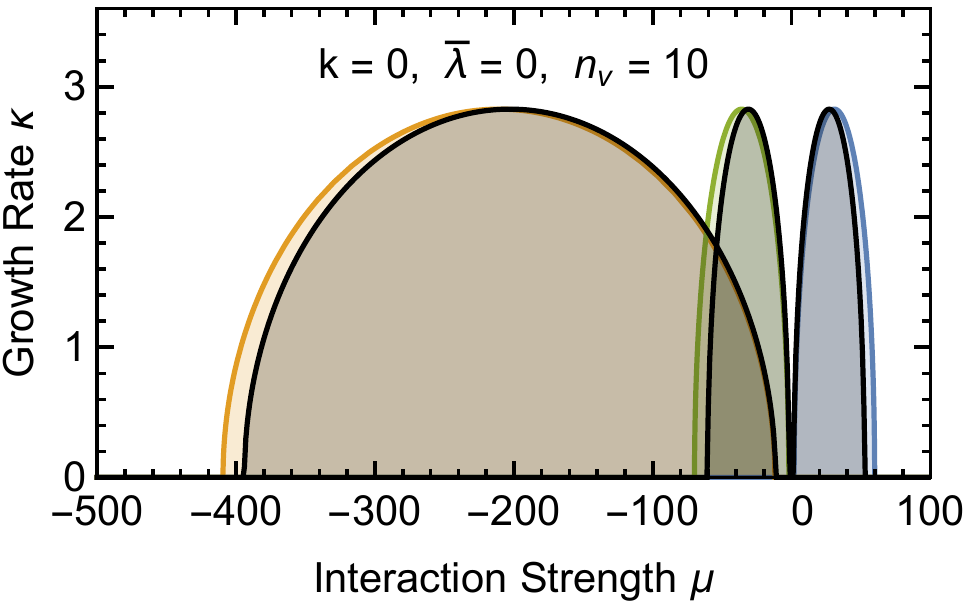}
\caption{Growth rate $\kappa$ for the unstable modes in the
  homogeneous ($k=0$) 1D case with $\alpha=1/2$, where $\mu<0$
  corresponds to normal and $\mu>0$ to inverted hierarchy.
  (Both $\kappa$ and $\mu$ are in units of~$\omega_0$.) The
  colored curves common to all panels are the three instabilities
  which obtain after velocity integration ($-1\leq v\leq+1$) with the
  instability ranges of equation~\eqref{eq:1Dinstabilityranges}. The
  overlaid black instability curves are for discrete velocity bins,
  where the number of bins $n_v=1$, 2, and 10 as indicated in the
  panels. The overlay in the top panel
  ($n_v=1$) is the colliding-beam example with $v=\pm 1$.  With
  increasing $n_v$ (top to bottom), discrete velocity bins approximate
  a uniform distribution.}
\label{fig:kappa-1D-multi-angle}
\end{figure}

The growth rates of all three intabilities as functions of $\mu$ are shown in
figure~\ref{fig:kappa-1D-multi-angle} as blue, green, and orange lines, all of
them having the same maximum. In the top panel, we overlay the two
instability curves for the original colliding-beam example, where $v=\pm1$.
Therefore, this panel directly illustrates the effect of the velocity
integration (``multi-angle effects'' in SN terminology) in that the velocity
integration takes us from the two black curves to the three colored ones.

We have also studied the equations using a set of discrete velocities, where
the $v=\pm 1$ case is the simplest example with $n_v=1$ bins. (We count the
number of bins in the range $0<v\leq 1$, i.e., there are equally many bins
for negative velocities, and the total number doubles for our two frequencies
$\omega=\pm\omega_0$.) Adding the intermediate values $v=\pm1/2$ takes us to
$n_v=2$, shown in the second panel of figure~\ref{fig:kappa-1D-multi-angle}. It
reveals that the symmetry between the hierarchies ($\mu\to-\mu$ symmetry) is
broken as soon as the velocity range is non-trivial and that there are two
normal-hierarchy solutions. Increasing $n_v$ eventually approximates a
uniform $v$ distribution. A fairly small number of velocity bins is enough to
achieve good agreement. We will see shortly that, including non-zero $k$
and/or~$\bar\lambda$, changes the picture because spurious instabilities
appear.

\subsubsection[Inhomogeneous modes ($k>0$) without matter effects
($\bar\lambda=0$)] {Inhomogeneous modes (\boldmath{$k>0$}) without matter
effects (\boldmath{$\bar\lambda=0$})}

Non-vanishing matter effects ($\bar\lambda\not=0$) and non-vanishing
inhomogeneities ($k\not=0$) modify the eigenvalue equation in similar ways:
The range of effective oscillation frequencies given by $\half\bar\lambda
v^2+k v+\omega$ increases considerably if $k/\omega_0\gg1$ and/or
$\bar\lambda/\omega_0\gg1$. Roughly one would suspect that significant
collective phenomena require a neutrino-neutrino coupling exceeding this
range of frequencies, i.e., a $\mu$ range exceeding something like the rms
spread of this range. In this sense, one would expect that the $\mu$-range of
unstable solutions would be shifted roughly linearly with $\bar\lambda$
and/or~$k$.

\begin{figure}[!b]
\centering
\hbox{\includegraphics[width=0.5\textwidth]{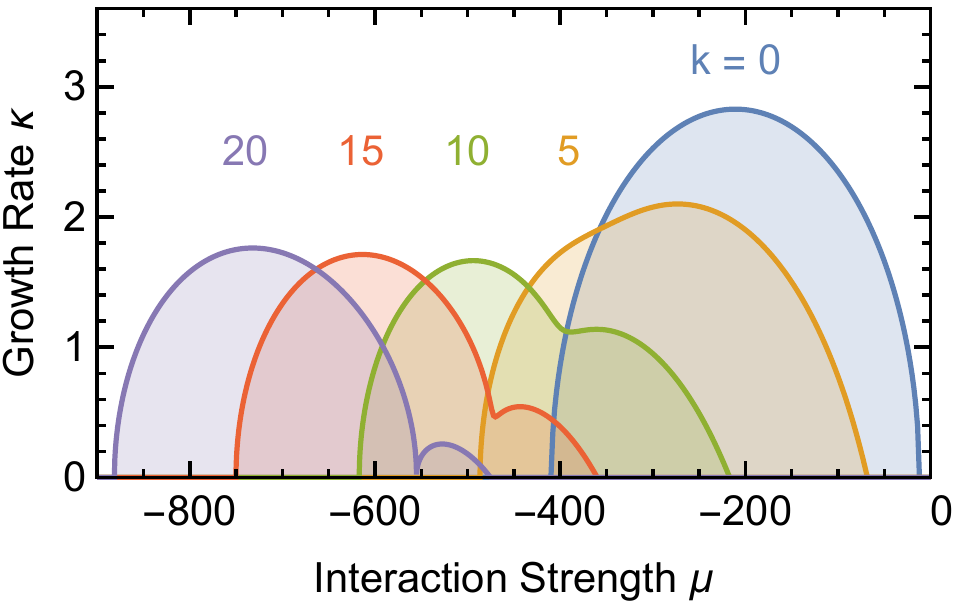}\includegraphics[width=0.5\textwidth]{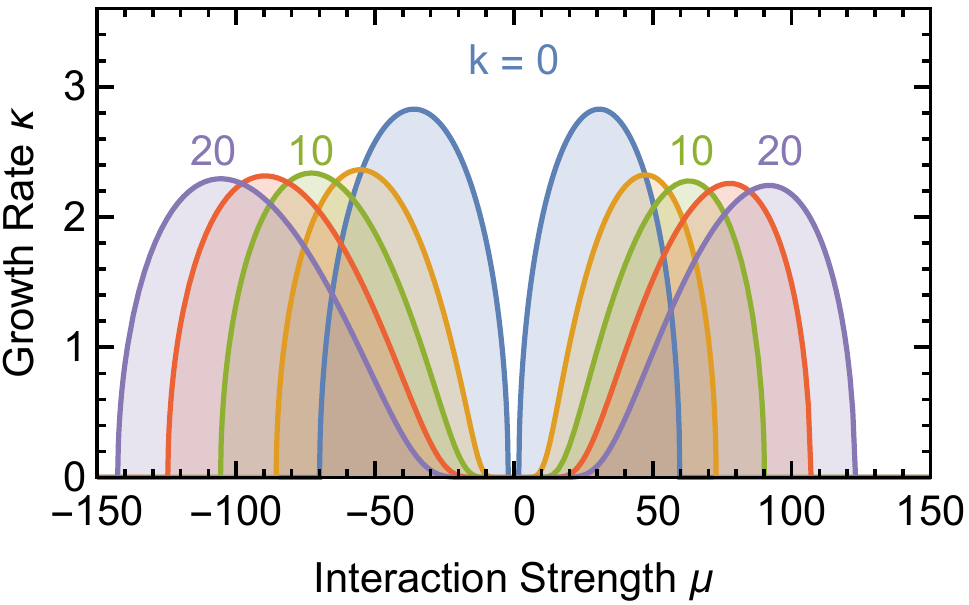}}
\vskip10pt
\hbox{\includegraphics[width=0.5\textwidth]{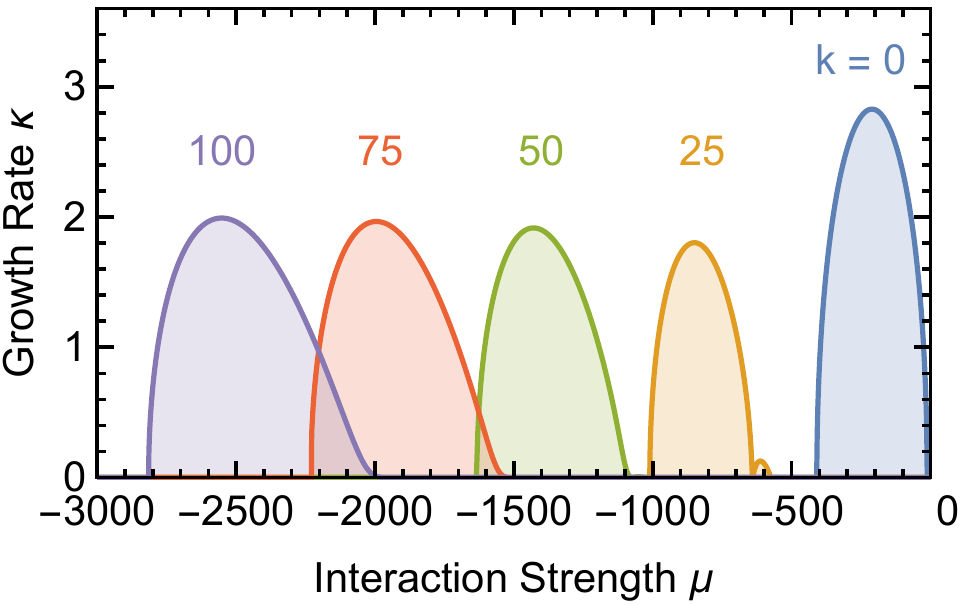}\includegraphics[width=0.5\textwidth]{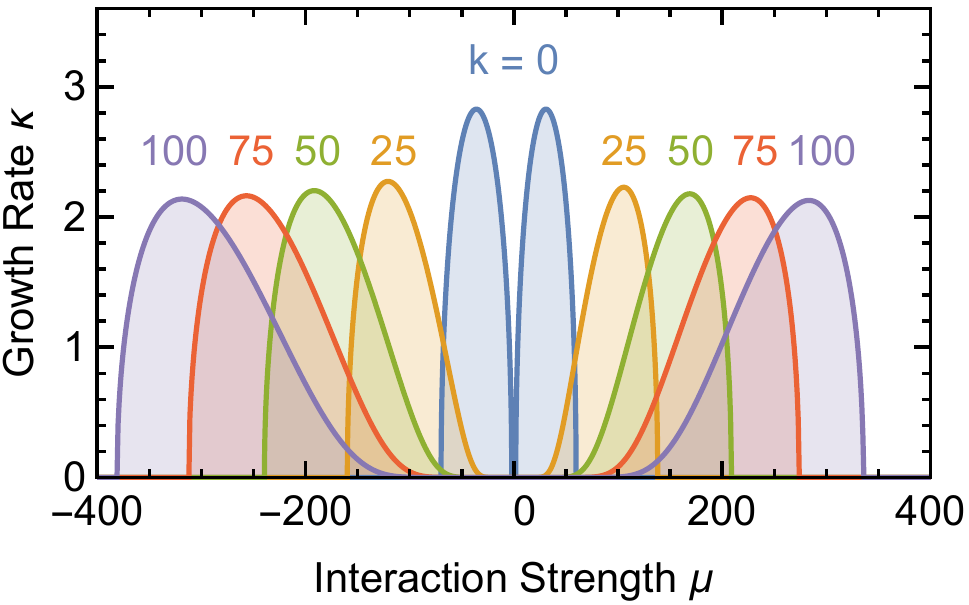}}
\caption{Growth rates $\kappa$ for different wave numbers $k$ as indicated
above the curves. The other parameters are $\bar\lambda=0$ and $\alpha=1/2$.
The left panels show only the MZA mode, in the right panels
we show the MAA mode ($\mu<0$) and the bimodal mode ($\mu>0$). The effect of non-zero $k$ is
essentially to shift the curves and for large $k$, the instability curves are similar
as a function of $\mu/k$.} \label{fig:kshift-1D}
\end{figure}

\begin{figure}[!b]
\centering
\includegraphics[width=0.5\textwidth]{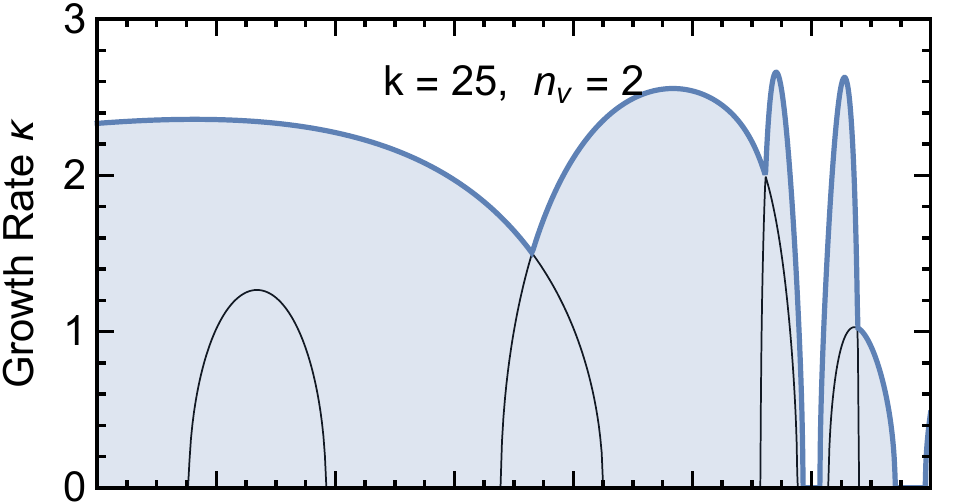}
\includegraphics[width=0.5\textwidth]{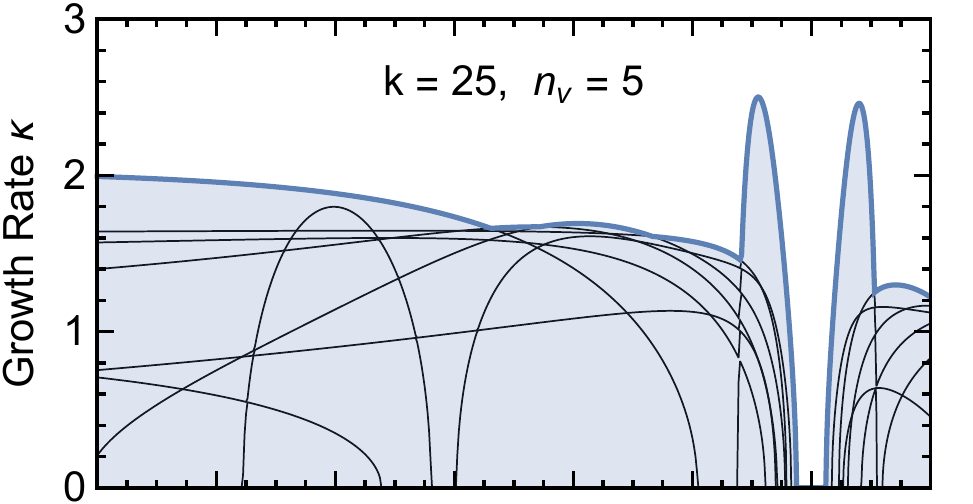}
\includegraphics[width=0.5\textwidth]{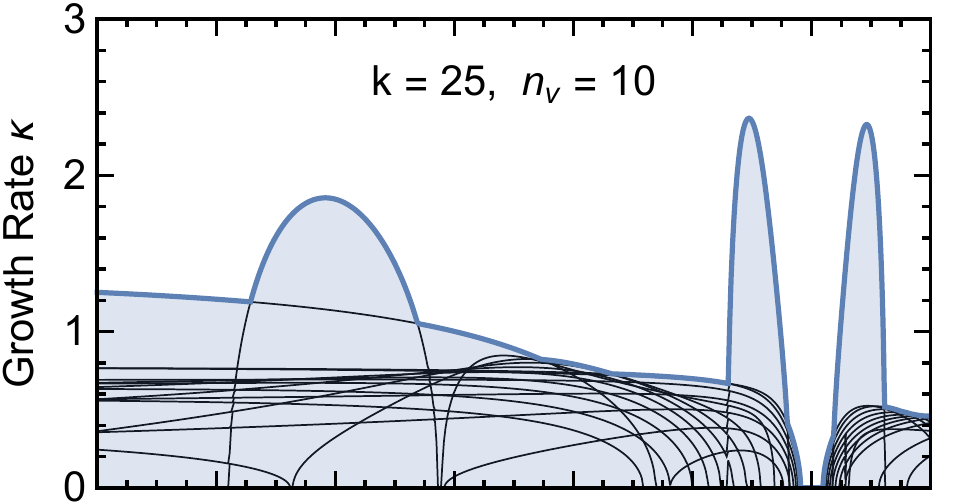}
\includegraphics[width=0.5\textwidth]{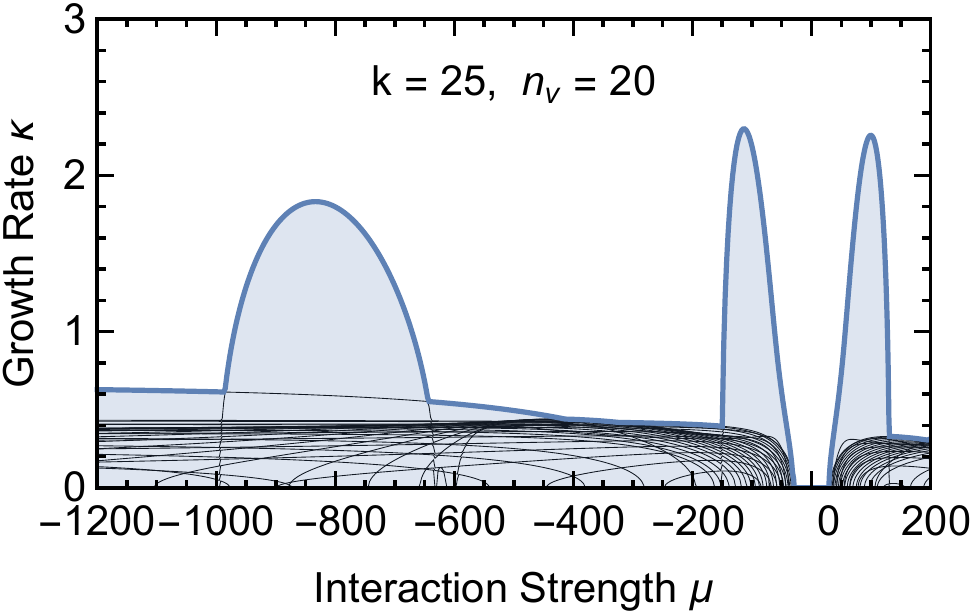}
\caption{Growth rates for all modes that appear when we consider $n_v$ positive and
$n_v$ negative-$v$ bins, using $k=25\,\omega_0$, $\bar\lambda=0$, and $\alpha=1/2$.
For growing $n_v$, more and more spurious modes appear, but their growth rates become
smaller and the physical modes begin to stick out.} \label{fig:1D-spurious}
\end{figure}

We first test this picture with $\bar\lambda=0$ and $k>0$. The $v$-integrals
in equation~\eqref{eq:In-definition} can be performed analytically
(Appendix~\ref{sec:integrals}) and the $\omega$-integration amounts to
summing over two terms with $\omega=\pm\omega_0$. However, finding the
eigenvalues of equation~\eqref{eq:1D-matrix-equation} requires numerical
tools. We use {\sc Mathematica} and show the result in
figure~\ref{fig:kshift-1D} for $0\leq k/\omega_0\leq100$ as indicated above
the curves. In the left panels, we only show the MZA mode. For relatively
small $k$, the function shifts left and deforms somewhat, whereas for larger
$k$ values, it shifts left nearly linearly with $k$. We have checked that
this linear behavior obtains numerically even for very large $k$ values---we
have tested values up to $10^7$. In other words, for large $k$, the
instability curves are very similar as a function of $\mu/k$, although they
narrow somewhat. In contrast to the earlier two-beam example, there does not
seem to be quite a universal function for large $k$. Moreover, in the earlier
case, the scaling was with $\mu/\sqrt{\omega_0 k}$, i.e., the nontrivial $v$
range has qualitatively changed the results with regard to the $k$-scaling.

In the right panels of figure~\ref{fig:kshift-1D}, we show analogous results
for the MAA mode ($\mu<0$) and the bimodal mode ($\mu>0$), which show
analogous behavior.

One can study the same case by solving the eigenvalue equation in terms of
velocity bins, in analogy to what one would do if one were to solve the EoMs
numerically instead of performing only a linear stability analysis. For
$k=25\,\omega_0$ we show the growth rates for all solutions in
figure~\ref{fig:1D-spurious} for different choices for the number $n_v$ of
velocity bins. (We recall that in our convention, the total number of
positive and negative $v$ bins is $2n_v$.) The large number of spurious modes
is a conspicuous feature of these plots, although for sufficiently large
$n_v$, the physical modes stick out.

With non-vanishing $k$ and/or $\bar\lambda$, the functional form of the
eigenfunctions in equation~\eqref{eq:eigenfunctionansatz} no longer factorizes as
a function of $v$ and one of $\omega$. It is conceivable that spurious modes
can be avoided, or their impact reduced, if one were to find a better way of
discretizing the neutrino modes than by simple bins in velocity and frequency
\cite{Duan:2014mfa}.

It is noteworthy that for all modes, spurious or physical, the growth rates
are of order $\omega_0$, i.e., they do not inherit a larger frequency scale
from $k$ or, in later cases, from $\bar\lambda$. Even for huge values of $k$
and $\bar\lambda$, this conclusion does not change and agrees with our
explicit result in the two-beam example.

\subsection[Including matter($\bar\lambda\not=0$)]
{Including matter (\boldmath{$\bar\lambda\not=0$)}}

Including matter in our ``multi-zenith-angle'' case has the effect of
introducing both $\bar\lambda$ and $k$ in the denominator of the integrals of
equation~\eqref{eq:In-definition}. They can still be done analytically
(Appendix~\ref{sec:integrals}), but lead to transcendental functions. Of
course, numerically one can find the eigenvalues without much problem. The
parameter $\bar\lambda$, like $k$, has the effect of broadening the effective
range of oscillation frequencies and of shifting the unstable collective modes to larger
values of $|\mu|$. We study the homogeneous and inhomogeneous cases
separately.

\subsubsection[Homogeneous mode ($k=0$)]{Homogeneous mode (\boldmath{$k=0$)}}

\begin{figure}[ht]
\centering
\includegraphics[width=1.0\textwidth]{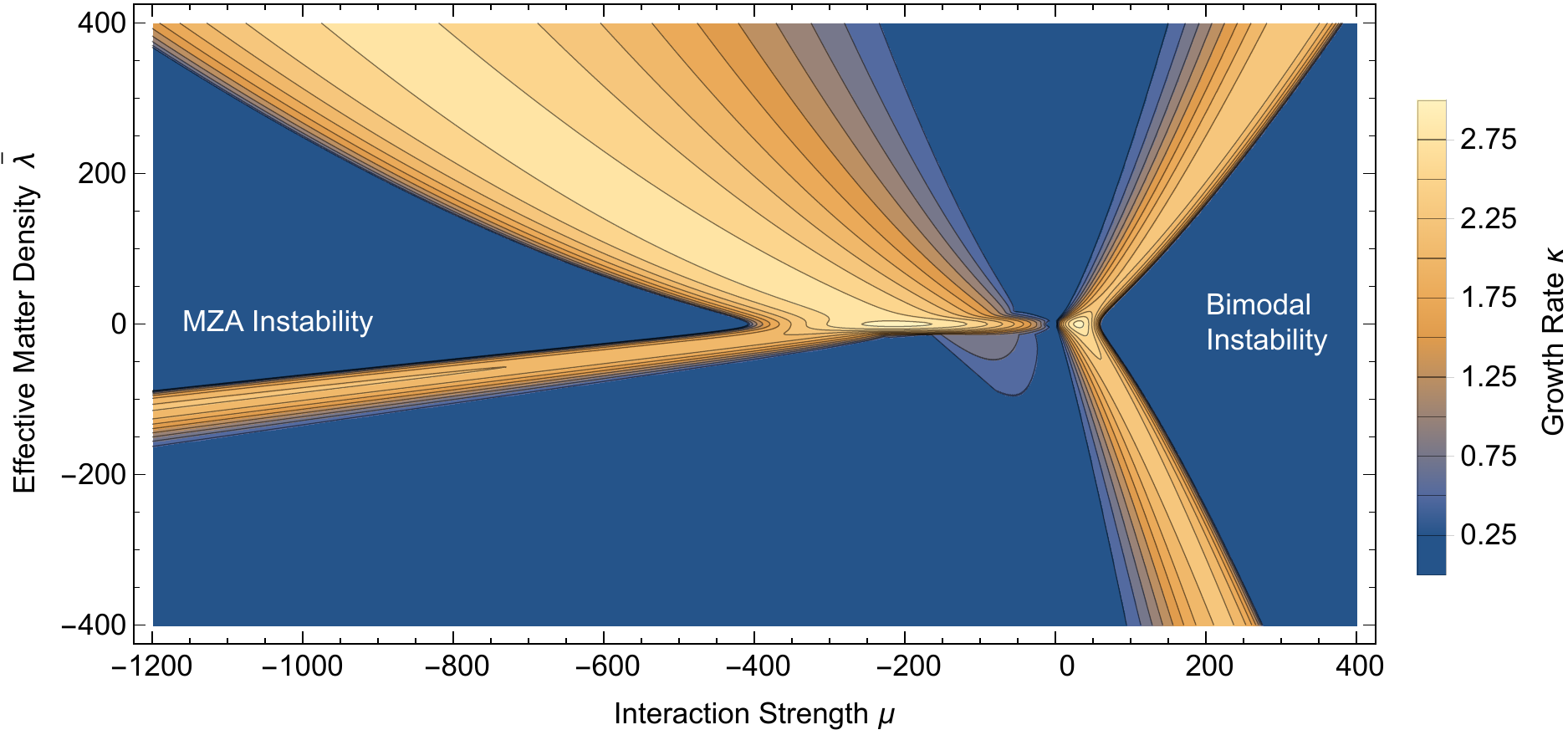}
\vskip12pt
\includegraphics[width=1.0\textwidth]{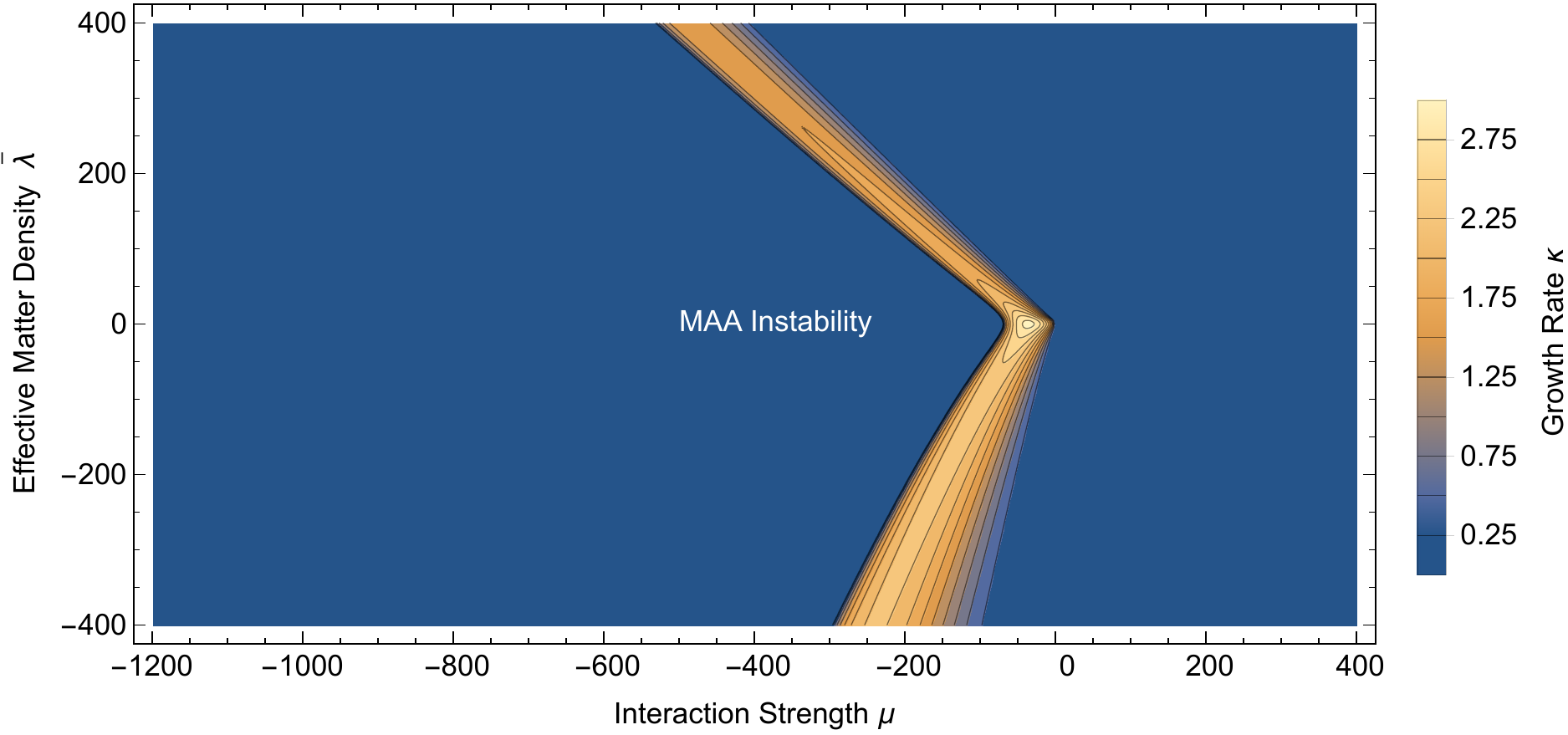}
\caption{Growth rate $\kappa$ of the 1D instabilities as a function of
  $\mu$ and $\bar\lambda$, assuming $\alpha=1/2$.
  {\em Upper panel:}
  The first block in the eigenvalue equation~\eqref{eq:1D-determinants} yields
  the bimodal instability for $\mu>0$ and the
  multi-zenith-angle (MZA) instability for $\mu<0$.
  {\em Lower panel:} The second block in equation~\eqref{eq:1D-determinants} provides the
  multi-azimuth-angle (MAA) instability for $\mu<0$.
  Notice that the first quadrant ($\mu,\bar\lambda>0$) represents
  IH, the third quadrant  ($\mu,\bar\lambda<0$) represents NH.} \label{fig:1D-contour}
\end{figure}

For the homogeneous mode ($k=0$), the eigenvalue equation \eqref{eq:1D-matrix-equation}
simplifies considerably because in this case $I_1=I_3=0$ and we are left with
\begin{equation}\label{eq:1D-matrix-equation-homogeneous}
\[1-\begin{pmatrix}I_2&0&I_4\\0&-2I_2&0\\I_0&0&I_2\end{pmatrix}\]
\begin{pmatrix}A_0\\A_1\\A_2\end{pmatrix}=0\,.
\end{equation}
We need to solve the two equations
\begin{equation}\label{eq:1D-determinants}
(I_2-1)^2=I_0 I_4
\quad\hbox{and}\quad
I_2=-1/2\,.
\end{equation}
As an
overview, we show in figure~\ref{fig:1D-contour} a contour plot of the growth
rate $\kappa$ in the two-dimensional parameter space of the interaction
strength $\mu$ and the effective matter density
$\bar\lambda=\lambda+\epsilon\mu$.

As explained earlier, the first quadrant ($\mu>0$ and $\bar\lambda>0$)
corresponds physically to inverted mass ordering (IH), whereas the third
quadrant ($\mu<0$ and $\bar\lambda<0$) corresponds to normal mass ordering
(NH). The other quadrants would be relevant, for example, for a background
medium of antimatter. Mathematically, $\mu$ and $\bar\lambda$ are simply
parameters which we leave unconstrained by physical considerations. As usual,
we use $\alpha=1/2$ and therefore $\epsilon=1-\alpha=1/2$ so that matter-free
space ($\lambda=0$) corresponds to the line $\bar\lambda=\mu/2$.

For $\bar\lambda=0$, the growth rates as a function of $\mu$ were shown in
figure~\ref{fig:kappa-1D-multi-angle} in the form of the colored curves. The
effect of increasing $|\bar\lambda|$ is to shift the unstable regions to
larger values of $|\mu|$,
creating the butterfly image seen in figure~\ref{fig:1D-contour}.
For $\mu>0$ we obtain the bimodal instability
which exists even in a single-angle treatment. For $\mu<0$ we have two
multi-angle instabilities as indicated. The solutions shown in the upper
panel derive from the first block in the eigenvalue equation~\eqref{eq:1D-determinants},
providing the bimodal and MZA instabilities. The MAA solution shown in the
lower panel derives from the second block in equation~\eqref{eq:1D-determinants}.

The contours for the different instabilities are quite different.  In
the regime of large $\bar\lambda$, one can expand the eigenvalue
equations in powers of $1/\bar\lambda$ and identify analytically the
behaviour of the footprint of the instability regions in the
$\mu$-$\bar\lambda$-plane (Appendix~\ref{sec:asymptotic}). We show the
footprint of the contours of figure~\ref{fig:1D-contour} on a
logarithmic scale in Appendix~\ref{sec:asymptotic} in
figure~\ref{fig:footprint-1D-asymp} together with the asymptotic
large-$\bar\lambda$ expansion.

\subsubsection[Inhomogeneous mode ($k>0$)]{Inhomogeneous mode (\boldmath{$k>0$)}}

\begin{figure}[!b]
\centering
\includegraphics[width=1.\textwidth]{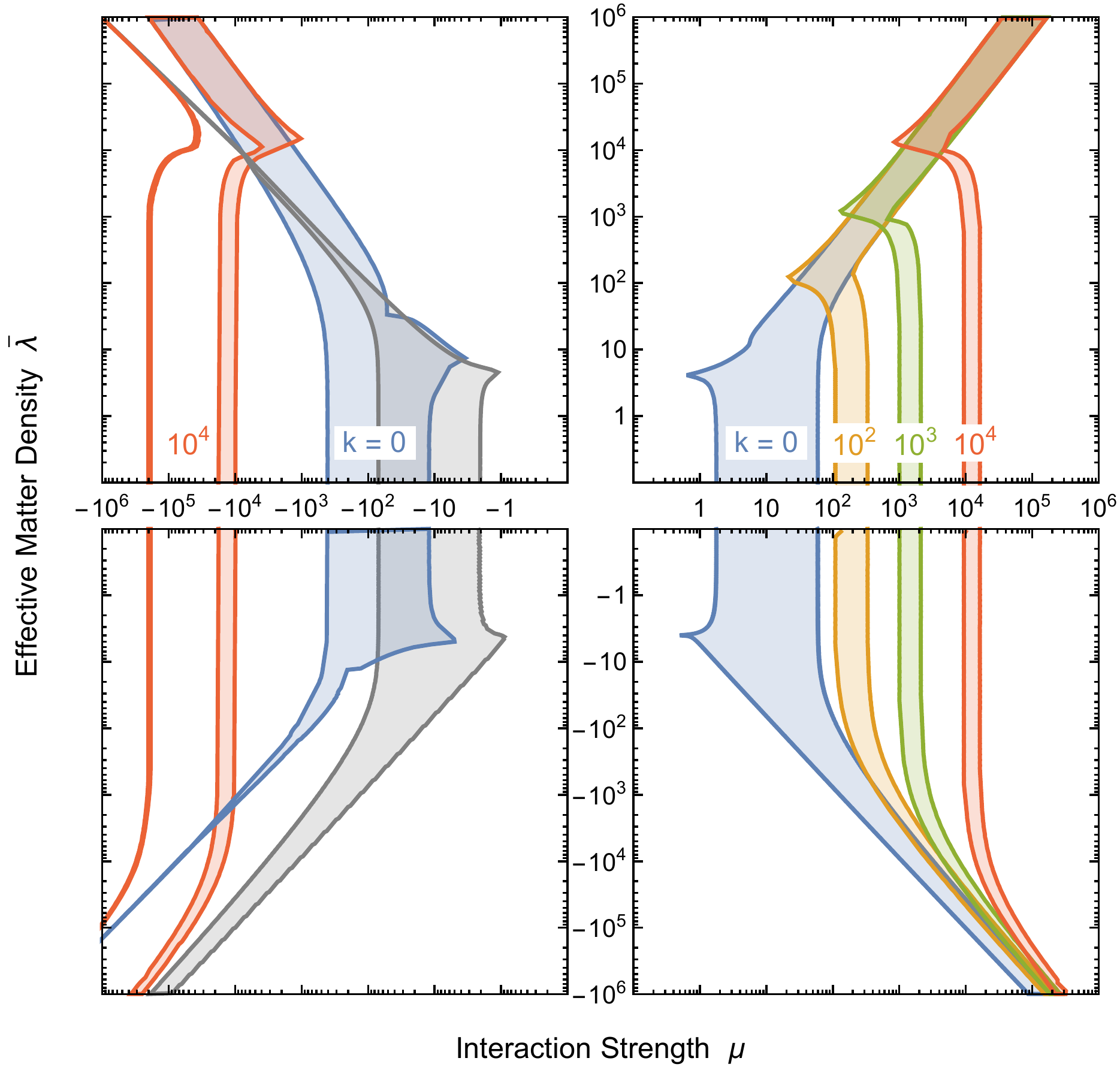}
\caption{Footprint of the 1D instabilities ($\kappa > 10^{-2}$)  in the
  $\mu$-$\bar\lambda$-plane for $\alpha=1/2$ and the indicated values
  of $k$. The homogeneous case ($k=0$) is the footprint of the contour
  plot of figure~\ref{fig:1D-contour}, here on a logarithmic scale. The
  corresponding large-$\bar\lambda$ asymptotic results are shown in
  Appendix~\ref{sec:asymptotic}, figure~\ref{fig:footprint-1D-asymp}.  }
\label{fig:footprint-1D-k}
\end{figure}

We next determine numerically the same instability footprints for
non-vanishing wave numbers $k$. We already know the impact of nonzero
$k$ for $\bar\lambda=0$, i.e., the instability is shifted to larger
$\mu$ values. We do not expect a big difference
for $\bar\lambda\gg k$ relative to the $k=0$ case. These
expectations are borne out by our results shown in
figure~\ref{fig:footprint-1D-k}. Considering first the simpler $\mu>0$
half of the plot, we show the instability footprints for $k=0$,
$10^2$, $10^3$ and $10^4$ as indicated in the plot. We can now easily
diagnose the impact of small-scale instabilities: Essentially they
fill in the entire space between the $k=0$ footprints between the
quadrant with positive and negative $\bar\lambda$, whereas the region
above the $k=0$ footprint in the upper quadrant, and the space below
in the lower quadrant remains stable.  The only small caveat is that
in the upper quadrant, for $k\sim \bar\lambda$, there are ``noses'' of
the footprint sticking into the previously stable region. So there is
a narrow sliver of parameters above the $k=0$ footprint, the envelope
of the noses, which becomes unstable due to small-scale instabilities.

\newpage

In the left half of the plot ($\mu<0$) the situation is somewhat more
complicated because of the presence of two instabilities. For large $k$, the
footprint actually connects asymptotically to the $k=0$ instabilities in a
crossed-over way which we illustrate only by the $k=10^4$ case. In other
words, for large $k$, the MAA and MZA instabilites strongly mix with each
other.

%%%%%%%%%%%%%%%%%%%%%%%%%%%%%%%%%%%%%%%%%%%%%%%%%%%%%%%%%%%%%%%%%%%%%%%%%%%%%%
\section{Two-dimensional system}
\label{sec:2D}
%%%%%%%%%%%%%%%%%%%%%%%%%%%%%%%%%%%%%%%%%%%%%%%%%%%%%%%%%%%%%%%%%%%%%%%%%%%%%%

We now turn to a 2D system, corresponding to the SN example where neutrinos
propagate within the ``expanding transverse sheet'' moving outward. The
radial motion is parameterized by our ``time'' variable, whereas ``space'' is
represented by two transverse directions. This example corresponds, with properly
scaled variables $\mu$ and $\lambda$, to the usual treatment of self-induced
flavor conversion in SNe, except that now we can include small-scale instabilities
with nonvanishing wavenumber $k$.

\subsection[Single angle ($|\bv|=1$)]{Single angle (\boldmath{$|\bv|=1$})}

\subsubsection{Eigenvalue equation}

We begin with the ``single
zenith angle case,'' meaning that the rescaled neutrino speed within the
transverse sheet is $|\bv|=1$ and the matter effect can be rotated away. Our
velocity phase space is the unit circle, described by an angle variable
$\varphi$ which we can measure relative to $\bk$. As the system is initially
prepared axially symmetric, all vectors $\bk$ are equivalent---the
eigenvalues depend only on $k=|\bk|$. The eigenvalue
equation~\eqref{eq:master} therefore reads
\begin{equation}\label{eq:master3}
\(k\,c_\varphi+\omega-\Omega\)Q_{\Omega,k,\omega,\varphi}
=\mu\int_{-\infty}^{+\infty} d\omega'\,h_{\omega'}\,\frac{1}{2\pi}
\int_{-\pi}^{+\pi} d\varphi'\,(1-c_\varphi c_{\varphi'}-s_\varphi s_{\varphi'})
\,Q_{\Omega,k,\omega',\varphi'}\,,
\end{equation}
where $c_\varphi=\cos\varphi$ and $s_\varphi=\sin\varphi$.

Proceeding as in our earlier cases, we notice that the r.h.s.\ of
equation~\eqref{eq:master3} has the form $A_1+A_{\rm c} \cos\varphi+A_{\rm
s}\sin\varphi$, i.e., a superposition of three linearly independent functions
on the interval $-\pi\leq\varphi\leq+\pi$. Therefore, the l.h.s.\ must be of
that form as well and we may use the eigenfunction ansatz
\begin{equation}\label{eq:eigenfunctionansatz2}
Q_{\Omega,k,\omega,\varphi}=
\frac{A_1+A_{\rm c}c_\varphi+A_{\rm s}s_\varphi}{k\,c_\varphi+\omega-\Omega}\,.
\end{equation}
We may insert this form on both sides, leading to three linearly independent
equations, corresponding to the coefficients of the three functions $1$,
$\cos\varphi$ and $\sin\varphi$. We may write the three equations in compact form
\begin{equation}\label{eq:2D-matrix-equation-1}
\[1-\begin{pmatrix}I_1&I_{\rm c}&0\\-I_{\rm c}&-I_{\rm cc}&0\\0&0&-I_{\rm ss}\end{pmatrix}\]
\begin{pmatrix}A_0\\A_{\rm c}\\A_{\rm s}\end{pmatrix}=0\,,
\end{equation}
where
\begin{equation}\label{eq:In-defintion-2}
I_{a}=\mu\int_{-\infty}^{+\infty} d\omega\,h_{\omega}\frac{1}{2\pi}
\int_{-\pi}^{+\pi} d\varphi\,\frac{f_a(\varphi)}{k\,c_\varphi+\omega-\Omega}\,.
\end{equation}
Here, $f_1(\varphi)=1$, $f_{\rm c}(\varphi)=\cos\varphi$, $f_{\rm
cc}(\varphi)=\cos^2\varphi$, and $f_{\rm ss}(\varphi)=\sin^2\varphi$. We have
used that such integrals vanish if they involve a single power of
$\sin\varphi$ because this is anti-symmetric on the integration interval,
explaining the zeroes in the matrix in equation~\eqref{eq:2D-matrix-equation-1}.
We also note that $I_{\rm ss}=I_1-I_{\rm cc}$, so we need only three
different integrals.

\subsubsection[Homogeneous mode ($k=0$)] {Homogeneous mode (\boldmath{$k=0$})}
\label{sec:Hm2DSingAng}

We first consider homogeneous solutions ($k=0$) where the angle integrals in
equation~\eqref{eq:In-defintion-2} can be performed explicitly. Using the
monochromatic frequency spectrum of equation~\eqref{eq:monochromaticspectrum}, the
eigenvalue equation becomes
\begin{equation}\label{eq:simplematrix-2}
{\rm det}\[\omega_0^2-\Omega^2-\mu
\begin{pmatrix}1&0&0\\0&-\frac{1}{2}&0\\
0&0&-\frac{1}{2}\end{pmatrix}
\[(1+\alpha)\,\omega_0+(1-\alpha)\,\Omega\]\]=0\,.
\end{equation}
These are three independent quadratic equations of the now-familiar
form of equation~\eqref{eq:quadratic1}. The first line corresponds to
the usual bimodal solution, the second and third line to two
degenerate multi-azimuth-angle (MAA) solutions which are unstable for
negative $\mu$ (normal hierarchy). For these modes, the instability
range is a factor of 2 larger.

\subsubsection[Inhomogeneous modes ($k>0$)]{Inhomogeneous modes (\boldmath{$k>0$})}

For the inhomogeneous modes ($k>0$), the required integrals entering
the eigenvalue equation are of the form
\begin{equation}\label{eq:In-3}
I_{a}=\frac{\mu}{k}\int_{-\infty}^{+\infty}
d\omega\,h_{\omega}F_a\(\frac{\omega-\Omega}{k}\)
\quad\hbox{where}\quad
F_a(w)=\frac{1}{2\pi}
\int_{-\pi}^{+\pi} d\varphi\,\frac{f_a(\varphi)}{\cos \varphi+w}\,.
\end{equation}
With our monochromatic spectrum equation~\eqref{eq:monochromaticspectrum}
we arrive at
\begin{equation}\label{eq:In-4}
I_{a}=\frac{\mu}{k}\[F_a\(\frac{\omega_0-\Omega}{k}\)
-\alpha\,F_a\(\frac{-\omega_0-\Omega}{k}\)\]\,.
\end{equation}
We define the auxiliary function of a complex argument $w$
\begin{equation}
s(w)=\sqrt{w-1}\sqrt{w+1}\,,
\end{equation}
which, for complex numbers, is in general not equal to
$\sqrt{w^2-1}$. We then find
\begin{equation}
F_1=\frac{1}{s(w)}\,,\quad
F_{\rm c}=1-\frac{w}{s(w)}\,,\quad
F_{\rm cc}=-w+\frac{w^2}{s(w)}\,,\quad
F_{\rm ss}=F_1-F_{\rm cc}=w-s(w)\,.
\end{equation}
These expressions allow us to write the eigenvalue equations
explicitly, involving only polynomials and square-root expressions.

\begin{figure}
\centering
\includegraphics[width=0.8\textwidth]{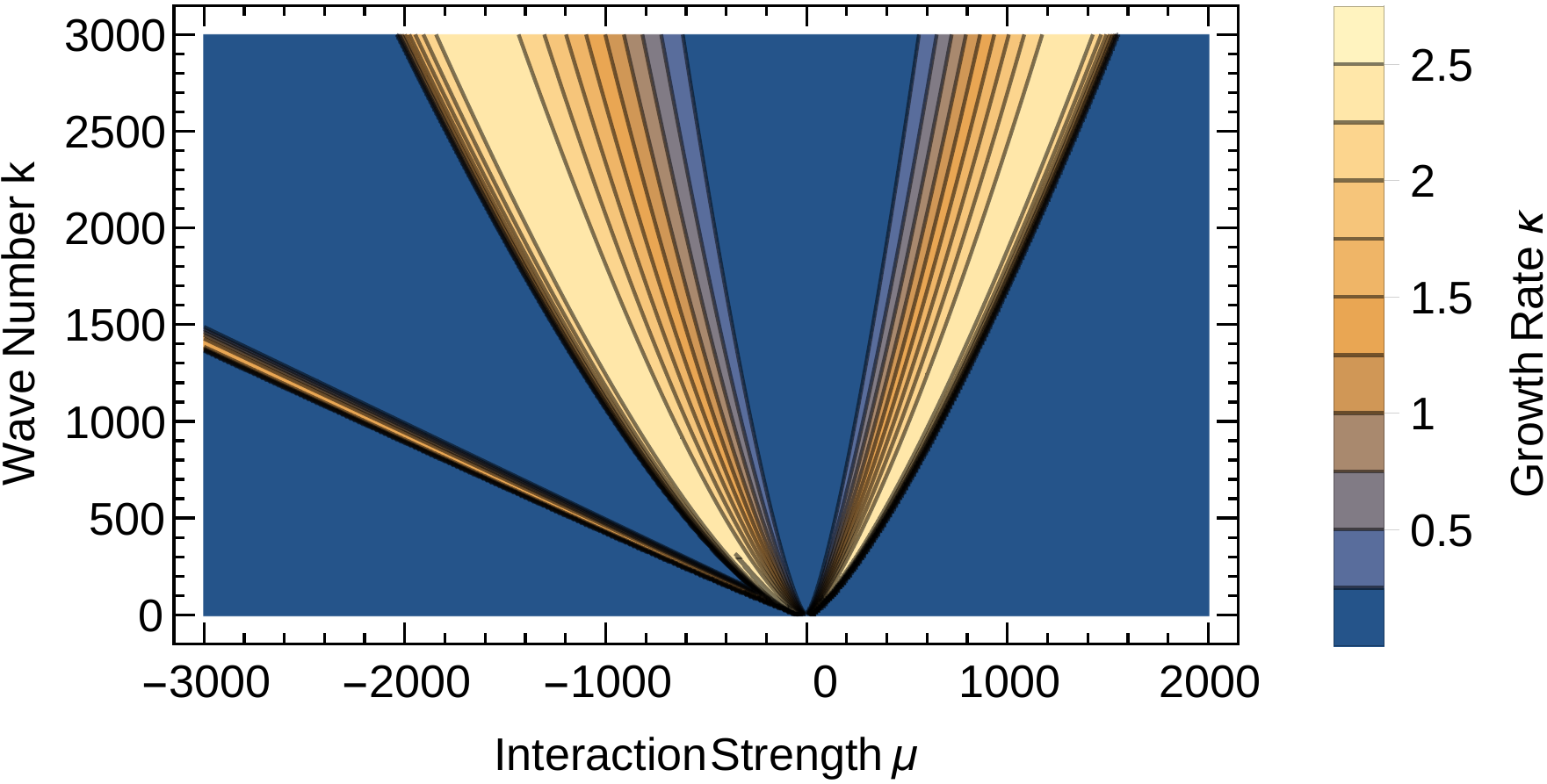}
\caption{Growth rates for the 2D case with $|\bv|=1$ (``single zenith
  angle''), using $\alpha=1/2$. The two quasi-symmetric regions
  correspond to the two instabilities which already appear in the 1D
  case \cite{Duan:2014gfa}, although here the unstable $\mu$ range
  shifts with $k^{3/4}$.  The third instability is genuinely 2D, it
  has no counterpart in the ``colliding beam'' examples, and its
  unstable $\mu$-range shifts linearly with
  $k$.} \label{fig:2D-contour-1}
\end{figure}

We now have three non-degenerate solutions, in contrast to the original 1D
example of Duan and Shalgar \cite{Duan:2014gfa}, one for positive $\mu$
(inverted hierarchy) and two for negative $\mu$. We show contour plots of the
growth rate $\kappa$ as a function of $\mu$ and $k$ for our usual example
$\alpha=1/2$ in figure~\ref{fig:2D-contour-1}. The two quasi-symmetric
regions correspond to the two
solutions which correspond to those of the 1D case, i.e., the left-right
symmetric and anti-symmetric cases also shown in
reference~\cite{Duan:2014gfa} in a similar plot. In addition we see a third
solution which is genuinely a result of the spatial 2D geometry with non-vanishing
wave-vector $\bk$. The eigenfunctions in this case
are proportional to $\sin\varphi$ where
$\varphi$ is the angle between $\bk$ and the velocity $\bv$ of a given mode.

The eigenvalue equation for the decoupled ss-block of
equation~\eqref{eq:2D-matrix-equation-1}, the new genuine 2D solution, is
explicitly
\begin{eqnarray}\label{eq:2D-ss-eigenvalue}
(1+\alpha)\,\omega_0-(1-\alpha)\,\Omega
&+&\alpha\sqrt{-k-\omega_0-\Omega}\sqrt{k-\omega_0-\Omega}
\nonumber\\
&&\kern2em{}-\sqrt{-k+\omega_0-\Omega}\sqrt{k+\omega_0-\Omega}
=-\frac{k^2}{\mu}\,,
\end{eqnarray}
which is a quartic equation.
It has unstable solutions for $\mu<0$. In analogy to the discussion of the 1D
case, we can extract the large-$k$ limiting solution with the substitution
$\Omega=\tilde\Omega-k$ and $\mu=-k/(1-\alpha)-\sqrt{4\,m\,\omega_0\,k\,
(1+\alpha)/(1-\alpha)^3}$, where $m$ is a dimensionless variable. The
detailed factors are not crucial and are the result of some tinkering. After the
substitution one expands equation~\eqref{eq:2D-ss-eigenvalue} for large $k$ and
keeps only the term proportional to the highest power of $k$. One finds
unstable solutions for $0\leq m\leq 1$ and the growth rate
$\kappa/\omega_0=4\sqrt{m(1-m)}\,\alpha/(1-\alpha^2)$ with $\kappa_{\rm
max}/\omega_0=2\alpha/(1-\alpha^2)$ which is $4/3$ for our example
$\alpha=1/2$.
The unstable $\mu$-range scales linearly with $k$. Notice
that the asymptotic solution obtains only for very large $k$-values because
the dominant term in the expansion of equation~\eqref{eq:2D-ss-eigenvalue} is
proportional to $k^{3/2}$, the next one proportional to $k$, and becomes
relatively unimportant only very slowly.

The $2{\times}2$ block in equation~\eqref{eq:2D-matrix-equation-1} leads to a
much more complicated equation, but still consists of polynomials and
square-roots. We follow a similar approach and substitute
$\Omega=\tilde\Omega-k$ and $\mu=m\,\omega_0^{1/4} k^{3/4}/\sqrt{1-\alpha}$,
expand the eigenvalue equation in powers of $k$ and keep only the largest
term, leading to
\smash{$\sqrt{2}\,\sqrt{-\omega_0-\tilde\Omega}\sqrt{\omega_0-\tilde\Omega}
=m^2\sqrt{\vphantom{2}\omega_0}\bigl(\sqrt{-\omega_0-\tilde\Omega}-\alpha\sqrt{\omega_0-\tilde\Omega}\bigr)$}.
This quartic equation provides the large-$k$ solution, which is symmetric for
both hierarchies, i.e., symmetric under $m\to-m$. Because the second-largest
power of $k$ is only a power 1/4 smaller than the largest, the asymptotic
solution requires extremely large $k$-values. Notice that in the 1D beam
example, the unstable range scaled with $k^{1/2}$, while here it is
$k^{3/4}$.

\subsection[Multi-angle effects ($0\leq v\leq1$)]
{Multi-angle effects (\boldmath{$0\leq v\leq1$})}

\subsubsection{Eigenvalue equation}

We finally turn to the ``multi zenith angle'' case of transverse
velocities within the full 2D disk described by $|\bv|\leq 1$, meaning
that multi-angle matter effects are now included. We describe the
velocity phase space by the speed $v=|\bv|$ and an angle variable
$\varphi$ which we measure relative to $\bk$ as before. Noting that
$(1/\Gamma_\bv)\int d\bv=(1/\pi)\int_{-\pi}^{+\pi}d\varphi\int_0^1
dv\,v$, the eigenvalue equation~\eqref{eq:master} becomes
\begin{eqnarray}\label{eq:master4}
&&\(\half\bar\lambda v^2+k\,v\,c_\varphi+\omega-\Omega\)Q_{\Omega,k,\omega,v,\varphi}
\nonumber\\[2ex]
&&=\mu\int_{-\infty}^{+\infty} d\omega'\,h_{\omega'}
\int_{-\pi}^{+\pi} \,\frac{d\varphi'}{\pi}\int_0^1 dv'\,v'\,
\[\half(v'^2+v^2)-vv'\,(c_\varphi c_{\varphi'}+s_\varphi s_{\varphi'})\]
\,Q_{\Omega,k,\omega',v',\varphi'}\,,\kern3em
\end{eqnarray}
where $c_\varphi=\cos\varphi$ and $s_\varphi=\sin\varphi$. The
r.h.s.\ as a function of $v$ and $\varphi$ is
$A_0+A_2 v^2+A_{\rm c} v\,c_\varphi+A_{\rm s} v\,s_\varphi$, so
the eigenfunctions are of the form
\begin{equation}\label{eq:eigenfunctionansatz3}
Q_{\Omega,k,\omega,v,\varphi}=
\frac{A_0+A_2 v^2+A_{\rm c}v\,c_\varphi+A_{\rm s}v\,s_\varphi}
{\half\bar\lambda v^2+k\,v\,c_\varphi+\omega-\Omega}\,.
\end{equation}
As usual, we insert this form on both sides, leading to four linearly independent
equations, corresponding to the coefficients of the four functions
$1$, $v^2$, $v\,c_\varphi$ and $v\,s_\varphi$. We may write the equations in compact
form
\begin{equation}\label{eq:2D-matrix-equation-2}
\[1-\begin{pmatrix} I^1_3& I^1_5& I^{\rm c}_4&0\\
 I^1_1& I^1_3& I^{\rm c}_2&0\\
-2I^{\rm c}_2&-2I^{\rm c}_4&-2I^{\rm cc}_3&0\\
0&0&0&-2I^{\rm ss}_3\end{pmatrix}\]
\begin{pmatrix}A_0\\A_2\\A_{\rm c}\\A_{\rm s}\end{pmatrix}=0\,,
\end{equation}
where we have used that $\int d\varphi'$ vanishes for integrals
involving an odd power of $\sin\varphi'$.
The integrals are now
\begin{equation}\label{eq:In-defintion-3}
I^a_n=\mu\int_{-\infty}^{+\infty} d\omega\,h_{\omega}
\int_{-\pi}^{+\pi} \frac{d\varphi}{2\pi}\int_0^1 dv
\,\frac{v^n\,f_a(\varphi)}{\half\bar\lambda v^2+k\,v\,c_\varphi+\omega-\Omega}\,.
\end{equation}
Here, $f_1(\varphi)=1$, $f_{\rm c}(\varphi)=\cos\varphi$, $f_{\rm
cc}(\varphi)=\cos^2\varphi$, and $f_{\rm ss}(\varphi)=\sin^2\varphi$.
We also note that $I^{\rm ss}_3=I^1_3-I^{\rm cc}_3$.

\begin{figure}[!b]
\centering
\includegraphics[width=1.0\textwidth]{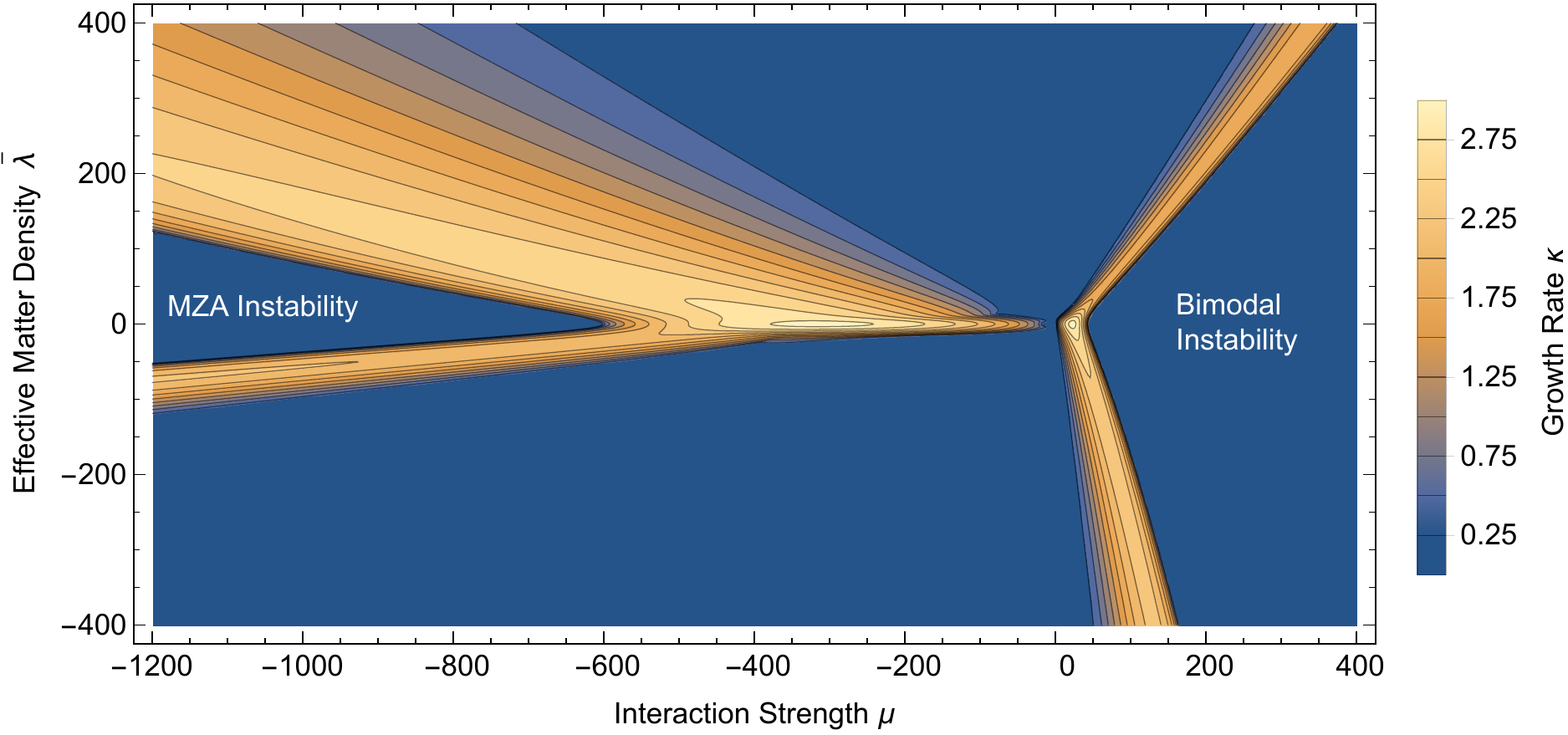}
\vskip12pt
\includegraphics[width=1.0\textwidth]{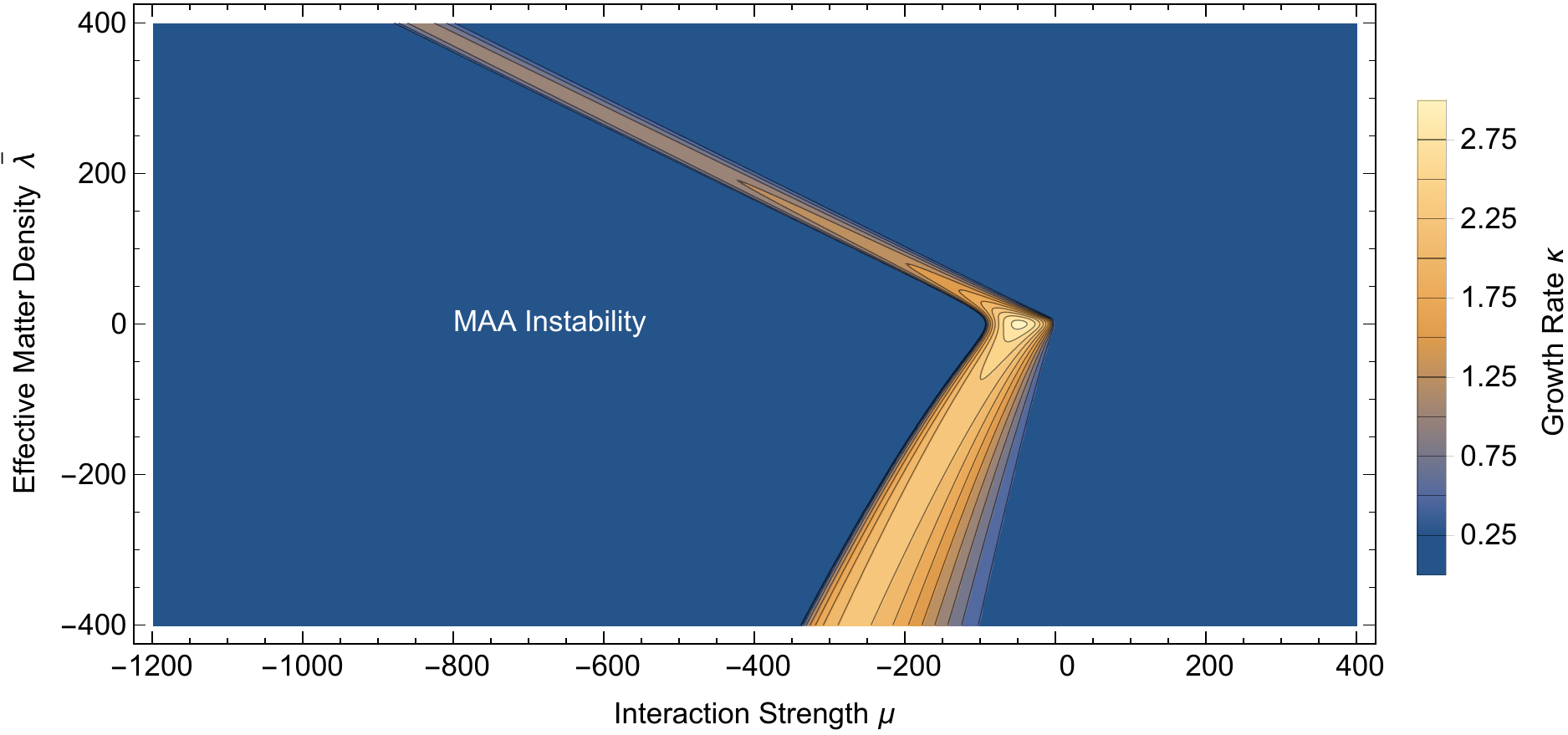}
\caption{Growth rate $\kappa$ of the 2D instabilities as a function of
  $\mu$ and $\bar\lambda$, assuming $\alpha=1/2$.
  {\em Upper panel:}
  The first block in the eigenvalue equation~\eqref{eq:2D-determinants} yields
  the bimodal instability for $\mu>0$ (inverted hierarchy) and the
  multi-zenith-angle (MZA) instability for $\mu<0$ (normal hierarchy).
  {\em Lower panel:} The second block in equation~\eqref{eq:2D-determinants} provides the
  multi-azimuth-angle (MAA) instability for $\mu<0$.
This figure is analogous to the corresponding 1D case shown in
  figure~\ref{fig:1D-contour}.} \label{fig:2D-contour}
\end{figure}

\subsubsection[Homogeneous mode ($k=0$) with matter ($\bar\lambda>0$)]
{Homogeneous mode (\boldmath{$k=0$) with matter (\boldmath{$\bar\lambda>0$})}}

In the homogeneous case ($k=0$), the function $\cos\varphi$ in the
denominator disappears, all angle integrals can be performed
analytically, and $I^{\rm c}_n=0$. Therefore,
equation~\eqref{eq:2D-matrix-equation-2} becomes
\begin{equation}\label{eq:2D-matrix-equation-3}
\[1-\begin{pmatrix} I_3& I_5&0&0\\
 I_1& I_3&0&0\\
0&0&-I_3&0\\
0&0&0&-I_3\end{pmatrix}\]
\begin{pmatrix}A_0\\A_2\\A_{\rm c}\\A_{\rm s}\end{pmatrix}=0\,,
\end{equation}
where the integral expressions after performing the $d\varphi$
integration are
\begin{equation}\label{eq:In-defintion-3a}
I_n=\mu\int_{-\infty}^{+\infty} d\omega\,h_{\omega} K_n\,,
\quad\hbox{where}\quad
K_n=\int_0^1 dv
\,\frac{v^n}{\half\bar\lambda v^2+\omega-\Omega}\,.
\end{equation}
The velocity integrals are explicitly
\begin{subequations}
\begin{eqnarray}
K_1&=&\frac{1}{\bar\lambda}\,\log\(1+\frac{\bar\lambda}{2(\omega-\Omega)}\)\,,
\\
K_3&=&\frac{1}{\bar\lambda}\,
\[1-\frac{2(\omega-\Omega)}{\bar\lambda}\,\log\(1+\frac{\bar\lambda}{2(\omega-\Omega)}\)\]
\,\\
K_5&=&\frac{1}{\bar\lambda}\,
\[\frac{1}{2}
-\frac{2(\omega-\Omega)}{\bar\lambda}
+\(\frac{2(\omega-\Omega)}{\bar\lambda}\)^2
\log\(1+\frac{\bar\lambda}{2(\omega-\Omega)}\)\]\,.
\end{eqnarray}
\end{subequations}
The previous ss and cc blocks in
equation~\eqref{eq:2D-matrix-equation-2} have now decoupled from the rest
and are degenerate, leading to the MAA instability. The remaining
$2{\times}2$ block provides the bimodal and MZA instability. In other words, we now need
to solve
\begin{equation}\label{eq:2D-determinants}
(I_3-1)^2=I_1 I_5
\quad\hbox{and}\quad
I_3=-1\,,
\end{equation}
in analogy to reference~\cite{Raffelt:2013rqa} with a slightly
different notation. This entire development is very similar to the 1D
case.

In the limit $\bar\lambda\to0$, $K_1$, $K_3$ and $K_5$
approach 1/2,
1/4 and 1/6 times $1/(\omega-\Omega)$, which one can also find by
setting $\bar\lambda=0$ before doing the $v$-integrations.  The matrix
can be diagonalized and we find three independent quadratic equations
of the form of equation~\eqref{eq:quadratic1} where we need to
substitute $\mu\to-\mu/4$ and
$\mu\to\mu\,(3\pm2\sqrt{3})/12$. Following the same steps as in
Sec.~\ref{sec:1D-homogeneous-nomatter}, the instability ranges for our
usual example $\alpha=1/2$ are found to be
\begin{subequations}\label{eq:2Dinstabilityranges}
\begin{eqnarray}
1.27<&\mu/\omega_0&<43.28
\,,\\
-93.25<&\mu/\omega_0&<-2.75
\,,\\
-602.81<&\mu/\omega_0&<-17.75
\,.
\end{eqnarray}
\end{subequations}
These results are numerically similar to equation~\eqref{eq:1Dinstabilityranges} for
the corresponding 1D-case.

In analogy to the butterfly diagram of the 1D case
(figure~\ref{fig:1D-contour}) we show in figure~\ref{fig:2D-contour} a
contour plot of the growth rate $\kappa$ in the two-dimensional
parameter space of the interaction strength $\mu$ and the effective
matter density $\bar\lambda=\lambda+\epsilon\mu$. The result looks
qualitatively similar to the 1D case. Again, for $\mu>0$ (inverted
mass ordering), we obtain the bimodal instability while for $\mu<0$
(normal ordering) we find the MZA and MAA instabilities.

\subsubsection[Inhomogeneous mode ($k>0$) without matter ($\bar\lambda=0$)]
{Inhomogeneous mode (\boldmath{$k>0$) without matter (\boldmath{$\bar\lambda=0$})}}

Next we consider the relatively simple case of $k>0$ without matter. We may write
the integrals of equation~\eqref{eq:In-defintion-3} in the form
\begin{equation}\label{eq:In-defintion-4}
I^a_n=\frac{\mu}{k}\int_{-\infty}^{+\infty} d\omega\,h_{\omega}\,K^a_n\,,
\quad\hbox{where}\quad
K^a_n=\int_{-\pi}^{+\pi} \frac{d\varphi}{2\pi}\int_0^1 dv
\,\frac{v^n\,f_a(\varphi)}{v\,c_\varphi+w}
\end{equation}
and we have introduced $w=(\omega-\Omega)/k$. We find explicitly
\begin{subequations}\label{eq:In-defintion-5}
\begin{eqnarray}
K^1_1&=&w+\frac{\sqrt{1-w}\sqrt{-w(1+w)}}{\sqrt{w}}\,,\\
K^1_3&=&\frac{2w^3}{3}+\frac{\sqrt{-w}\sqrt{1-w^2}\,(1+2w^2)}{3\sqrt{w}}\,,\\
K^1_5&=&\frac{8w^5}{15}+\frac{\sqrt{-w}\sqrt{1-w^2}\,(3+4w^2+8w^4)}{15\sqrt{w}}\,,\\
K^{\rm c}_2&=&\frac{1}{2}-w^2-\sqrt{-w^2}\sqrt{1-w^2}\,,\\
K^{\rm c}_4&=&\frac{1}{4}-\frac{2w^4+\sqrt{-w^2}\sqrt{1-w^2}\,(1+2w^2)}{3}\,,\\
K^{\rm cc}_3&=&-w\,\(\frac{1}{2}-w^2-\sqrt{-w^2}\sqrt{1-w^2}\)\,,\\
K^{\rm ss}_3&=&\frac{w(3-2w^2)}{6}+\frac{\sqrt{-w}(1-w^2)^{3/2}}{3\sqrt{w}}\,.
\end{eqnarray}
\end{subequations}
Notice that $K^{ss}_3+K^{cc}_3=K_3^1$.

\begin{figure}[ht]
\centering
\includegraphics[width=0.8\textwidth]{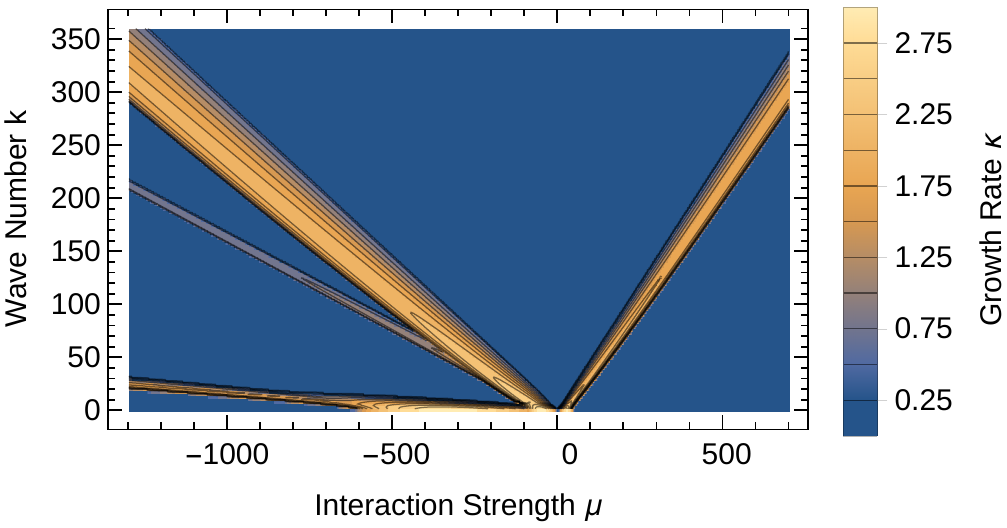}
\caption{Growth rates for the 2D case with $0<|\bv|=1$ (``multi zenith angle''),
using $\alpha=1/2$. This figure is analogous to the single-angle case shown in
figure~\ref{fig:2D-contour-1}, but now we have three instabilities for $\mu<0$
(normal mass ordering) and the usual bimodal one for $\mu>0$ (inverted ordering).
For all four instabilities, the unstable $\mu$ range scales linearly with $k$
as discussed in the text.} \label{fig:2D-contour-2}
\end{figure}

With the help of these analytic integrals it is relatively easy to solve the
eigenvalue equation numerically. We show a contour plot of the growth rate
$\kappa$ in the $\mu$-$k$-plane in figure~\ref{fig:2D-contour-2}. The $3{\times}3$ block
in equation~\eqref{eq:2D-matrix-equation-2} provides three different solutions, i.e.,
one for $\mu>0$ (the usual bimodal solution in inverted mass ordering) and
two solutions for $\mu<0$ (normal ordering). The $1{\times}1$ block provides a further
solution for $\mu<0$.  Figure~\ref{fig:2D-contour-2} corresponds to
figure~\ref{fig:2D-contour-1} in the single-angle case. In comparison, we have one more
instability, now a total of four, of which three are for $\mu<0$ (normal mass ordering).

One may also extract the large-$k$ asymptotic behavior
(Appendix~\ref{sec:2D-k-asymp}). For the $3{\times}3$ block
in equation~\eqref{eq:2D-matrix-equation-2} one finds that
the system is unstable for
\begin{equation}
\mu=a_i\(k+m\sqrt{k}\)
\qquad\hbox{for}\qquad
0<m<m_{{\rm max},i},
\qquad\hbox{where~~}
i=1,\ldots,3\,,
\end{equation}
i.e., the unstable $\mu$ region scales linearly with $k$, in contrast
to the corresponsing 1D case. The values of the coefficients $a_i$ and
$m_{{\rm max},i}$ are given in Appendix~\ref{sec:2D-k-asymp}.

For the $1{\times}1$ block
in equation~\eqref{eq:2D-matrix-equation-2} one finds an instability on a
very narrow strip around $\mu=-6\,k$. However, the maximum growth rate
decreases with $k^{-1/2}$ so that in the limit $k\to\infty$ this
instability disappears and we are left with those arising from
the $3{\times}3$ block.

\subsubsection[Inhomogeneous mode ($k>0$) with matter ($\bar\lambda>0$)]
{Inhomogeneous mode (\boldmath{$k>0$) with matter (\boldmath{$\bar\lambda>0$})}}

As a grand finale, we now turn to the most general 2D case with matter
($\bar\lambda>0$) and inhomogeneities ($k>0$). We need to find the
zeroes of the determinant in
equation~\eqref{eq:In-defintion-3} and write the integrals in the form
\begin{equation}\label{eq:In-defintion-5a}
I^a_n=\frac{\mu}{\bar\lambda}\int_{-\infty}^{+\infty} d\omega\,h_{\omega}\,K^a_n\,,
\quad\hbox{where}\quad
K^a_n=\int_{-\pi}^{+\pi} \frac{d\varphi}{2\pi}\int_0^1 dv
\,\frac{v^n\,f_a(\varphi)}{v^2/2+q\,v\,c_\varphi+w}\,,
\end{equation}
where $q=k/\bar\lambda$ and $w=(\omega-\Omega)/\bar\lambda$. These
integrals can be performed analytically; we provide our results in
Appendix~\ref{sec:integrals}.

\begin{figure}[ht]
\centering
\includegraphics[width=1.0\textwidth]{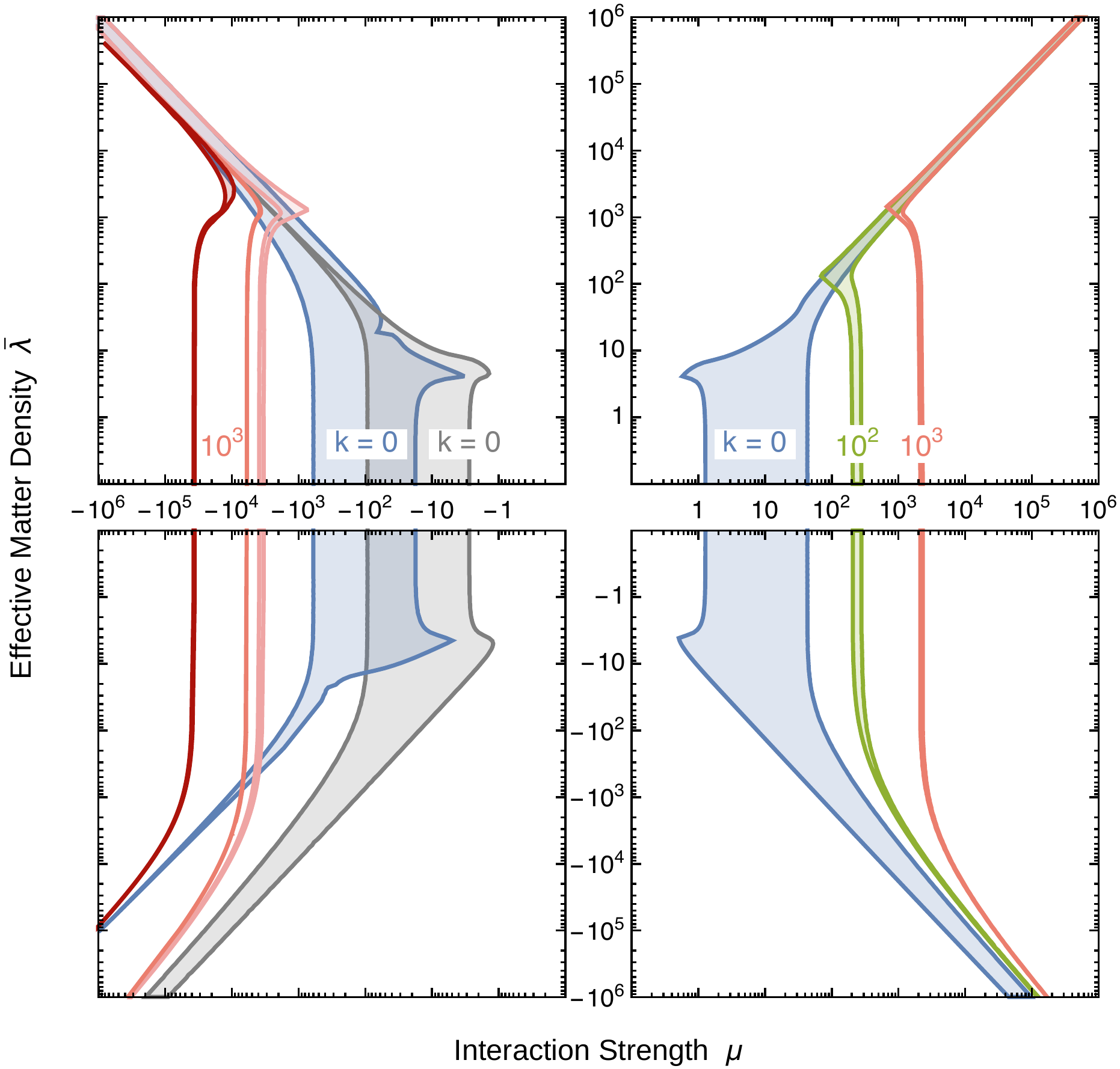}
\caption{Footprint of the 2D instabilities ($\kappa > 10^{-2}$) in the
  $\mu$-$\bar\lambda$-plane for $\alpha=1/2$ and the indicated values
  of $k$. The homogeneous case ($k=0$) is the footprint of the contour
  plot of figure~\ref{fig:2D-contour}, here on a logarithmic scale. The
  corresponding large-$\bar\lambda$ asymptotic results are shown in
  Appendix~\ref{sec:asymptotic-2D}, figure~\ref{fig:footprint-2D-asymp}.  }
\label{fig:footprint-2D-k}
\end{figure}

We next determine numerically the instability footprints for non-vanishing
wave numbers $k$ and show the result in figure~\ref{fig:footprint-2D-k},
which looks qualitatively similar to the corresponding 1D case that was shown
in figure~\ref{fig:footprint-1D-k}.  For the simpler $\mu>0$ half of the
plot, we show the instability footprints for $k=0$, $10^2$ and $10^3$ as
indicated in the plot. These $k>0$ footprints fill the space between the
$k=0$ footprint and the horizontal axis. In addition, in the upper panel,
there are small ``noses'' of the $k>0$ footprints which slightly extend in
the space above the $k=0$ footprint, but this is a very small effect.

For $\mu<0$ the situation is more complicated because there are three
instabilities. As in the 1D case, for large $k$ the footprint connects
asymptotically to the $k=0$ instabilities in a crossed-over way, which we
illustrate for $k=10^3$. The main novelty of the 2D case is the appearance of
another instability, which merges with one of the others when $\bar\lambda\gg
k$. In other words, one of the instabilities somewhat splits into two
unstable ranges. We also recall that for very large $k$, the maximum growth
rate of one of them decreases and vanishes for $k\to\infty$, in which case we
are back to a total of three instabilities. In the unphysical third quadrant,
we notice somewhat pronounced ``noses'' of the $k>0$ footprints.

In all cases the main message is the same as in the 1D case: The small-scale
instabilities fill the space between the $k=0$ instability and the horizontal
axis, whereas the space between the  $k=0$ instability and the vertical axis
remains stable.

%%%%%%%%%%%%%%%%%%%%%%%%%%%%%%%%%%%%%%%%%%%%%%%%%%%%%%%%%%%%%%%%%%%%%%%%%%%%%%
\section{Conclusions}
\label{sec:conclusions}
%%%%%%%%%%%%%%%%%%%%%%%%%%%%%%%%%%%%%%%%%%%%%%%%%%%%%%%%%%%%%%%%%%%%%%%%%%%%%%

Several recent papers \cite{Raffelt:2013isa, Duan:2014gfa, Mangano:2014zda,
Mirizzi:2015fva} have studied the phenomenon of spatial spontaneous
symmetry breaking in the ``colliding beam'' model of neutrinos interacting
with each other refractively. One important finding was that for any neutrino
density (or in our nomenclature for any value of the neutrino-neutrino
interaction energy $\mu$) there is some range of spatial wave vectors $k$
where the system is unstable with regard to self-induced flavor conversion.
As a consequence, it seemed that an interacting neutrino gas would never be
stable for any conditions, with potentially far-reaching consequences for SN
physics.

We have studied similar models, but including the multi-angle matter effect.
We concur with the previous results in that smaller-scale modes are unstable
at larger values of $\mu$ for a given matter density. This means that on a
``footprint plot'' such as figure~\ref {fig:footprint-2D-k}, modes with $k>0$
fill the space between the traditional footprint for $k=0$ and the horizontal
axis, but not the space towards the vertical axis. If we show the instability
footprint in a plot like figure~\ref{fig:firstfootprint}, adapted to more
physical SN parameters, the large-$k$ modes extend the instability region in
the direction of larger neutrino density for fixed matter density.

Therefore, if the instability footprint of the traditional homogeneous (or
rather spherically symmetric) mode does not intersect with the SN density
profile, the large-$k$ modes are safe as well. In this sense, the traditional
large-scale mode remains the most sensitive stability probe. 
On the other hand, if the physical SN profile of density and neutrino fluxes
intersects any instability region, instabilities on a large range of spatial
scales will occur and one would not expect any simple outcome of the flavor
conversion process.

Our analysis is based on a linearized stability analysis of a model which we
have developed in section~\ref{sec:EOM}. We have formulated the problem is
such a way that it includes, as special cases, the ``colliding beam''
examples of the previous literature, allows the inclusion of multi-angle
effects by a simple modification of phase-space integration, and formulates
the SN case as a 2D system evolving in time, i.e., the non-trivial dynamical
evolution is in the 2D expanding sheet of neutrinos as a function of SN
distance. All cases of the previous literature are covered in a single simple
formulation. 

In particular, our homogeneous, multi-angle, 2D
  system corresponds to the usual scenario described in the previous
  literature \cite{Raffelt:2013rqa, Chakraborty:2014lsa}, where both
  the zenith and azimuthal multi-angle effects are included. Note that
  the usual single zenith angle description of the early days of the
  collective oscillation discussions \cite{Fogli:2007bk} cover only a
  small part of the parameter space, for example the bimodal
  instabilities shown in our figure~\ref{fig:2D-contour-1}. In
  addition, for the first time we have have considered a scenario
  where inhomogeneous modes are included in the description of SN
  neutrino evolution.  We have found that the homogeneous mode remains
  the dominant source of instability.

Still, in many ways our investigation is only a mathematical case
study that may or may not apply to a realistic SN. Our main
simplification is that we assume the flavor content of the SN neutrino
stream to be stationary and to evolve only as a function of
distance. Moreover, we assume a uniform boundary condition at the
neutrino sphere, i.e., global spherical symmetry of neutrino emission.
In other words, our case study still contains substantial and
nontrivial simplifying assumptions to reduce the complexity of the
full problem. Future work will have to go beyond some of these
simplifications to develop a more realistic understanding of what
really happens to neutrino flavor in the dense SN environment.

\section*{Acknowledgments}

We acknowledge partial support by the Deutsche Forschungsgemeinschaft through
Grant No.\ EXC 153 (Excellence Cluster ``Universe'') and by the European
Union through the Initial Training Network ``Invisibles,'' Grant No.\
PITN-GA-2011-289442 and a Marie Curie Fellowship for S.C., Grant No.\
PIIF-GA-2011-299861. R.H.\ acknowledges MPP for hospitality at the beginning
of this project. S.C.\ and G.R.\ acknowledge the Mainz Institute for
Theoretical Physics (MITP) for hospitality and partial support during the
completion of this work.

\appendix

%%%%%%%%%%%%%%%%%%%%%%%%%%%%%%%%%%%%%%%%%%%%%%%%%%%%%%%%%%%%%%%%%%%%%%%%%%%%%%
\section{Analytic integrals}
\label{sec:integrals}
%%%%%%%%%%%%%%%%%%%%%%%%%%%%%%%%%%%%%%%%%%%%%%%%%%%%%%%%%%%%%%%%%%%%%%%%%%%%%%

When searching for the complex eigenvalues $\Omega$ we need various integrals
that can be found easily with Wolfram's {\sc Mathematica}. There can
be issues about the validity of the analytic expressions in the
complex plane, so we here give the integrals explicitly.

In the 1D case, for a non-vanishing $k$ and in the absence of matter
effects ($\bar\lambda=0$), we need integrals of the form
\begin{equation}
f_n(w)=\int_{-1}^{+1}dv\, \frac{v^n}{v + w}\,,
\end{equation}
where $w$ is a complex number. We first define two auxiliary functions
\begin{subequations}
\begin{eqnarray}
L(w)&=&\log\(\frac{w+1}{w-1}\)\,,
\\
A(w)&=&2+w\,\bigl[i\,\pi~{\rm sign~Im}(w) -2\, {\rm arctanh}(w)\bigr]\,.
\end{eqnarray}
\end{subequations}
The required integrals are found to be
\begin{subequations}\label{eq:analyticfunctions1}
\begin{eqnarray}
f_0(w)&=&L(w)
\,,\\
f_1(w)&=&A(w)
\,,\\
f_2(w)&=&-2w+w^2L(w)
\,,\\
f_3(w)&=&{\textstyle \frac{2}{3}}+w^2A(w)
\,,\\
f_4(w)&=&-{\textstyle \frac{2}{3}}\(w+3w^3\)+w^4 L(w)
\,.
\end{eqnarray}
\end{subequations}
The actual argument will be of the form $w=(\omega-\Omega)/k$.

For non-vanishing matter effects ($\bar\lambda\not=0$) and non-vanishing $k$,
we need integrals of the form
\begin{equation}
g_n(p,w)=\int_{-1}^{+1}dv\, \frac{v^n}{v^2+p\,v+w}\,,
\end{equation}
where $w$ is a complex number and $p$ is real. Again we define two auxiliary
functions
\begin{subequations}
\begin{eqnarray}
K(p,w)&=&\frac{1}{2}\log\(\frac{w+1+p}{w+1-p}\)\,,
\\
B(p,w)&=&\frac{1}{\sqrt{4w-p^2}}\,
\[{\rm arctan}\(\frac{2-p}{\sqrt{4w-p^2}}\)
+{\rm arctan}\(\frac{2+p}{\sqrt{4w-p^2}}\)\]\,.
\end{eqnarray}
\end{subequations}
The required integrals are found to be
\begin{subequations}\label{eq:analyticfunctions2}
\begin{eqnarray}
g_0(p,w)&=&2\,B(p,w)
\,,\\
g_1(p,w)&=&-p\,B(p,w)+K(p,w)
\,,\\
g_2(p,w)&=&2+\(p^2-2w\) B(p,w)-p\,K(p,w)
\,,\\
g_3(p,w)&=&-2p-p\(p^2-3w\) B(p,w)+\(p^2-w\) K(p,w)
\,,\\
g_4(p,w)&=&\textstyle{\frac{2}{3}}\(1+3p^2-3w\)+\(p^4-4p^2w+2w^2\)B(p,w)\
-\(p^3-2pw\)K(p,w)
\,.\kern2em
\end{eqnarray}
\end{subequations}
The actual arguments are going to be $p=2k/\bar\lambda$ and
$w=2(\omega-\Omega)/\bar\lambda$.

In the 2D case to solve equation~\eqref{eq:In-defintion-3} we need the
integrals defined in equation~\eqref{eq:In-defintion-5}, i.e., integrals
of the form
\begin{equation}
K^a_n=\int_{-\pi}^{+\pi} \frac{d\varphi}{2\pi}\int_0^1 dv
\,\frac{v^n\,f_a(\varphi)}{v^2/2+q\,v\,c_\varphi+w}\,,
\end{equation}
where $f_1(\varphi)=1$, $f_{\rm c}(\varphi)=\cos\varphi$, $f_{\rm
cc}(\varphi)=\cos^2\varphi$, and $f_{\rm
  ss}(\varphi)=\sin^2\varphi$. These integrals can be found
analytically with the help of {\sc Mathematica}. We first define
two auxiliary functions
\begin{subequations}
\begin{eqnarray}
A_{q,w}&=&\arctan\(\frac{q^2-w}{\sqrt{-w^2}}\)+\arctan\(\frac{1-2q^2+2w}{\sqrt{4q^2-(1+2w)^2}}\)
\,,\\
B_{q,w}&=&2\sqrt{-w^2}-\sqrt{4q^2-(1+2w)^2}
\end{eqnarray}
\end{subequations}
Our desired integrals are then found to be
\begin{subequations}\label{eq:analyticfunctions4}
\begin{eqnarray}
K^1_1&=&A_{q,w}\,S_w
\,,\\[1ex]
K^1_3&=&\[B_{q,w}+2\(q^2-w\)A_{q,w}\]\,S_w
\,,\\[1ex]
K^1_5&=&\[-2\sqrt{-w^2}+\(6q^2-6w+1\)B_{q,w}
+4\(3q^4 - 6 q^2 w + 2 w^2\) A_{q, w}\]\frac{S_w}{2}
\,,\\[1ex]
K^{\rm c}_2&=&\frac{1-\(B_{q,w}+2q^2A_{q,w}\)S_w}{2q}
\,,\\[1ex]
K^{\rm c}_4&=&\frac{1+\[2\sqrt{-w^2}-\(6q^2-2w+1\)B_{q,w}
-4q^2\(3q^2-4w\)A_{q,w}\]S_w}{4q}
\,,\\[1ex]
K^{\rm cc}_3&=&\frac{-1-4w-\[2\sqrt{-w^2}-\(6q^2+2w+1\)B_{q,w}
-4q^2\(3q^2-2w\)A_{q,w}\]S_w}{8q^2}
\,,
\end{eqnarray}
\end{subequations}
where $S_w=-i\,{\rm sign}({\rm Im}\,w)$.

%%%%%%%%%%%%%%%%%%%%%%%%%%%%%%%%%%%%%%%%%%%%%%%%%%%%%%%%%%%%%%%%%%%%%%%%%%%%%%
\section{Frequently encountered eigenvalue equations}
\label{sec:quadratic}
%%%%%%%%%%%%%%%%%%%%%%%%%%%%%%%%%%%%%%%%%%%%%%%%%%%%%%%%%%%%%%%%%%%%%%%%%%%%%%

Based on our ``monochromatic'' neutrino spectrum
equation~\eqref{eq:monochromaticspectrum} with vacuum oscillation
frequencies $\pm\omega_0$ and $\alpha$ the number of antineutrinos
relative to neutrinos, we constantly encounter eigenvalue equations of
the form
\begin{equation}
F\(\frac{2\tilde\mu}{\omega_0-\Omega}\)-\alpha\, F\(\frac{2\tilde\mu}{-\omega_0-\Omega}\)
=1-\alpha\,,
\end{equation}
where $\tilde\mu$ is an interaction energy. In the simplest case, $F(x)=x$,
this is the traditional eigenvalue equation for the flavor pendulum if we
notice that our $\tilde\mu=\mu(1-\alpha)/2$, whereas $\mu$ is the traditional
interaction energy. We also encounter $F(x)=\sqrt{x}$ and $F(x)=\log(x)$. To
study these equations, we transform them to dimensionless variables by the
substitutions
\begin{equation}
\tilde\mu=m\,\omega_0\frac{1+\alpha}{1-\alpha}
\quad\hbox{and}\quad
\Omega=w\,\omega_0\frac{1+\alpha}{1-\alpha}\,,
\end{equation}
leading to
\begin{equation}\label{eq:F-equation-2}
F\(\frac{2\,m}{(1-\alpha)/(1+\alpha)-w}\)
-\alpha\, F\(\frac{2\,m}{-(1-\alpha)/(1+\alpha)-w}\)
=1-\alpha\,.
\end{equation}
These substitutions allow us to easily take the limit of a
``symmetric'' neutrino distribution with $\epsilon=1-\alpha\to0$.

\begin{figure}
\centering
\includegraphics[width=0.5\textwidth]{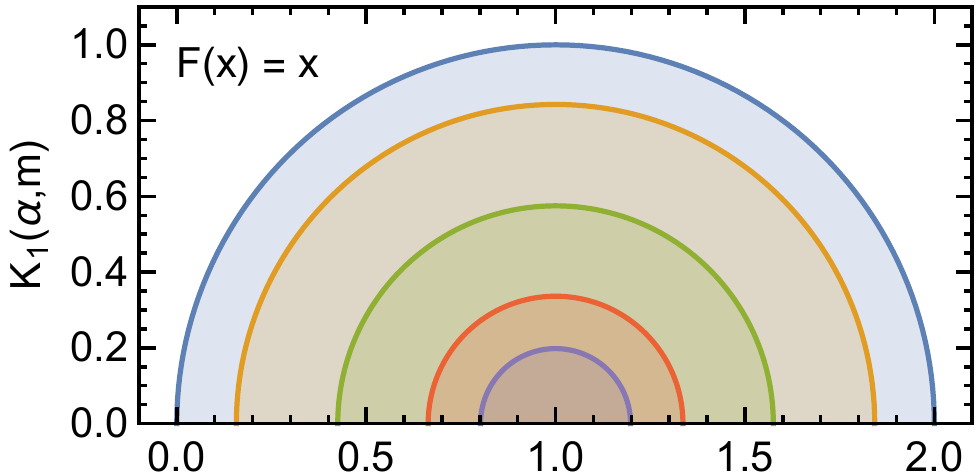}
\vskip12pt
\includegraphics[width=0.5\textwidth]{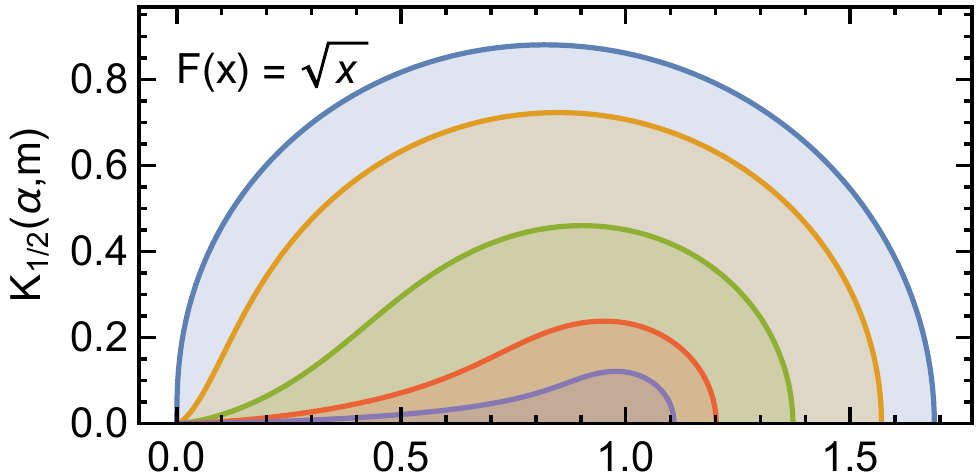}
\vskip12pt
\includegraphics[width=0.5\textwidth]{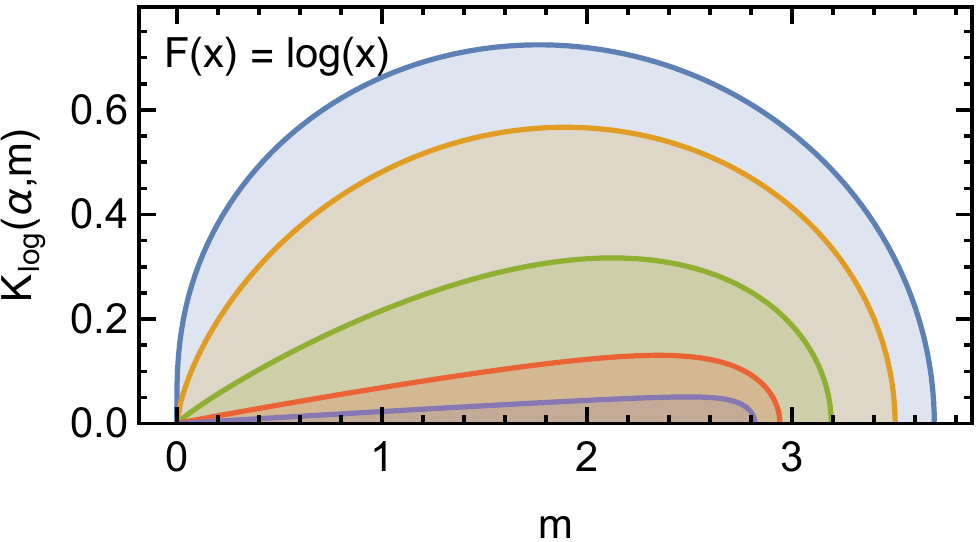}
\caption{The functions $K(\alpha,m)$ as defined in the text for
the cases $F(x)=x$, $\sqrt{x}$, and $\log(x)$ as indicated, for
$\alpha=1$, 0.3, 0.1, 0.03, and 0.01, in each case from outside in.}
\label{fig:appendixfigure}
\end{figure}

For the simplest case of the linear function $F(x)=x$,
equation~\eqref{eq:F-equation-2} becomes
a quadratic equation. It has the solution
\begin{equation}
w=-m\pm\sqrt{-(2-m)m+\(\frac{1-\alpha}{1+\alpha}\)^2}\,.
\end{equation}
It has a nonvanishing imaginary part in the range
\begin{equation}
\frac{(1-\sqrt{\alpha})^2}{1+\alpha}<m<\frac{(1+\sqrt{\alpha})^2}{1+\alpha}\,.
\end{equation}
We denote with $K_1(\alpha,m)$ the function which is the positive imaginary
part of $w$. In the present case, it simply is
\begin{equation}
K_1(\alpha,m)=\sqrt{(2-m)m-\(\frac{1-\alpha}{1+\alpha}\)^2}\,.
\end{equation}
As a function of $m$, it is a semi-circle with center at $m=1$ and radius
$2\sqrt{\alpha}/(1+\alpha)$, i.e.,
\begin{equation}
K_1^{\rm max}=\frac{2\sqrt{\alpha}}{1+\alpha}\,.
\end{equation}
This function is shown for several values of $\alpha$ in the top panel of
figure~\ref{fig:appendixfigure}.
In the limit of equal neutrino and antineutrino densities, i.e., for
$\alpha\to 1$, the imaginary part is $K(1,m) =\sqrt{(2-m)m}$. This limiting
result can be found from equation~\eqref{eq:F-equation-2} directly by substituting
$\epsilon=1-\alpha$, expanding the equation in powers of $\epsilon$, and
keeping only the lowest power, leading to the equation $2m+2mw+w^2=0$.

To compare with Sec.~\ref{sec:1D-beam-homogeneous}, notice that
$\tilde\mu=\mu(1-\alpha)/2=m\omega_0(1+\alpha)/(1-\alpha)$. Using this
relationship between $\mu$ and $m$ as well as the one between $\Omega$ and
$w$ reproduces the previous results. In particular, for $\alpha\to 1$ we find
the physical growth rate $\kappa=\sqrt{(2\mu-\omega_0)\omega_0}$ which grows
without limit for $\mu\to\infty$. It is interesting that in terms of the
scaled variable $m$ one obtains, as a limiting results for $\alpha\to1$, a
semi-circle for $K(1,m)$ as a function of $m$. So in principle one could study
all of our problems in the $\alpha\to 1$ limit, and yet obtain representative
results for $\alpha<1$, i.e., one could essentially eliminate the annoying
parameter $\alpha$ and still obtain meaningful results for the asymmetric
case.

The next more complicated case is the square-root function,
$F(x)=\sqrt{x}$, for which equation~\eqref{eq:F-equation-2}
takes on the form
\begin{equation}\label{eq:F-equation-3}
\(\frac{2\,m}{(1-\alpha)/(1+\alpha)-w}\)^{1/2}
-\alpha\, \(\frac{2\,m}{-(1-\alpha)/(1+\alpha)-w}\)^{1/2}
=1-\alpha\,.
\end{equation}
It can be transformed to a quartic equation, but the results are
too cumbersome to deal with and not particularly informative. Again we can expand this
equation in
powers of $1-\alpha$ and find, for the limit $\alpha\to 1$, the cubic equation
$2w^3+m(1+2w)^2=0$. The imaginary part of the solution is
\begin{equation}
K_{1/2}(1,m)=
\frac{8m\,(3-2 m)+(2m)^{2/3}\[8m\,(9-4 m)+3 \sqrt{81-48\, m}-27\]^{2/3}}
{4\sqrt{3}\,(2m)^{1/3}\[8m\,(9-4 m)+3 \sqrt{81-48\, m}-27\]^{1/3}}\,.
\end{equation}
This function is shown in the second panel of figure~\ref{fig:appendixfigure} and looks
almost like a semi-circle, but is not quite one. It is
is nonzero for $0<m<27/16=1.6875$ and takes on its maximum value of
$\frac{1}{16}\sqrt{\frac{3}{2}(69+11\sqrt{33})}=0.880086$ at
$m=\frac{3}{32}(3+\sqrt{33})=0.819803$.
In figure~\ref{fig:appendixfigure},
we show $K_{1/2}(\alpha,m)$ also for several other value of $\alpha$. The curves always
begin
at $m=0$, i.e., there is no lower threshold in this case.

We finally turn to the logarithmic case, $F(x)=\log(x)$, natural logarithm always
understood,
for which equation~\eqref{eq:F-equation-2} becomes
\begin{equation}\label{eq:F-equation-4}
\log\(\frac{2\,m}{(1-\alpha)/(1+\alpha)-w}\)
-\alpha\,\log \(\frac{2\,m}{-(1-\alpha)/(1+\alpha)-w}\)=1-\alpha\,.
\end{equation}
There is no general analytic solution, but again we can consider the $\alpha\to 1$
expansion where we find $1=w[1+\log(-w/2m)]$. The solution is $w(m)=1/W(-e/2m)$, where
$e$ is Euler's number and $W(z)$ is the Lambert $W$-function, i.e., $W(z)$ is the solution
of
$z=W\,e^W$. In {\sc Mathematica} it is implemented as $W(z)=\hbox{\tt ProductLog[z]}$.
The positive imaginary part of our $w(m)$, i.e., $K_{\rm log}(1,m)={\rm Im}[-1/W(-e/2m)]$,
is
nonzero for $0<m<e^2/2=3.69453$ and is shown in the bottom panel of
figure~\ref{fig:appendixfigure}. Again, it looks
deceivingly like a semi-circle, but is not exactly one. Its maximum occurs
at $m=1.76684$ and $K_{\rm log}^{\rm max}=0.724611$.

We usually consider $\alpha=1/2$ as our main example. In this case, the eigenvalue equation
can
be solved analytically with the solution
\begin{equation}
w_{\alpha=1/2}=\frac{e-3m\pm\sqrt{3m(3m-4e)}}{3 e}\,.
\end{equation}
The imaginary part is nonzero for $0<m<4e/3$, has its maximum at $m=2e/3$, and takes on
a maximum value of 2/3.

%%%%%%%%%%%%%%%%%%%%%%%%%%%%%%%%%%%%%%%%%%%%%%%%%%%%%%%%%%%%%%%%%%%%%%%%%%%%%%
\section{Asymptotic solutions for 1D and \boldmath{$\bar\lambda\to\infty$}}
\label{sec:asymptotic}
%%%%%%%%%%%%%%%%%%%%%%%%%%%%%%%%%%%%%%%%%%%%%%%%%%%%%%%%%%%%%%%%%%%%%%%%%%%%%%

We can derive analytic asymptotic solutions for the 1D case with matter, i.e.,
the large-$\bar\lambda$ continuation of the contour plot figure~\ref{fig:1D-contour}.
We begin with the bimodal and MZA instability for $\bar\lambda>0$ and consider
the first block of the eigenvalue equation \eqref{eq:1D-determinants}.
It is of the form
$1+C_1\mu+C_2\mu^2=0$, where the coefficients $C_1$ and $C_2$ depend on
$\alpha$, $\omega_0$, $\Omega$ and~$\bar\lambda$.
We have evaluated the integrals
according to the explicit transcendental functions given in
equation~\eqref{eq:analyticfunctions2}. Inspired by numerical
solutions, we assume that both the real and imaginary parts of the solutions
$\Omega$ remain of order $\omega_0$ and do not become large as
$\bar\lambda\to\infty$, an assumption that later bears out to be consistent
with the solutions. Therefore, we may expand $C_1$ and $C_2$ in powers of
$\bar\lambda^{-1}$ and find that the dominant terms are
$C_1\propto\bar\lambda^{-1}$ and $C_2\propto\bar\lambda^{-3/2}$. In terms
of a dimensionless interaction strength $\hat\mu$ of order unity we write
\begin{equation}\label{eq:mu-scaling-1}
\mu=\frac{\hat\mu}{1-\alpha}\,(6/\pi)^{1/2}\,(2\omega_0)^{1/4}\, \bar\lambda^{3/4}\,,
\end{equation}
where the exact coefficient was chosen for later convenience. The
lowest-order term in $C_2\mu^2$ no longer depends on $\bar\lambda$,
whereas the lowest-order term in $C_1\mu\propto \bar\lambda^{-1/4}$
and slowly becomes small as $\bar\lambda\to\infty$.
To lowest order in $\bar\lambda^{-1}$, the eigenvalue equation is found to be
\begin{equation}
\sqrt{\frac{\omega_0}{\omega_0-\Omega}}-\alpha\sqrt{\frac{\omega_0}{-\omega_0-\Omega}}
=\frac{1-\alpha}{\hat{\mu}^2}\,.
\end{equation}
This is an example of the type of equations that we always encounter
in this context and which are discussed in Appendix~\ref{sec:quadratic}.

\begin{figure}
\centering
\hbox to\textwidth{\includegraphics[height=4.9cm]{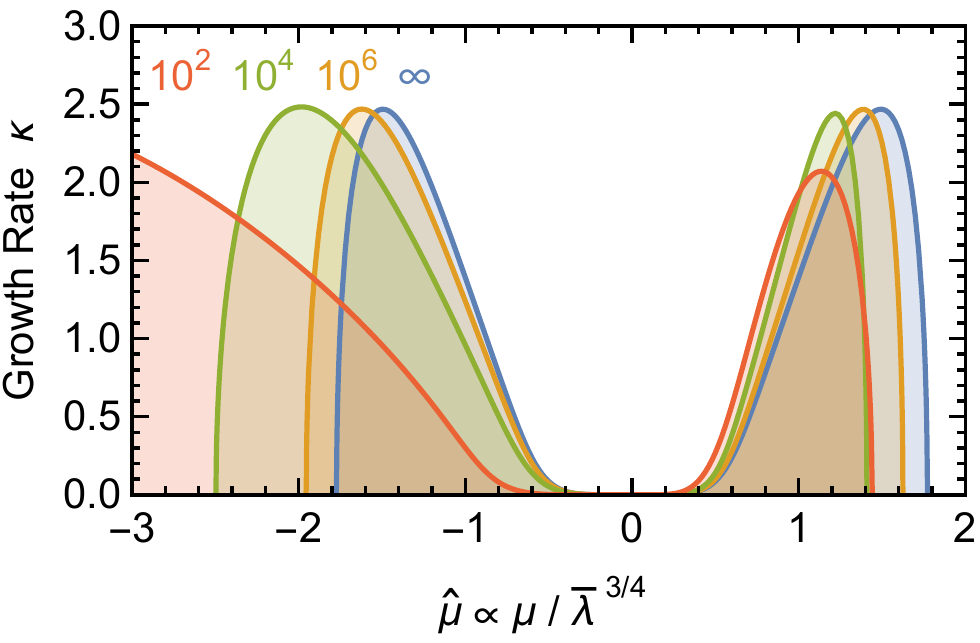}
\hfil\includegraphics[height=4.9cm]{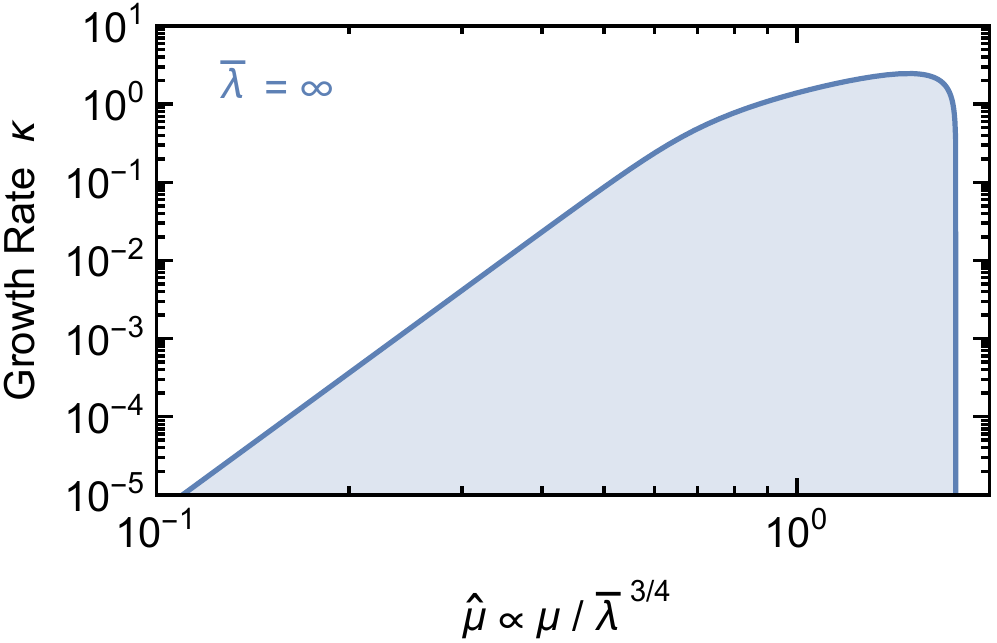}}
\caption{Growth rate $\kappa$ of the bimodal and MAA instabilities, using $\alpha=1/2$.
The interaction strength is scaled according to equation~\eqref{eq:mu-scaling-1}.
The blue curves show the asymptotic behavior for $\bar\lambda\to\infty$, the other curves
are for the indicated $\bar\lambda$ values.}
\label{fig:Asymptotic-bimodal}
\end{figure}

The asymptotic solution derives from the term quadratic in $\mu$ and thus
remains unchanged under $\mu\to-\mu$, i.e., it applies to both hierarchies.
We show the asymptotic solution as a blue curve in
figure~\ref{fig:Asymptotic-bimodal} on a linear and logarithmic scale. We also
show the growth rates for $\bar\lambda=10^2$, $10^4$ and $10^6$ where the
solution is not symmetric under $\mu\to-\mu$ because the linear term in $\mu$
kicks in. We have already noted that one needs very large $\bar\lambda$
values to obtain the asymptotic solution because the second-largest term only
scales with $\bar\lambda^{-1/4}$ relative to the dominant term. The
asymptotic behavior is achieved for much smaller $\bar\lambda$ values if
$\mu>0$. The growth rate vanishes completely above a certain $|\mu|$ value,
but obtains nonzero values otherwise, i.e., there is no lower $\hat\mu$
threshold. However, for $\hat\mu\lesssim0.5$, the growth rate is a steep
power-law of $\hat\mu$ and can be taken to be effectively zero.

\begin{figure}[!b]
\centering
\includegraphics[width=0.96\textwidth]{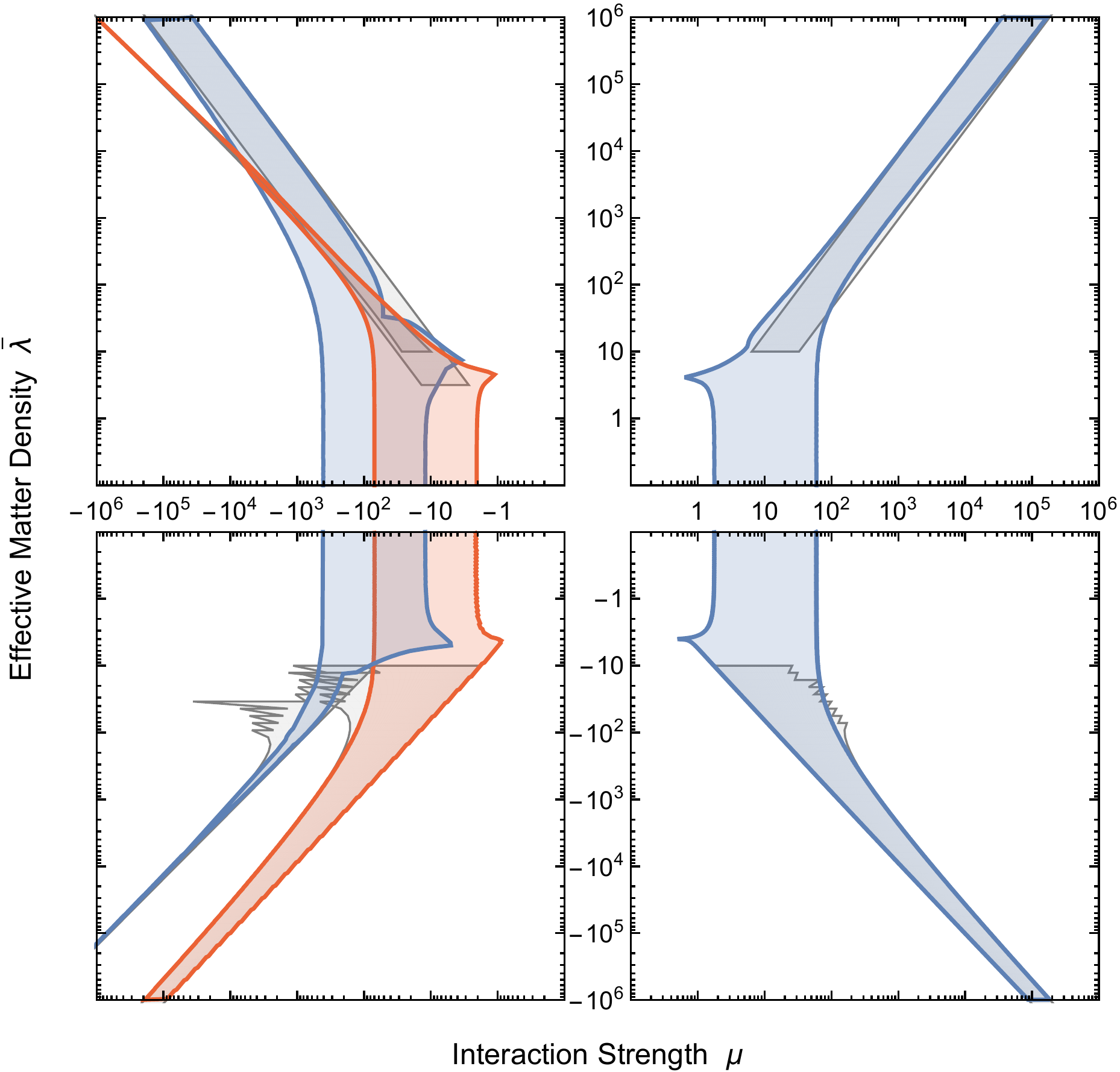}
\caption{Footprint of the 1D instabilities in the $\mu$-$\bar\lambda$
  plane for $k=0$ (homogeneous mode) and $\alpha=1/2$ as explained in
  the text. The colored regions derive from a numerical solution,
  where the blue footprints correspond to the $2{\times}2$ block in
  equation~\eqref{eq:1D-determinants}, the red solutions to the
  $1{\times}1$ block. The grey regions show the asymptotic solutions
  in the large-$\bar\lambda$ limit derived in this appendix.}
\label{fig:footprint-1D-asymp}
\end{figure}

For our usual example $\alpha=1/2$ we find numerically that the maximum growth
rate occurs for $\hat\mu=1.494$. Therefore, we find that
\begin{equation}
\mu=\pm4.911\,\omega_0^{1/4}\,\bar\lambda^{3/4}
\end{equation}
gives us the locus of the maximum growth
rate in the $\mu$-$\bar\lambda$ plane for the bimodal and MAA solutions.
The maximum value of $\hat\mu$ before the growth rate becomes zero is
1.7724. On the small-$\hat\mu$ side, the growth rate drops below $\kappa<1/100$, our
usual criterion, at $\hat\mu=0.3478$. Therefore, the footprint of the
instability is the region between the lines $\mu=1.143\,\lambda^{3/4}$ and
$5.826\,\lambda^{3/4}$, where both $\mu$ and $\lambda$ are given in units of the vacuum
oscillation frequency $\omega_0$. This footprint is shown in the first
quadrant (upper right) of figure~\ref{fig:footprint-1D-asymp}.
The corresponding footprint in the second quadrant (upper left) is also shown.

\begin{figure}[ht]
\centering
\includegraphics[width=0.5\textwidth]{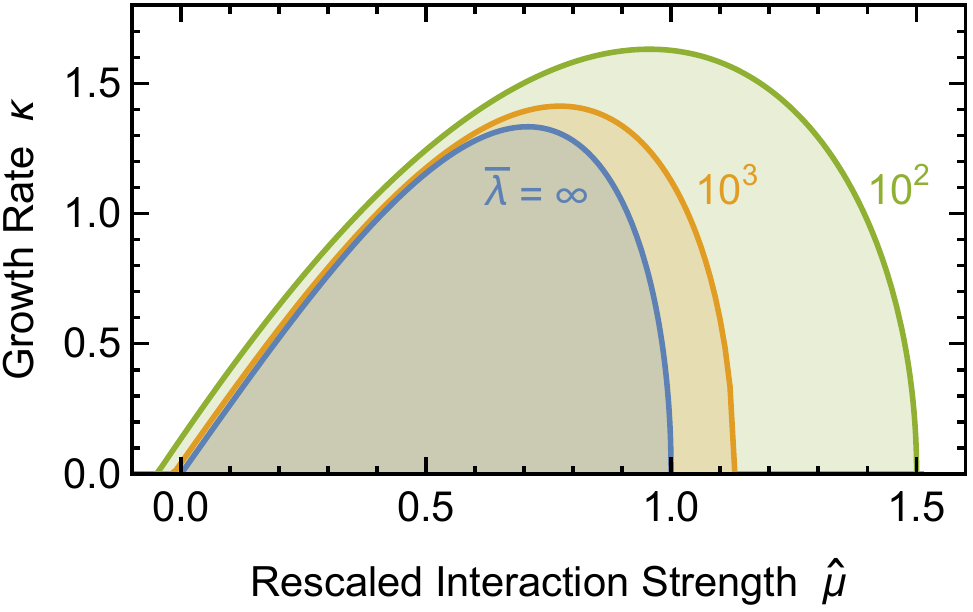}
\caption{Asymptotic growth rate $\kappa$ for the MZA instability, using $\alpha=1/2$.
The interaction strength is scaled according to equation~\eqref{eq:mu-scaling-2}.
The blue curve shows the asymptotic behavior for $\bar\lambda\to\infty$
according to equation~\eqref{eq:MZA-solution}, the other curves are
for $\bar\lambda=10^2$ and $10^3$, from outside in.}
\label{fig:Asymptotic-MZA}
\end{figure}

We next turn to the MZA solution which exists only in inverted hierarchy ($\mu<0$)
and we consider the second block in equation~\eqref{eq:1D-determinants}.
If we use
$\mu=-\bar\lambda/[2(1-\alpha)]$ the leading terms cancel,
leaving us with a leading term of order
$\bar\lambda^{-1/2}$.
To obtain the lowest-order equation, we introduce
another dimensionless parameter $\hat\mu$ and write
\begin{equation}\label{eq:mu-scaling-2}
\mu=\frac{-\bar\lambda}{2\,(1-\alpha)}
-\hat\mu\,\frac{\pi\sqrt{1-\alpha^2}}{2\,(1-\alpha)^2}\,\sqrt{\omega_0\bar\lambda}\,,
\end{equation}
where, of course, the detailed coefficients in the second term are chosen for
later convenience. One then finds a quadratic equation with solutions
\begin{equation}\label{eq:MZA-solution}
\frac{\Omega}{\omega_0}=\frac{1+\alpha^2-2\hat\mu^2\,(1+\alpha^2)}{1-\alpha^2}
\pm i\,\frac{4\alpha}{1-\alpha^2}\,\sqrt{\hat\mu^2(1-\hat\mu^2)}\,.
\end{equation}
Notice that these solutions require $0\leq\hat\mu\leq1$ and we have always
assumed $0\leq\alpha\leq1$. The imaginary part, as a function of $\hat\mu^2$,
has the familiar semi-circular shape. In figure~\ref{fig:Asymptotic-MZA} we
show it as a function of $\hat\mu$ (blue curve) and we also show the full
solution for $\bar\lambda=10^3$ and $10^2$. The asymptotic solution is quite
good for relatively small $\bar\lambda$ values. In contrast to the other
solutions, on a logarithmic scale the unstable range becomes very narrow as
$\bar\lambda\to\infty$.

The maximum growth rate obtains for $\hat\mu=1/\sqrt{2}$. Therefore, for our usual example
$\alpha=1/2$ we find that for the MZA solution,
\begin{equation}
\mu=-\bar\lambda-\pi\sqrt{3\omega_0\bar\lambda/2}
\end{equation}
gives us the locus of the maximum growth rate in the $\mu$-$\bar\lambda$ plane.
The growth rate becomes exactly zero for $\hat\mu\leq0$ and $\hat\mu\geq1$, so the footprint
(see upper-left quadrant in figure~\ref{fig:Asymptotic-MZA})
is delimited by the curves $\mu=-\bar\lambda$ and
$\mu=-\bar\lambda-\pi\sqrt{3\bar\lambda\omega_0}$.
The width of the footprint
scales with $\sqrt{\bar\lambda}$, i.e., on a logarithmic scale it becomes very narrow
for large $\bar\lambda$.

For $\bar\lambda<0$, the above approach does not lead to unstable solutions. Numerically
we observe that for $\bar\lambda\to-\infty$, the real part of the solutions
approaches ${\rm Re}(\Omega)\to\bar\lambda/2$, i.e., a large negative number.
Therefore, to be able to expand the equation, we express $\Omega=\bar\lambda/2+w\,\omega_0$
and seek self-consistent solutions with the dimensionless eigenvalue $w$
of order unity. After expansion for $\bar\lambda\to-\infty$, the
asymptotic eigenvalue equations are
\begin{equation}\label{eq:1D-asymp-eigenvalue1}
\frac{\log(1-w)-\alpha\log(-1-w)}{1-\alpha}=a
\end{equation}
where
\begin{equation}\label{eq:1D-asymp-a-expressions}
a=\log\(-\frac{2\bar\lambda}{e^2\omega_0}\)-\frac{1}{\hat\mu}
\qquad\hbox{or}\qquad
a=\log\(-\frac{2\bar\lambda}{e^2\omega_0}\)+\frac{3+\hat\mu^2}{(3-\hat\mu)\hat\mu}\,,
\end{equation}
where $e$ is Euler's number and as always the logarithm is with base $e$.
This equation is one example for the type discussed in Appendix~\ref{sec:quadratic}.

For our usual example $\alpha=1/2$, we can solve this equation analytically with the
explicit
result
\begin{equation}
w_{\alpha=1/2}=\frac{2-e^a\pm i\sqrt{e^a(8-e^a)}}{2}\,.
\end{equation}
It has a nonzero imaginary part for $-\infty<a<\log(8)=2.0794$, although it
becomes exponentially small for $a\ll -1$. The maximum imaginary part obtains for
$a=\log(4)=1.3863$ and the maximum is 2. Therefore, the maximum growth rate obtains for
\begin{equation}
A=\frac{1}{\hat\mu}
\qquad\hbox{or}\qquad
A=-\frac{3+\hat\mu^2}{(3-\hat\mu)\hat\mu}\,,
\end{equation}
where
\begin{equation}
A=-\log(4)-2+\log\(-\frac{2\bar\lambda}{\omega_0}\)=\log\(\frac{-\bar\lambda}{2
e^2\omega_0}\)\,.
\end{equation}
Therefore, we have altogether three solutions, corresponding to the three instabilities,
with
maximum growth
rates on the locus in the $\mu$-$\bar\lambda$ plane given by
\begin{subequations}
\begin{eqnarray}
\mu&=&-2\bar\lambda\,\frac{6}{3A+\sqrt{3(A+2)(3A-2)}}\to-2\bar\lambda\,\frac{1}{A}
\,,\\[0ex]
\mu&=&+2\bar\lambda\,\frac{1}{A}\,,\\[0ex]
\mu&=&+2\bar\lambda\,\frac{3A+\sqrt{3(A+2)(3A-2)}}{2\,(A-1)}\to 6\bar\lambda\,,
\end{eqnarray}
\end{subequations}
where the limiting behavior is understood for $A\to\infty$. Because $\bar\lambda\to-\infty$,
the
first solution corresponds to positive $\mu$ and thus to the bimodal solution, the second
and
third solutions
are the MAA and MZA instabilities, respectively.

To draw the footprints in the lower quadrants of
figure~\ref{fig:footprint-1D-asymp}, we notice that $\kappa=0$ for
$a>\log(8)$ and on the other side $\kappa<1/100$ for $a<\log(4 -
\sqrt{39999}/50)=-9.90348$.  Therefore, the asymptotic footprints are
limited by
\begin{equation}\label{eq:1D-limiting-a}
a_1=\log(8)
\quad\hbox{and}\quad
a_2=\log(4 - \sqrt{39999}/50)
\end{equation}
from which the limiting curves are extracted by solving
equation~\eqref{eq:1D-asymp-a-expressions} for $\hat\mu$.  Once more we
note that two of the footprints are ``wide'' and nearly symmetric
between $\mu\to-\mu$, whereas the third instability has a very narrow
footprint.

%%%%%%%%%%%%%%%%%%%%%%%%%%%%%%%%%%%%%%%%%%%%%%%%%%%%%%%%%%%%%%%%%%%%%%%%%%%%%%
\section{Asymptotic solutions for 2D with \boldmath{$\bar\lambda=0$} and
  \boldmath{$k\to\infty$}}
\label{sec:2D-k-asymp}
%%%%%%%%%%%%%%%%%%%%%%%%%%%%%%%%%%%%%%%%%%%%%%%%%%%%%%%%%%%%%%%%%%%%%%%%%%%%%%

We are looking for the large-$k$ solutions of the 2D case without matter ($\bar\lambda=0$).
We need to find the zeroes of the determinant of the matrix in
equation~\eqref{eq:2D-matrix-equation-2}. We first
look at the $3\times3$ block and calculate it according to the explicit
integrals that we have found. Next we substitute the variables as
$\omega=1$, $\alpha=1/2$, $\Omega=-k+x$, and
\begin{equation}
\mu=a\,(k+m\sqrt{k})\,,
\end{equation}
where $a$ is a coefficient to be determined and overall the substitution for
$\mu$ is an educated guess. Except for the choice $\alpha=1/2$,
everything is still completely general. The unknown frequency to be found
is $x$. Its imaginary part is the growth rate which we are looking for.
The parameter $m$ is an effective interaction strength because it gives us
$\mu$ in this parameterised form.

Next we expand the determinant as a power series for large $k$ and find to lowest
nontrivial order
\begin{eqnarray}
\hbox{det($3{\times3}$ block)}&=&\frac{2880-480\,a - 424\,a^2 - 11\,a^3}{2880}\nonumber\\
&-&\frac{3\,i}{320}\[\sqrt{2} a^2 (32 + a) \(2 \sqrt{x-1} - \sqrt{x+1}\)\]\frac{1}{\sqrt{k}}
+{\cal O}(1/k)\,.
\end{eqnarray}
For term proportional to $1/\sqrt{k}$ to dominate we demand the first
term to vanish, giving us three possible values for $a$ from the requirement
$2880-480\,a - 424\,a^2 - 11\,a^3=0$. The explicit results are quite complicated
expressions. Numerically one finds
\begin{eqnarray}
a_1&=&-37.1825\,,\\
a_2&=&-3.42115\,,\\
a_3&=&+2.05821\,.
\end{eqnarray}
In other words, we have three asymptotic solutions, where one is for positive
$\mu$ and two for negative $\mu$ as expected.

If we now imagine that $a$ is one of these solutions, the first term in the determinant
vanishes and in the second term we can substitute $a^3=(2880-480\,a - 424\,a^2)/11$ to
remove the $a^3$ term. In anticipation of the result we further introduce the quantity
\begin{equation}
m_{\rm max}=\frac{162\sqrt{3}\[120-a\(20+3a\)\]}{11\,\[1080-a\(120 + 53a\)\]}\,,
\end{equation}
which for our three possible $a$ values are numerically
\begin{eqnarray}
m_{\rm max,1}&=&1.23675\,,\\
m_{\rm max,2}&=&4.49396\,,\\
m_{\rm max,3}&=&2.77208\,.
\end{eqnarray}
Then we are left with the equivalent of the determinant equation
\begin{equation}
\sqrt{6}\,m=i\, m_{\rm max}\(2\sqrt{x-1}-\sqrt{x+1}\)\,.
\end{equation}
It has the explicit solutions
\begin{equation}
x=\frac{5}{3}-\frac{10\,m^2}{3m_{\rm max}^2}
\pm\frac{8\sqrt{m^2(m^2-m_{\rm max}^2)}}{3m_{\rm max}^2}\,.
\end{equation}
The solution has an imaginary part for $0<m<m_{\rm max}$.
Therefore, the large-$k$ footprint of the three instabilities is limited
by the lines
\begin{equation}
\mu=a_i\,k
\quad\hbox{and}\quad
\mu=a_i\,\(k+m_{{\rm max},i}\sqrt{k}\)\,.
\end{equation}
For $\mu$-values between these lines, the system is unstable.

Finally we turn to the $1{\times}1$ block in equation~\eqref{eq:2D-matrix-equation-2}.
We proceed with the same substitutions except for
\begin{equation}
\mu=a\,(k+b)\,,
\end{equation}
where for the moment we leave open what $b$ is supposed to mean. Expanding the
$1{\times}1$ block determinant in powers of large $k$, we here find
\begin{eqnarray}
\hbox{det($1{\times1}$
block)}&=&\frac{6+a}{6}+\frac{a}{6}\,(-9+b+3x)\,\frac{1}{k}\nonumber\\
&+&i\,\frac{2\sqrt{2}\,a}{3}\,\[2(x-1)^{3/2}-(x+1)^{3/2}\]\frac{1}{k^{3/2}}
+{\cal O}(1/k^2)\,.
\end{eqnarray}
Again we can get rid of the first term, this time by setting $a=-6$, i.e., the
footprint of this instability is for negative $\mu$. The remaining equation is
\begin{equation}
\hbox{det($1{\times1}$ block)}=(9-b-3x)\,\frac{1}{k}
-i\,4\sqrt{2}\,\[2(x-1)^{3/2}-(x+1)^{3/2}\]\frac{1}{k^{3/2}}
+{\cal O}(1/k^2)\,.
\end{equation}
The leading term does not provide an imaginary solution. In other words,
for very large $k$ we do not have an instability. If we keep both the leading
and next to leading term, we finally need to solve the equation
\begin{equation}
(9-b-3x)\,\sqrt{k}=i\,4\sqrt{2}\,\[2(x-1)^{3/2}-(x+1)^{3/2}\]\,.
\end{equation}
Solving this equation actually leads to an asymptotic solution where
the growth rate exists for a range of $b$-values. However, the maximum
growth rate decreases with $1/\sqrt{k}$.
Therefore, we have overall four instabilities, but for $k\to \infty$ the one from the single
block disappears.

%%%%%%%%%%%%%%%%%%%%%%%%%%%%%%%%%%%%%%%%%%%%%%%%%%%%%%%%%%%%%%%%%%%%%%%%%%%%%%
\section{Asymptotic solutions for 2D with \boldmath{$k=0$} and
\boldmath{$\bar\lambda\to\infty$}}
\label{sec:asymptotic-2D}
%%%%%%%%%%%%%%%%%%%%%%%%%%%%%%%%%%%%%%%%%%%%%%%%%%%%%%%%%%%%%%%%%%%%%%%%%%%%%%

We can derive asymptotic solutions for the 2D case with matter, i.e.,
the large-$\bar\lambda$ solutions of the eigenvalue
equation~\eqref{eq:2D-matrix-equation-3}, corresponding to the two
equations~\eqref{eq:2D-determinants}. We begin with the $2{\times}2$
block and $\bar\lambda\to+\infty$. As in the 1D case, we assume that
$\Omega$ remains of order $\omega_0$, an assumption which is confirmed
by the results.  We express the interaction strength in terms of a
dimensionless parameter $\hat\mu$ in the form
\begin{equation}\label{eq:2D-mu-parameter}
\mu=\frac{\hat\mu}{1-\alpha}\,\bar\lambda\,.
\end{equation}

\begin{figure}[ht]
\centering
\includegraphics[width=0.96\textwidth]{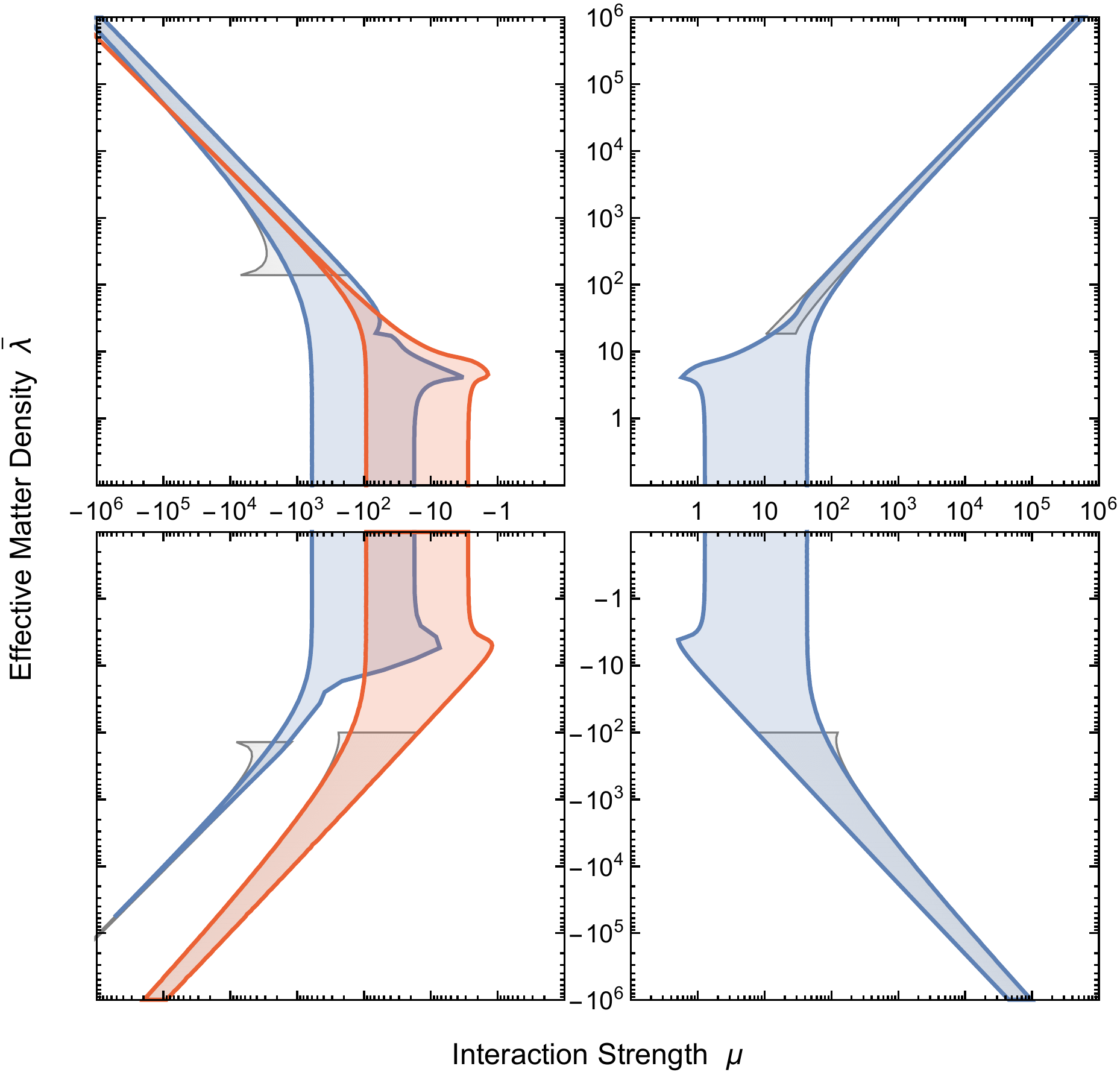}
\caption{Footprint of the 2D instabilities in the $\mu$-$\bar\lambda$
  plane for $k=0$ (homogeneous mode) and $\alpha=1/2$ as explained in
  the text. The colored regions derive from a numerical solution,
  where the blue footprints correspond to the $2{\times}2$ block in
  equation~\eqref{eq:2D-determinants}, the red solutions to the
  $1{\times}1$ block. The grey regions show the asymptotic solutions
  in the large-$\bar\lambda$ limit derived in this appendix.}
\label{fig:footprint-2D-asymp}
\end{figure}
To lowest order in $\bar\lambda^{-1}$ the eigenvalue equation is,
using $w=\Omega/\omega_0$,
\begin{equation}\label{eq:2D-asymp-eigenvalue1}
\frac{\log(1-w)-\alpha\log(-1-w)}{1-\alpha}=a\,,
\quad\hbox{where}\quad
a=\log\(\frac{\bar\lambda}{2\omega_0}\)-\frac{2(\hat\mu-1)^2}{\hat\mu^2}\,.
\end{equation}

This result is identical with equation~\eqref{eq:1D-asymp-eigenvalue1}, but
with a different expression for $a$.  To draw the asymptotic
footprints we simply need to solve for $\hat\mu$ using the limiting
$a$-values given in equation~\eqref{eq:1D-limiting-a}. The result is shown
in figure~\ref{fig:footprint-2D-asymp} as grey shaded regions in the
upper panels, to be compared with the blue regions which derive from a
numerical solution of the full eigenvalue equations.

\begin{figure}[b]
\centering
\includegraphics[width=0.5\textwidth]{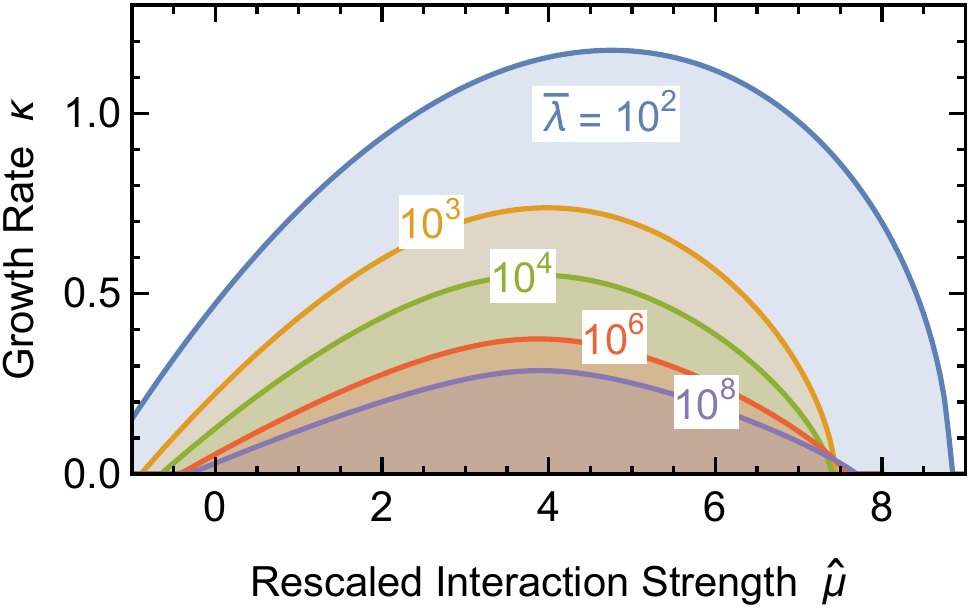}
\caption{Growth rate $\kappa$ for the instability
deriving from the $1{\times}1$ block in equation~\eqref{eq:2D-determinants},
using $\alpha=1/2$, and solving the full eigenvalue equation.
The interaction strength is scaled according to
equation~\eqref{eq:2D-mu-parameter-2}.
The curves are for the indicated values of $\bar\lambda$.}
\label{fig:Asymptotic-2D-1block}
\end{figure}

Next we turn to the $1{\times}1$ block for the limit
$\bar\lambda\to+\infty$ and express the interaction strength in the form
\begin{equation}\label{eq:2D-mu-parameter-2}
\mu=-\frac{\bar\lambda+\hat\mu\omega_0\log(\bar\lambda/2\omega_0)}{1-\alpha}\,.
\end{equation}
With $\hat\mu=0$ the eigenvalue equation is identically fulfilled to lowest
order in $\bar\lambda^{-1}$, i.e., to lowest order unstable solutions require
$\mu=-\lambda/(1-\alpha)$. This simple behavior indeed corresponds to the
very ``thin'' footprint shown in red in the upper left panel of
figure~\ref{fig:footprint-2D-asymp}. Including $\hat\mu\not=0$ leads to an
approximate eigenvalue equation which is not very simple and does not lead to
simple asymptotic solutions. Expressing $\mu$ in terms of $\hat\mu$ as in
equation~\eqref{eq:2D-mu-parameter-2} we can numerially find the growth rate
$\kappa$ as a function of $\hat\mu$ as shown in
figure~\ref{fig:Asymptotic-2D-1block}. It is clear that the instability
footprint in the logarithmic figure~\ref{fig:footprint-2D-asymp} will be very
narrow. We also notice that the maximum growth rate decreases with increasing
$\bar\lambda$. (For all of the other instabilities and for $\alpha=1/2$, the
maximum growth rate $\kappa_{\rm max}=2\,\omega_0$.)

For the next cases we turn to the limit $\bar\lambda\to-\infty$. In this
limit, we write $\Omega=\bar\lambda/2+w\,\omega_0$ in analogy to the 1D case.
In the $\bar\lambda\to-\infty$ limit, the eigenvalue is characterized by $w$
values of order unity. We also write the interaction strength again in the
form of equation~\eqref{eq:2D-mu-parameter}. The limiting eigenvalue equation
is the same as in equation~\eqref{eq:2D-asymp-eigenvalue1}, but now with
\begin{equation}
a=\log\(-\frac{\bar\lambda}{2\omega_0}\)-\frac{2(\hat\mu-1)^2}{(\hat\mu-4)\,\hat\mu}
\qquad\hbox{or}\qquad
a=\log\(-\frac{\bar\lambda}{2\omega_0}\)-\frac{\hat\mu+1}{\hat\mu}\,,
\end{equation}
where the first expression applies to the $2{\times}2$ block, the second to
the $1{\times}1$ block of the eigenvalue matrix. As before, to draw the
asymptotic footprints we solve for $\hat\mu$ using the limiting $a$-values
given in equation~\eqref{eq:1D-limiting-a}. The result is shown in
figure~\ref{fig:footprint-2D-asymp} as grey shaded regions in the lower
panels, to be compared with the blue and red regions which derive from a
numerical solution of the full equations.

\begingroup\raggedright

\endgroup

%%%%%%%%%%%%%%%%%%%%%%%%%%%%%%%%%%%%%%%%%%%%%%%%%%%%%%%%%%%%%%%%%%%%%%
\end{document}